\newcommand{\hkpc}{\mbox{ } {h}^{-1}~\rm{ kpc}}

\newcommand{\msol}{\mbox{ } {M}_{\odot}}

\newcommand{\msig}{{M}_{\rm bh}-\sigma}
\newcommand{\mhalo}{{M}_{\rm bh}-{M}_{\rm tot}}

\newcommand{\kpc}{{\rm kpc}}
\newcommand{\mg}{\mbox{ }{\rm \mu G}}

\newcommand{\mgmyr}{~\mu {\rm G}~{\rm Myr}^{-1}}

\newcommand{\eqref}[1]{Eq.\ #1}
\newcommand{\figref}[1]{Figure #1}
\newcommand{\tabref}[1]{Table #1}

\documentclass[useAMS,usenatbib]{mn2e}
\usepackage{epsfig,subfigure}

\bibpunct[; ]{(}{)}{;}{a}{}{,}

\title{An examination of magnetized outflows from active galactic nuclei in galaxy clusters}

\author[P. M. Sutter et al.]
  {P.~M.~Sutter$^1$ \thanks{Email: psutter2@illinois.edu},
   H.-Y.~Karen~Yang$^{2}$, 
          P.~M.~Ricker$^{2}$, 
          G.~Foreman$^{2}$, 
          and D.~Pugmire$^{3}$
  \\
$^{1}$Department of Physics,
	  University of Illinois at Urbana-Champaign,
      Urbana, IL 61801-3080
  \\
$^{2}$Department of Astronomy,
      University of Illinois at Urbana-Champaign,
      Urbana, IL 61801
  \\
$^{3}$Oak Ridge Leadership Computing Facility, 
       Oak Ridge National Laboratory, 
       Oak Ridge TN 37831}

\begin{document}

         
\maketitle
         
\label{firstpage}
            
\begin{abstract} 
We present 3D adaptive mesh refinement MHD simulations of an isolated galaxy
cluster that include injection of kinetic, thermal, and magnetic energy via a
central active galactic nucleus (AGN) in order to study and evaluate the role
that AGN may play in producing the observed cluster-wide magnetic fields.
Using the MHD solver in FLASH 3.3, we compare several sub-resolution approaches
to the evolution of AGN, specifically focusing on large-scale jet and bubble
models.  We examine the effects of magnetized outflows on the accretion history
of the black hole and cluster thermodynamic properties, discuss the ability of
various models to magnetize the cluster medium, and assess the sensitivity of
these models to their underlying subgrid parameters.  We find that magnetized
jet-based models suffer a severe reduction in accretion rate compared to
hydrodynamic jets; however, bubble models remain largely unaffected.  While
both jets and sporadically-placed bubbles have difficulty reproducing the
observed strength and topology of cluster magnetic fields, models based on
centrally-located bubbles come closest to observations.  Finally, whereas jet
models are relatively insensitive to changes in their subgrid parameters, the
accretion rate and average magnetic field produced by the bubbles vary by as
much as an order of magnitude  depending on the grid resolution and accretion
strength.  
\end{abstract}


\section{Introduction}
\label{sec:agn_introduction}

Radio observations of clusters of galaxies indicate that they host large-scale,
volume-filling diffuse magnetic fields of strength $0.1-10 \mu$G, while
additional rotation measure observations suggest that these fields are tangled
with auto-correlation lengths of $10$-$20$~kpc (see~\citet{Carilli2002a} for a
review) and have complex topology~\citep{Falceta-Goncalves2010}.  These fields
are extremely important in understanding cluster astrophysics: they provide a
source of non-thermal pressure support~\citep{Dolag2000}, allow for synchrotron
emission from cosmic
rays~\citep{Miniati2001,Pfrommer2007,Brunetti2007,Skillman2008}, potentially
suppress or modify thermal
conduction~\citep{Balbus2000,Balbus2001,Parrish2009a}, and modify turbulence in
the cluster atmosphere~\citep{Narayan2001a,Chandran2004a,Shukurov2006}.
However, despite their ubiquity and importance we do not at present understand
their origins and evolution or the precise correlations of these fields with
other cluster properties.

While exotic processes in the early universe may generate these large-scale
fields~\citep[e.g.,][]{Baym1996,Bamba2008,Battefeld2008}, magnetic fields may
have difficulty surviving the radiation-dominated era due to very high
diffusion rates~\citep{Lesch1998}.  Alternatively, we may turn to astrophysical
mechanisms, and especially active galactic nuclei (AGN), for seeding and
amplifying magnetic fields. In this scenario, weak seed fields ($\sim
10^{-18}$~G) are generated via a plasma process, such as the Biermann battery
mechanism~\citep{BiermannL.1950,Widrow2002a}.  Dynamo action in supermassive
black hole (SMBH) accretion disks can quickly amplify these weak fields and
expel them into the intracluster medium via relativistic
jets~\citep{Koide1999}.  The jets can propagate tens of kiloparsecs into the
intracluster medium~\citep{Kirkpatrick2011} and inflate bubbles when these jets
slow and reach approximate pressure balance with their
surroundings~\citep{Colbert1996}.  The bubbles rise into the intracluster
medium (ICM) and eventually disperse, distributing heat~\citep{Voit2005a} and
potentially magnetic fields within cluster cores. AGN outflows are more than
powerful enough to match the observed magnetic energy in
clusters~\citep{Colgate2000} and are natural carriers of magnetic
flux~\citep{Daly1990}.  Rotation measure observations of jets indicate the
presence of magnetic fields in them~\citep{Contopoulos2009a}.  AGN feedback is
also a strong candidate for preventing excessive cooling in cluster
cores~\citep{McNamara2007}.

However, the complex physics of AGN accretion disks and jets coupled with the
small scales involved ($\sim 100$~pc) makes it difficult to include AGN in
cluster and cosmological simulations, which typically resolve scales greater
than $2$~kpc. Thus, we must include the accretion and feedback processes as
subgrid models.  Subgrid models have been developed and employed by many
authors with varying levels of sophistication.  Accretion rate calculations can
vary from the simple Bondi rate~\citep{BondiH.1952} to estimates of viscous
angular momentum transport~\citep{DeBuhr2010} to fully stochastic
models~\citep{Pope2007}.  If the resolution and computing resources permit it,
large-scale jets can be placed on the simulation grid with many modifications,
including simple fluxes at cell boundaries~\citep{Gaspari2011},
limited-lifetime jets~\citep{Morsony2010}, extended jets~\citep{Cattaneo2007a},
wide-angle jets~\citep{Sternberg2007}, and precessing
jets~\citep{Sternberg2008,Falceta-Goncalves2010b}.  Since jets eventually
inflate bubbles, it is easier computationally to simply place already-formed
bubbles~\citep[e.g.,][]{Sijacki2007,Gardini2007,DiMatteo2008,booth_cosmological_2009}
or to slowly inflate them~\citep[e.g.,][]{Jones2005}.  Whereas jets
continuously feed back energy and momentum onto the simulation grid, bubbles
are modeled as discrete events.

All of the above models have been discussed using purely hydrodynamic
simulations. While there has been some research into the effects of AGN-driven
turbulence in a magnetized cluster~\citep{Dubois2008a}, magnetized outflows
themselves have only been studied by a few authors.  The stability of AGN-blown
bubbles with a predetermined magnetic field configuration has been somewhat
well examined~\citep[e.g.,][]{Robinson2004a,Jones2005,
Ruszkowski2007,Dursi2008,Xu2008b,Gourgouliatos2010b}.
Additionally,~\citet{Xu2010a} and~\citet{ONeill2010} have followed the
evolution of magnetized jets and bubbles in the cosmological formation of a
large cluster.  However, no simulations of magnetized AGN outflows have
included feedback processes coupled to their surroundings; these studies have
presumed a fixed energy in the magnetized bubbles. This link between the
feedback energy and the accretion rate is essential in order to reproduce the
observed cosmic SMBH mass density~\citep{Hopkins2006} and the correlation
between SMBH mass and galactic bulge velocity
dispersion~\citep{booth_cosmological_2009}.  Furthermore, there has been no
systematic comparison study of the various models of AGN feedback when magnetic
energy is included (see~\citet{Dubois2011} and our companion
paper,~\citet{Yang2011}, for a comparison study of purely hydrodynamic
outflows).

In this work, we use FLASH MHD simulations of a mock isolated cluster to
examine several different AGN subgrid models, exploring variations of their
parameters for a variety of resolutions (we have independently studied subgrid
models of SMBH formation and merging in~\citet{Sutter2010}).  Since we cannot
perform an exhaustive study of all the models presented above, we select a
representative bubble model~\citep{Sijacki2007} and a representative jet
model~\citep{Cattaneo2007a}.  We add the magnetic field injection model
of~\citet{Li2006a}.  We perform a systematic study of these models by varying
some of the parameters available for each model in otherwise identical
simulation setups.  Note that for this study, we do not evaluate the models in
terms of their respective abilities to reproduce cluster observables; our goal
in this manuscript is to take the models as read in their respective papers and
apply magnetic injection to them.  By choosing jet and bubble models we are
able to compare and contrast continuous, centrally located small-volume
injections (jets) with sporadic, randomly located large-volume injections
(bubbles).  We also vary the parameters associated with accretion rate models,
since these are tightly coupled to the feedback properties and hence the
resulting magnetic field.  We choose to simulate both bubble and jet models
because they are very distinct: they differ in the form of injected
hydrodynamic energy, the shape of the injection region, and the periodicity of
feedback events. These two approaches allow us to bracket a wide range of
plausible feedback models and explore the roles that their injected magnetic
fields can play in the evolution of their host clusters.

We have three main goals in this work: to examine the effect that introducing
magnetic fields into the AGN injection region has on the accretion and feedback
properties of the SMBH, to evaluate the ability of the various studied models
to magnetize an initially unmagnetized cluster, and to examine the robustness
of these models to changes in their subgrid parameters.  We will characterize
the resulting magnetic fields in terms of field morphology, growth rates, and
radial profiles.  We begin with a discussion of our simulation code and mock
cluster setup in Section~\ref{sec:agn_numerical} and a description of the
subgrid models we study in Section~\ref{sec:agn_subgrid}.  We examine the role
that magnetic fields play in modifying the feedback properties of the modeled
AGN in Section~\ref{sec:agn_effects} and the ability of our fiducial jet and
bubble models to magnetize a cluster in Section~\ref{sec:agn_growth}. We then
examine the magnetic topology of the produced fields for the fiducial runs via
streamlines and rotation measure maps in Section~\ref{sec:agn_topology}.
Finally, we perform a parameter survey in Section~\ref{sec:agn_survey}.  For
this survey, we focus mainly on the changes to the accretion rate and magnetic
properties of the outflows.  We conclude and offer recommendations for
constructing reliable subgrid models in Section~\ref{sec:agn_conclusion}.

\section{Numerical approach}
\label{sec:agn_numerical}

Using the adaptive mesh refinement code FLASH 3.3
~\citep{Fryxell2000b,Dubey2008}, we performed three-dimensional
magnetohydrodynamic simulations with radiative cooling and AGN feedback within
an isolated mock cluster sitting in a $2048$~kpc box.  We constructed the mock
cluster gas using an ideal-gas equation of state in hydrostatic equilibrium
with an NFW gravitational profile~\citep{Navarro1996}.  We assumed a cluster
mass of $1.5 \times 10^{14} \msol$, concentration $5.53$, gas fraction $0.1$,
and Hubble constant $h=0.65$.  For computing accretion and feedback properties,
we placed a $3 \times 10^9 \msol$ black hole in the cluster center.  Note that
we did not include the black hole mass in the computation of the gravitational
potential.  We computed radiative cooling rates using the collisional
ionization equilibrium tables of~\citet{Sutherland1993} assuming $1/3$ solar
metallicity. This is the same cluster as used in the analysis
of~\citet{Cattaneo2007a}, and is motivated by observations of M87.

We maintained a minimum resolution of $32$~kpc throughout the simulation volume
and enabled progressive nested refinement centered on the black hole.  We
varied the peak refinement (see below) from $0.5$ to $16$~kpc.  We defined the
maximally-refined region as a box of width $80$~kpc for jets and $160$~kpc for
bubbles. We did not include any other refinement criteria.

\section{Subgrid models}
\label{sec:agn_subgrid}

\subsection{Accretion rate}
\label{sec:agn_accrate}

The most basic component of the accretion model is the black hole mass.  While
this mass will change with time as gas accretes onto the SMBH, we must select
an initial seed value.  The accretion rate is strongly dependent on mass, and
hence the the feedback energies in the early cluster evolution can vary
greatly.  Ideally, we would perform cosmological simulations and include the
formation and evolution of black holes along with our simulated
clusters~\citep[e.g.,][]{booth_cosmological_2009,Sutter2010}.  Since we are
simulating an already-formed mock cluster, we must select some value.  While
some authors~\citep[e.g.,][]{Sijacki2007} have placed low-mass ($\sim 10^5
\msol$) black holes in massive clusters to study their growth, this is not
consistent with the known $\mhalo$ relation~\citep{bandara_relationship_2009}.
We will choose a value consistent with an already-formed cluster, $M_{\rm bh} =
3 \times 10^9 \msol$ 

For the purposes of this study, we restrict our survey to the 
somewhat simple, but widely used, so-called $\alpha$-model 
of estimating the SMBH accretion rate.
The $\alpha$-model is a simple modification to the 
Bondi-Hoyle-Lyttleton~\citep{BondiH.1952} accretion rate:
\begin{equation}
  \dot{M}_{\rm Bondi} = 4 \pi G^2 M_{\rm bh}^2
                        \frac{\rho}{{c_{\rm s}}^3},
  \label{eq:agn_accretionBondi}
\end{equation}
where the sound speed $c_{\rm s}$ and the density $\rho$ are measured 
on the simulation mesh, and $M_{\rm bh}$ is the black hole mass.
This approximates well the observed accretion rate 
when the Bondi rate is measured at parsec scales~\citep{Allen2006}. 
However, simulations of clusters and 
larger structures cannot obtain sufficient resolution to reach these 
parsec scales, so we must include some compensating factor. 
If we assume that we are underestimating the true accretion rate 
at the scales we typically resolve, we can simply multiply the 
calculated Bondi rate by a constant:
\begin{equation}
  \dot{M}_{\rm bh} = \alpha \dot{M}_{\rm Bondi}.
  \label{eq:agn_accretion}
\end{equation}
We are free to choose the value of $\alpha$, and we will examine 
values from $1$ (i.e., regular Bondi accretion) to $300$, both 
at fixed resolution and by allowing the $\alpha$ parameter to scale 
with resolution, under the assumption that as we lower the 
resolution we must make greater enhancements to the 
calculated Bondi rate.  
While more sophisticated alternatives exist, such as the 
$\beta$-model of~\citet{booth_cosmological_2009} 
or the stochastic model of~\citet{Pope2007},
we currently do not have a complete understanding of the 
complex physics involved in accretion flows. Hence, for this study we 
will keep this portion of the subgrid model as simple as possible. 
 
We must also choose the region in our simulation mesh where we will measure the
density and sound speed for computing the accretion rate.  While we may simply
sample the immediate zones around the SMBH, this approach may suffer from undue
variability, especially in the case of jet-based feedback where the primary
effects of the feedback are felt in those same zones. However, if we choose too
large a radius, we will poorly sample the gas properties and incorrectly
estimate the accretion rate.

Similarly, we must remove gas from the simulation mesh as we accrete it onto
the black hole. As before, we have the freedom to choose from where we will
remove the gas.  If we choose too small a depletion region, then in the case of
strong accretion events we may potentially remove too much gas from the central
zones, leading to numerical instability. On the other hand, removing gas from
zones far away from the SMBH is clearly unphysical, since this gas does not
actually accrete onto the black hole.  To achieve a balance, many
authors~\citep[e.g.,][]{Dubois2010} choose a minimum gas removal threshold so
that the size of the depletion region is chosen to ensure that no more than
some fraction of the gas in the depleted cells (for example, $10 \%$) is
removed in any one timestep.

Since the accretion rate is proportional to the gas density and the gas density
is directly affected by our depletion mechanism, our choices for the accretion
and depletion regions influence each other and thereby the feedback properties
of the AGN.  We include in our study a spherical region with radius $1$ to $4$
zones, varying the accretion and depletion radii jointly and separately. We
also maintain a maximum depletion fraction of $10 \%$ with the \emph{minimum}
depletion radius set as described above.

For all accretion models, we impose an upper limit on the accretion rate corresponding to the Eddington rate, 
\begin{equation}
  \dot{M}_{\rm Edd} = \frac{4 \pi G M_{\rm bh} 
                      m_{\rm p}}{\epsilon_{\rm f} \sigma_{\rm T} c},
  \label{eq:agn_accretionEdd}
\end{equation}
where $m_{\rm p}$ is the mass of the proton, $\sigma_{\rm T}$ is the Thompson cross-section, and $\epsilon_{\rm f} $ is the radiative efficiency.  

In~\tabref{\ref{tab:agn_accParms}} we summarize the parameters included in the
accretion rate model and the values included in our survey: the initial black
hole mass $M_{\rm bh}$, the Bondi multiple $\alpha$, the maximum accretion rate
$\dot{M}_{\rm max}$, the accretion radius $R_{\rm acc}$, the minimum depletion
radius $R_{\rm dep}$, and the maximum gas depletion fraction, $f_{\rm dep}$.

\begin{table}
  \centering
  \caption{Parameters of the accretion rate model.}
\begin{tabular}{ccc}
\hline
\hline
Parameter & Description & Value(s) \\
\hline
$M_{\rm bh}$ & Initial BH mass ($M_\odot$) & $3 \times 10^9$ \\
$\alpha$ & Bondi multiple & $1-300$ \\
$\dot{M}_{\rm max}$ & Maximum accretion rate & $\dot{M}_{\rm Edd}$ \\
$R_{\rm acc}$ & Accretion radius (zones) & $1-4$ \\
$R_{\rm dep}$ & Minimum depletion radius (zones) & $1-4$ \\
$f_{\rm dep}$ & Maximum gas depletion fraction & $0.1$ \\
\hline
\end{tabular}
\label{tab:agn_accParms}
\end{table}

\subsection{Jet-based feedback}
\label{sec:agn_jets}

We follow the general prescription of~\citet{Cattaneo2007a} for building our
jet-based feedback models. This particular model does not simulate the
relativistic jet immediately after its launch from the AGN accretion system,
which we do not have the resolution to accurately simulate, but rather the
large-scale non-relativistic outflow as the jet extends to kpc scales and
begins to entrain ICM material. This jet imparts thermal and kinetic energy to
the ICM as well as a small amount of mass fed back from the accretion disk.

The injection rates of the mass, momentum, and energy onto the grid are treated as source terms in the hydrodynamic equations,
where the energy injection rate is
\begin{equation}
  \dot{E} = \epsilon_{\rm f} \dot{M_{\rm bh}} c^2 
            \left( 1-\eta \right) | \Psi |.
  \label{eq:feedbackJetEner}
\end{equation}
Similarly, the momentum injection rate is
$\dot{\bf{P}} = \sqrt{2 \epsilon_{\rm f}} \dot{M} c \Psi$ ,
and the mass injection rate is
$\dot{M}_{\rm inj} = \eta \dot{M} | \Psi |$.
In the above, $\epsilon_{\rm f}$ is a feedback efficiency and 
$\eta$ is the jet mass loading factor, which is a parameterization 
of the entrainment of gas as the jet propagates.
A jet mass loading factor of $\eta=1$ corresponds to the case 
in which there is no entrainment and the feedback energy is entirely kinetic. 
Higher values allow for the deposition of thermal energy. 

The window function $\Psi$, 
which provides a mapping onto the mesh, is
\begin{equation}
  \Psi({\bf x}) = \frac{1}{2 \pi r_{\rm ej}^2} 
         \exp{\left(-\frac{x^2+y^2}{2 r_{\rm ej}^2}\right)}  
         \frac{z}{h_{\rm ej}^2}.
  \label{eq:agn_psi}
\end{equation}
We cut off injection at $z = h_{\rm ej}$ and $r = r_{\rm ej}$. 
Note that we normalize the injected energy within the window 
function by dividing $\Psi$ by its integral 
$\| \Psi \| = 1 - \exp{(-0.5 r_{\rm ej}^{-2})}$.
The injection region is oriented 
along the $z$-axis. 
There is no threshold associated with activating jet feedback; the 
jet operates continuously, 
as suggested by observations~\citep{Peterson2006}.

Note that our specific implementation differs in several ways from that
of~\citet{Cattaneo2007a}. First, we normalize our value of $\Psi$. Subsequent
papers based on the same model, such as~\citet{Dubois2010}, do this as well.
Also, their model fixed the mass of the SMBH for purposes of calculating the
accretion rate. For jets, this is a reasonable approximation, since the
accretion rate does not reach high values.  We also simulate a larger volume
and maintain a larger maximally-refined central region. Furthermore, we use a
smaller $R_{\rm acc}$ and allow depletion of gas from the ICM. When accounting
for all these difference we have been able to match their results. 

As do~\citet{Cattaneo2007a} we fix the values of $\eta = 100$ and
$\epsilon_{\rm f}=0.1$, which are chosen to match observed jet velocities. We
vary the jet size, both exploring larger jets with fixed resolution and
maintaining a fixed ratio of jet to grid size.  While larger jets may be
somewhat unphysical, in cosmological simulations we can only reach resolutions
of $\Delta x \sim 2$-$4$~kpc, and so this may be the only option available.
While a jet height of $0$ is a valid model, corresponding to a simple flux at a
cell boundary, we will not study it here since we require some spatial extent
for the injection of magnetic flux (see below).

\tabref{\ref{tab:agn_jetParms}} summarizes the aspects of the jet model and our
chosen values: the feedback efficiency $\epsilon_{\rm f}$, the jet mass loading
factor $\eta$, the jet height $h_{\rm ej}$, and jet radius $r_{\rm ej}$. 

\begin{table}
  \centering
  \caption{Parameters of the jet-based feedback model.}
\begin{tabular}{ccc}
\hline
\hline
Parameter & Description & Value(s)\\
\hline
$\epsilon_{\rm f}$ & Feedback efficiency & $0.1$ \\
$\eta$ & Jet mass loading factor & $100$ \\
$h_{\rm ej}$ & Jet height (kpc) & $2-16$ \\
$r_{\rm ej}$ & Jet radius (kpc) & $2.5-20$ \\
\hline
\end{tabular}
\label{tab:agn_jetParms}
\end{table}

\subsection{Bubble-based feedback}
\label{sec:agn_bubbles}

For bubble-based feedback we follow the method outlined 
in~\citet{Sijacki2007}, which places over-pressurized 
bubbles displaced from the SMBH location. 
In their model we have only thermal 
energy injection:
\begin{equation}
  \dot{E} = \epsilon_{\rm m} \epsilon_{\rm f} c^2 \Delta M_{\rm bh},
  \label{eq:agn_feedbackBub}
\end{equation}
where $\Delta M_{\rm bh}$ is the increase in BH mass since the 
last bubble event,
$\epsilon_{\rm f}$ is the feedback efficiency and 
$\epsilon_{\rm m}$ is the mechanical heating 
efficiency.
We distribute this energy on a per-mass basis in a sphere with radius 
determined by
\begin{equation}
  R_{\rm bub} = R_{\rm 0} \left( \frac{\dot{E} \Delta t}{E_{\rm 0}} 
                     \frac{\rho_{\rm 0}}{\rho} \right)^{1/5},
  \label{eq:agn_rbub}
\end{equation}
where we define the radial scaling $R_{\rm 0} = 30 \hkpc$, 
energy scaling $E_{\rm 0} = 10^{55}$~ergs, density scaling
$\rho_{\rm 0} = 10^4 h^{-2} \msol~{\rm kpc}^{-3}$, 
$\Delta t$ is the time since the last bubble injection event.  
These scalings are chosen to ensure that 
a bubble in a typical cluster environment will have 
a realistic size when it has reached pressure equilibrium 
with the ICM. We have used the same values as in the 
cosmological runs of~\citet{Sijacki2007}.
Since bubble events are episodic, we must select some criterion 
for forming bubbles. We only form bubbles when the black hole 
has increased its mass since the previous bubble event by 
$\Delta M_{\rm bh}/M_{\rm bh}  > 0.01\%$, which was chosen 
by~\citet{Sijacki2007} to produce the observed 
$\msig$ relation. 

Since observed bubbles are displaced away from the cluster
centers~\citep[.e.g.,][]{Voit2005a}, we randomly place each bubble with a
maximum displacement $R_{\rm dis}$ equal to the calculated bubble radius
$R_{\rm bub}$. However, we also examine the case in which we fix the bubble
radius at $40$~kpc and do not displace the bubble centers away from the central
SMBH, as has been used in several cosmological
simulations~\citep[e.g.,][]{DiMatteo2008,Battaglia2010}.  This setup should be
an intermediate case between bubbles and jets.

In~\tabref{\ref{tab:agn_bubbleParms}} we summarize our bubble model parameters
and the values we examine in our survey: the minimum black hole mass increase
to trigger a bubble event $\Delta_{\rm bh}$, the feedback efficiency
$\epsilon_{\rm f}$, the mechanical heating efficiency $\epsilon_{\rm m}$, the
bubble radial scale $R_{\rm 0}$, the bubble energy scale $E_{\rm 0}$, the
bubble density scale $\rho_{\rm 0}$, and finally the maximum displacement of
the bubble center $R_{\rm dis}$.

\begin{table}
  \centering
  \caption{Parameters of the bubble-based feedback model.}
\begin{tabular}{ccc}
\hline
\hline
Parameter & Description & Values(s)\\
\hline
$\Delta_{\rm bh}$ & BH mass increase& $0.01\%$ \\
$\epsilon_{\rm f}$ & Feedback efficiency & $0.1$ \\
$\epsilon_{\rm m}$ & Mechanical heating efficiency & $0.2$ \\
$R_{\rm 0}$ & Radial scale ($h^{-1}$~kpc) & $30$ \\
$E_{\rm 0}$ & Energy scale (ergs) & $10^{55}$ \\
$\rho_{\rm 0}$ & Density scale ($h^{-2} \msol~{\rm kpc}^{-3}$) & $10^4$ \\
$R_{\rm dis}$ & Maximum center displacement & $0.0-R_{\rm bub}$ \\
\hline
\end{tabular}
\label{tab:agn_bubbleParms}
\end{table}

As opposed to the jet feedback mechanisms, which as we will see maintain 
low accretion rates due to the more concentrated heating of the core, 
episodic bubble-based models can occasionally allow the accretion rate to 
reach large fractions of the Eddington rate. Indeed,  
observations indicate that  
the SMBH accretion rate is very high, at 
least at high redshift~\citep{Fan2006}. When the accretion rate, and hence 
the available feedback energy, reaches large values the feedback takes 
the form of 
pure radiation~\citep{Fender1999,Gallo2003}. In this case, 
instead of mechanically inflating bubbles, the outflows from the 
AGN simply heat the nearby gas:
\begin{equation}
  \dot{E} = \epsilon_{\rm f} \epsilon_{\rm r} c^2 \dot{M},
  \label{eq:agn_feedbackQso}
\end{equation}
where $\epsilon_{\rm f}$ is the feedback efficiency 
and $\epsilon_{\rm r}$ is the QSO heating efficiency. 
We must choose a threshold to 
switch to this feedback mode, and we follow~\citet{Sijacki2007} 
with a value of $0.01 \dot{M}_{\rm Edd}$. We are free here to 
choose our radius for depositing the energy, and for 
numerical stability we choose a feedback radius of 
$8$~zones.

\tabref{\ref{tab:agn_qsoParms}} lists the parameters and values for the
quasar-mode feedback: the two-mode feedback threshold $\chi_{\rm radio}$, the
feedback efficiency $\epsilon_{\rm f}$, the QSO heating efficiency
$\epsilon_{\rm r}$, and the radius of feedback $R_{\rm radio}$.  We fix all
these values and do not include any variance in our parameter survey here.

\begin{table}
  \centering
  \caption{Parameters of the quasar-based feedback model.}
\begin{tabular}{ccc}
\hline
\hline
Parameter & Description & Value(s)\\
\hline
$\chi_{\rm radio}$ & Two-mode feedback threshold & $0.01 \dot{M}_{\rm Edd}$ \\
$\epsilon_{\rm f}$ & Feedback efficiency & $0.1$ \\
$\epsilon_{\rm r}$ & QSO heating efficiency & $0.05$ \\
$R_{\rm radio}$ & Feedback radius (zones) & $8$ \\
\hline
\end{tabular}
\label{tab:agn_qsoParms}
\end{table}

\subsection{Magnetic field injection}
\label{sec:agn_magfield}

We take the form of injected magnetic fields to be the ``tower'' 
model of~\citet{Li2006a}, which assumes an underlying collimated 
jet extending away from the accretion system:
\begin{eqnarray}
  B_r(r',z')    & = & 2 B_{\rm 0} z' r' \exp{\left( -{r'}^2 - {z'}^2 \right)} \\
  B_z(r',z')    & = & 2 B_{\rm 0} \left( 1 - {r'}^2 \right)
                 \exp{\left( -{r'}^2 - {z'}^2 \right)} \\ 
  B_\phi(r',z') & = & B_{\rm 0} \alpha_{\rm B} r' 
                 \exp{\left( -{r'}^2 - {z'}^2 \right)}, 
  \label{eq:mag}
\end{eqnarray}
where $r'=\sqrt{x^2+y^2}/r_{\rm 0}$ and $z'=z/r_{\rm 0}$.
Here, we set the scale radius $r_0$ to be $0.5 R_{\rm inj}$, so 
that $r_0 = 0.5 R_{\rm bub}$ for bubbles and $r_0 = 0.5 R_{\rm ej}$ for jets, 
so that the entire injected magnetic structure fits inside the given 
feedback region. 
The magnetic field injected via the above prescription 
is constructed so that it has 
zero divergence, 
and this scaling ensures that we minimize divergences caused by the 
artificial cutoff.
$\alpha_{\rm B}$ is the ratio 
of poloidal to toroidal flux, which we choose 
to be $\alpha_{\rm B}=\sqrt{10}$ 
for a field with minimum initial Lorentz force, as suggested by \cite{Li2006a}. 
We determine the scale $B_{\rm 0}$ by giving some fraction of the 
available feedback energy to the magnetic field: either 
$0.0$ for purely hydrodynamic outflows, $0.5$ for an equipartition 
case, or $1.0$ for purely magnetic outflows. 
Note that for purely magnetic outflows in jets, we ignore the details 
of the window 
function $\Psi$ in~\eqref{\ref{eq:agn_psi}} and use it simply 
to determine the extent of the magnetic feedback region.
For jet models, we align the axis of the magnetic field with the 
axis of the jet, and for bubbles we align the magnetic field 
axis with a vector pointing from the bubble center to the position 
of the SMBH.

We summarize the parameters of our magnetic field injection
in~\tabref{\ref{tab:agn_magParms}}: the radial scale $r_0$, the ratio of
poloidal to toroidal flux $\alpha_{\rm B}$, and the fraction of feedback energy
available to magnetic fields $E_{\rm B}$. Aside from the energy of the magnetic
injection, we do not vary these parameters.

\begin{table}
  \centering
  \caption{Parameters of the magnetic injection model.}
\begin{tabular}{ccc}
\hline
\hline
Parameter & Description & Values(s)\\
\hline
$r_0$ & Radial scale & $1/2 r_{\rm inj}$ \\
$\alpha_{\rm B}$ & Poloidal/toroidal ratio & $\sqrt{10}$ \\
$E_{\rm B}$ & Fraction of total energy in $B$ & $0.0-1.0$ \\
\hline
\end{tabular}
\label{tab:agn_magParms}
\end{table}

\subsection{Parameter survey}
\label{sec:agn_parameters}

In Tables~\ref{tab:agn_jetParmSurvey} and~\ref{tab:agn_bubbleParmSurvey} we
detail our parameter survey for jets and bubbles, respectively.  We only
include in the tables those parameters that we vary in our survey.  We have
collected each set of parameter changes into several groups.  Each group is
given a unique numerical identifier, and each set of parameters within that
group is given an alphabetical label.  For later plots, we label purely
hydrodynamic simulations as ``h'', equipartition models (i.e., $E_{\rm B} = 0.5
E_{\rm tot}$) as ``e'', and purely magnetic feedback runs as ``m''.  Thus, the
label ``J3Bm'' will designate purely magnetic outflow (``m'') from the second
parameter set (``B'') of the third grouping (``3'') of the jet models (``J'').

For jets, in our first group we study changing the peak grid resolution with a
fixed jet size of $h_{\rm ej}=2.5$~kpc and $r_{\rm ej}=2.0$~kpc.  In the next
group we also vary the resolution but allow the jet size to scale with the
resolution such that $h_{\rm ej}=5$~zones and $r_{\rm ej}=4$~zones. In the
third group we vary the accretion strength $\alpha$. Finally, in the last group
we investigate the effects of changing the accretion and depletion radii on the
resulting feedback. We do this because the jet model as described
in~\citep{Cattaneo2007a} does not include any gas depletion, and changing these
values can have significant consequences for jets since the feedback is applied
very close to the central black hole.  Note that for most of the jet groups we
do not use our peak resolution of $0.5$~kpc, but rather the larger $1.0$~kpc.
We chose this so that we would have enough computational resources to complete
the study.  We only include equipartition magnetic fields in the first group,
where we vary the grid resolution.  Note that the model set ``J1B'' serves as
our fiducial jet case. 
 
\begin{table*}
\caption{Jet model parameter survey.}
\begin{center}
\begin{tabular}{ccccccc}
\hline
\hline
Designation & $\Delta x$~(kpc) & $\alpha$ & $ h_{\rm ej}$~(kpc) & $ r_{\rm ej}$~(kpc) & $ R_{\rm acc} / \Delta x$ & $ R_{\rm dep} / \Delta x$ \\
\hline
\\
\multicolumn{7}{c}{Varying Resolution} \\
\hline
J1A &  0.50 &  1 &  2.0 &  2.5 &  2.0 &  2.0 \\
J1B &  1.00 & - & - & - & - & - \\
\\
\multicolumn{7}{c}{Scaling Jet Size with Resolution} \\
\hline
J2A & - & - &  4.0 &  5.0 & - & -  \\
J2B &  2.00 & - &  8.0 &  10.0 & - & -  \\
J2C &  4.00 & - &  16.0 &  20.0 & - & -  \\
J2D & - & - &  8.0 &  10.0 & - & -  \\
\\
\multicolumn{7}{c}{Varying Alpha} \\
\hline
J3A &  0.50 &  100 &  2.0 &  2.5 & - & -  \\
J3B &  1.00 & - & - & - & - & -  \\
J3C & - &  300 & - & - & - & -  \\
\\
\multicolumn{7}{c}{Varying Accretion and Depletion Radii} \\
\hline
J4A & - &  1 & - & - &  1.0 &  1.0  \\
J4B & - & - & - & - &  4.0 &  4.0  \\
J4C & - & - & - & - &  2.0 &  0.0  \\
\hline
\end{tabular}
\end{center}
\label{tab:agn_jetParmSurvey}
\end{table*}

For the first group of our bubble model survey, we vary the resolution while
keeping the other bubble parameters fixed. Since the bubbles are much larger
than the jets ($50$-$200$~kpc), we may lower our resolution much more than in
the jet runs.  In the next two groups we vary the accretion multiple $\alpha$
both with fixed resolution and by scaling $\alpha$ with the grid resolution.
For our last group, we fix the bubble position on the SMBH. Note that for this
case, we also fix the bubble radius to $R_{\rm 0} \approx 40$~kpc.  This
represents an intermediate case between jets and bubbles and is used frequently
in cosmological simulations~\citep[e.g.,][]{DiMatteo2008,Battaglia2010}.  We
include the equipartition magnetic feedback mode only with the bubble run
(``B1B'') which is at the same resolution as the jet runs, and for the fixed
bubble run. We do not examine the role that the accretion and depletion radii
play in this bubble survey, since the bubble feedback is not as sensitive to
these criteria.  Note that the model set ``B1B'' serves as our fiducial bubble
case. 

\begin{table}
\caption{Bubble model parameter survey.}
\begin{center}
\begin{tabular}{cccc}
\hline
\hline
Designation & $\Delta x$~(kpc) & $\alpha$ & $ R_{\rm dis} / R_{\rm bub} $ \\
\hline
\\
\multicolumn{4}{c}{Varying Resolution} \\
\hline
B1A &  0.50 &  1  & 1.0  \\
B1B &  1.00 & -  & -  \\
B1C &  2.00 & -  & - \\
B1D &  4.00 & -  & - \\
B1E &  8.00 & -  & - \\
\\
\multicolumn{4}{c}{Varying Alpha} \\
\hline
B2A &  2.00 &  100  & -  \\
B2B & - &  300  & -  \\
\\
\multicolumn{4}{c}{Scaling Alpha with Resolution} \\
\hline
B3A &  0.50 &  50  & -  \\
B3B &  4.00 &  300  & -  \\
\\
\multicolumn{4}{c}{Fixing Bubble Position} \\
\hline
B4A &  1.00 &  1  & 0.0 \\
\hline
\end{tabular}
\end{center}
\label{tab:agn_bubbleParmSurvey}
\end{table}

\section{The effects of magnetic injection on AGN feedback}
\label{sec:agn_effects}

Before we fully examine the growth and structure of magnetic fields using the
models described above, we must first see what, if any, effects the presence of
injected fields has on the accretion and feedback properties of the AGN as well
as on some of the resulting hydrodynamic characteristics of the cluster medium.
We choose three of our models to examine: J1B and B1B, representing our
fiducial cases for jets and bubbles, respectively, and model B4A, which uses
bubbles with fixed centers and radii.  For each model we examine purely
hydrodynamic injection, fully magnetic injection, and an intermediate
equipartition case in which the available injection energy is evenly split
between magnetic and hydrodynamic components.  For the plots below, models J1B
and B1B have plot titles of ``Jets'' and ``Bubbles'', respectively, while model
B4A is called ``Fixed Bubbles.''

\begin{figure}
  \centering {\includegraphics[width=\columnwidth]{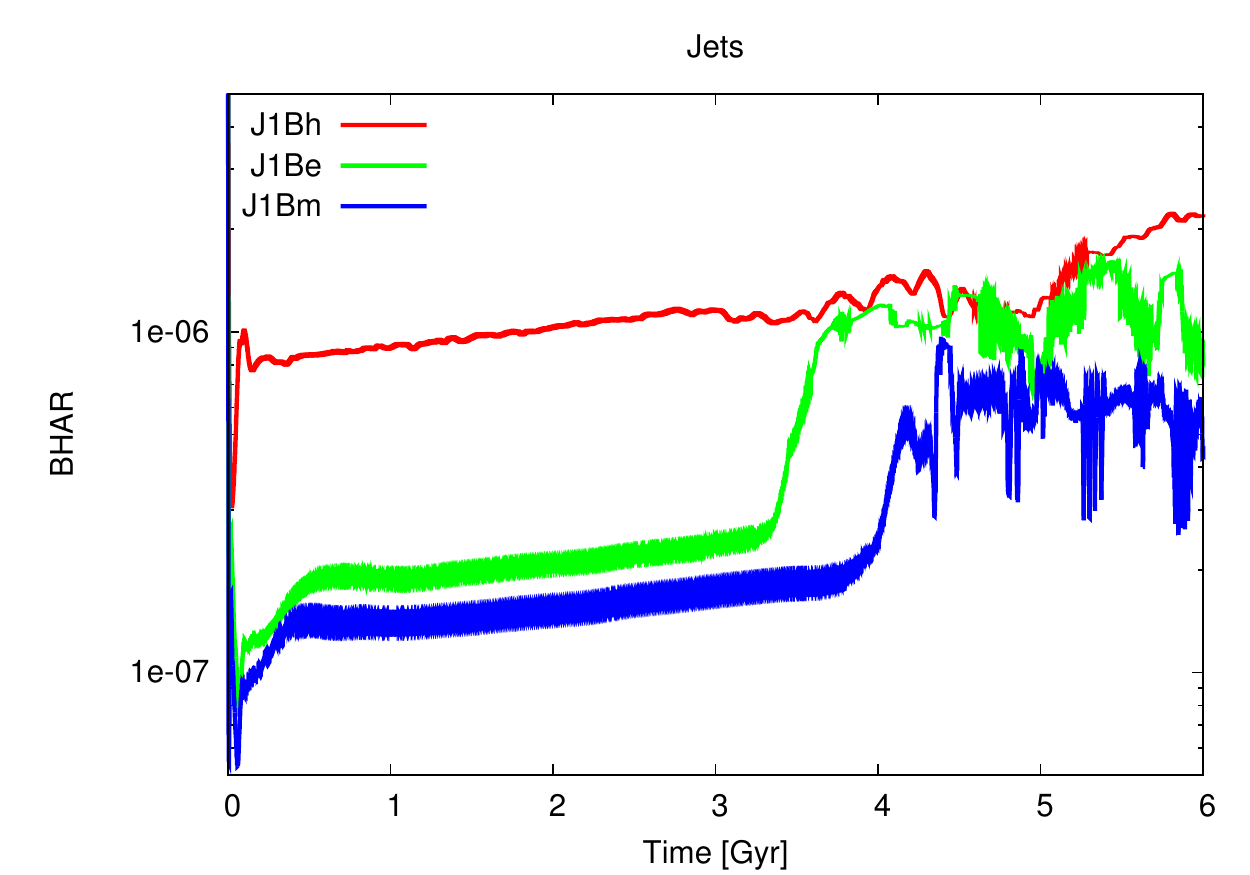}}
  \caption[Effects of magnetization on black hole accretion rate.]
           {Effects of magnetic injection on the 
            black hole accretion rate. 
            Shown is BHAR ($\dot{M}/\dot{M}_{\rm Edd}$) versus time for purely 
            hydrodynamic, equipartition, and fully magnetic modes of
            the J1B (Jets) model}
\label{fig:agn_effects_accrate_jets}
\end{figure}

~\figref{\ref{fig:agn_effects_accrate_jets}} shows the black hole accretion
rate as a fraction of the Eddington limit ($\dot{M}/\dot{M}_{\rm Edd}$) for the
jet model.  We find that simply the presence of an injected magnetic field
drastically reduces the accretion rate: for $\sim 3$~Gyr the magnetic fields
suppress the accretion rate by a factor of five, with much smaller differences
between the equipartition and fully magnetic runs. This has two causes. First,
the shape of the magnetic injection, which is initially a torus rather than the
axial jet of the hydrodynamic feedback, does somewhat prevent accretion onto
the black hole. However, we performed tests where we injected thermal energy
with the same distribution as the magnetic injection and found only small
differences.  The main cause of the reduced accretion rate is the
``unspringing'' of the highly tense injected fields. As the fields unfold after
injection they efficiently drive gas away from the central zones. Eventually,
however, the gas is able to cool sufficiently and overcome this tension and the
accretion rate correspondingly jumps. However, the magnetized outflows still
maintain an accretion rate a factor of two lower than in the purely
hydrodynamic case.  The accretion rate for the magnetized cases is also highly
variable with a short characteristic timescale of $1$-$10$~Myr. This is
indicative of the complex relationship between the unfolding magnetic fields
and the gas which is attempting to accrete onto the black hole.

After $6$~Gyr, the purely hydrodynamic mode undergoes a period of intense
activity characterized by a rapid rise in accretion rate as the gas cools
sufficiently and an associated enhanced feedback
phase~\citep[see][]{Cattaneo2007a}. At this point the strong magnetic fields in
the core become highly tangled, and the combination of strong fields
surrounding the core and cold, dense gas at the core results in very low plasma
$\beta$ values and hence severely lowers our magnetohydrodynamical timestep.
Thus we can not efficiently carry the simulation further and simultaneously
perform our intended parameter survey. For this run and all others, we evolve
the simulations until the timestep drops too low to continue effectively.
However, we are still able to perform these simulations across billions of
years, and the timing of the onset of complex fields can yield useful
information about the relationship between the magnetic fields and the cluster
gas.

\begin{figure}
  \centering  {\includegraphics[width=\columnwidth]{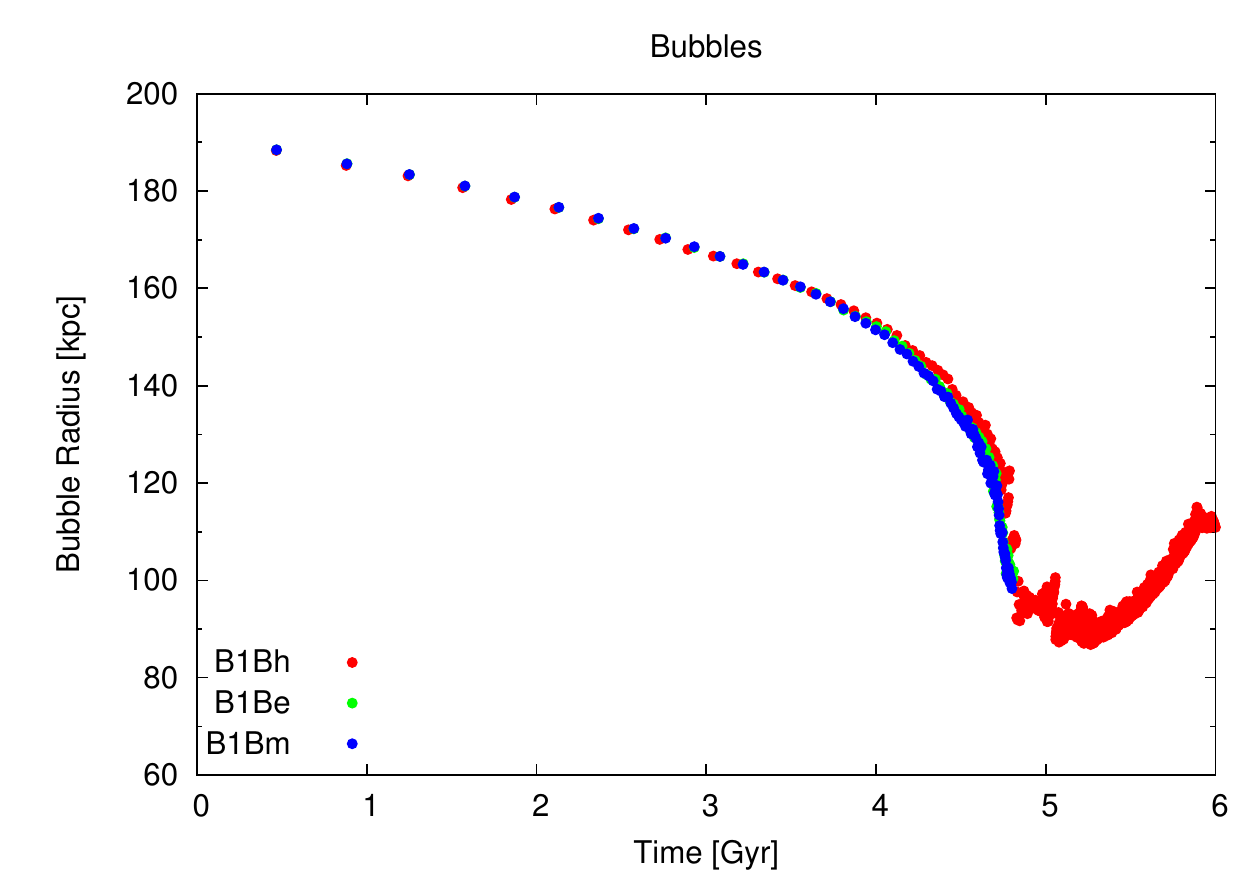}}
  \caption[Effects of magnetization on bubble size.]
           {Effects of magnetic injection on the bubble size. Shown is
            bubble radius at each injection event as a function of time
            for model B1B.}
\label{fig:agn_effects_bubsize}
\end{figure}

\begin{figure*}
  \centering
  {\includegraphics[width=\columnwidth]{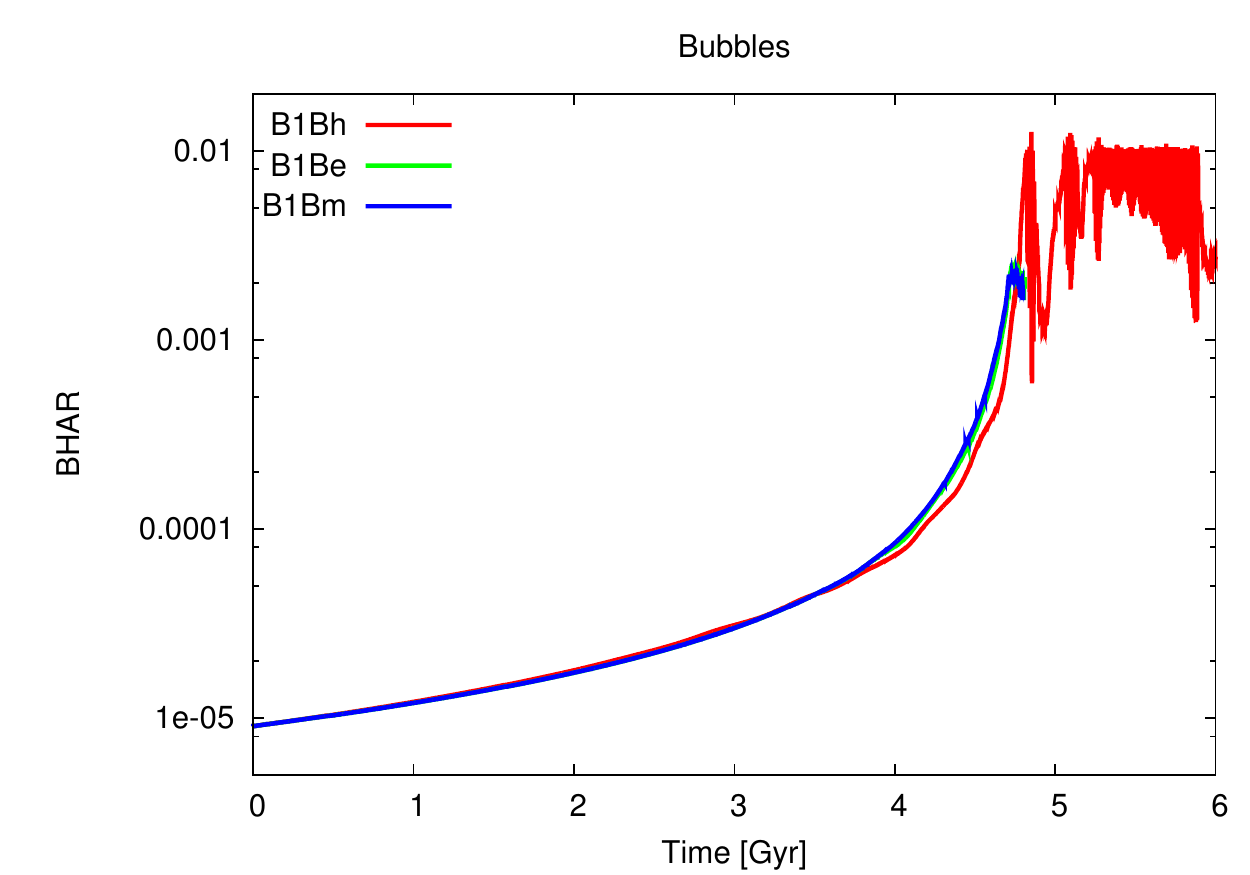}}
  {\includegraphics[width=\columnwidth]{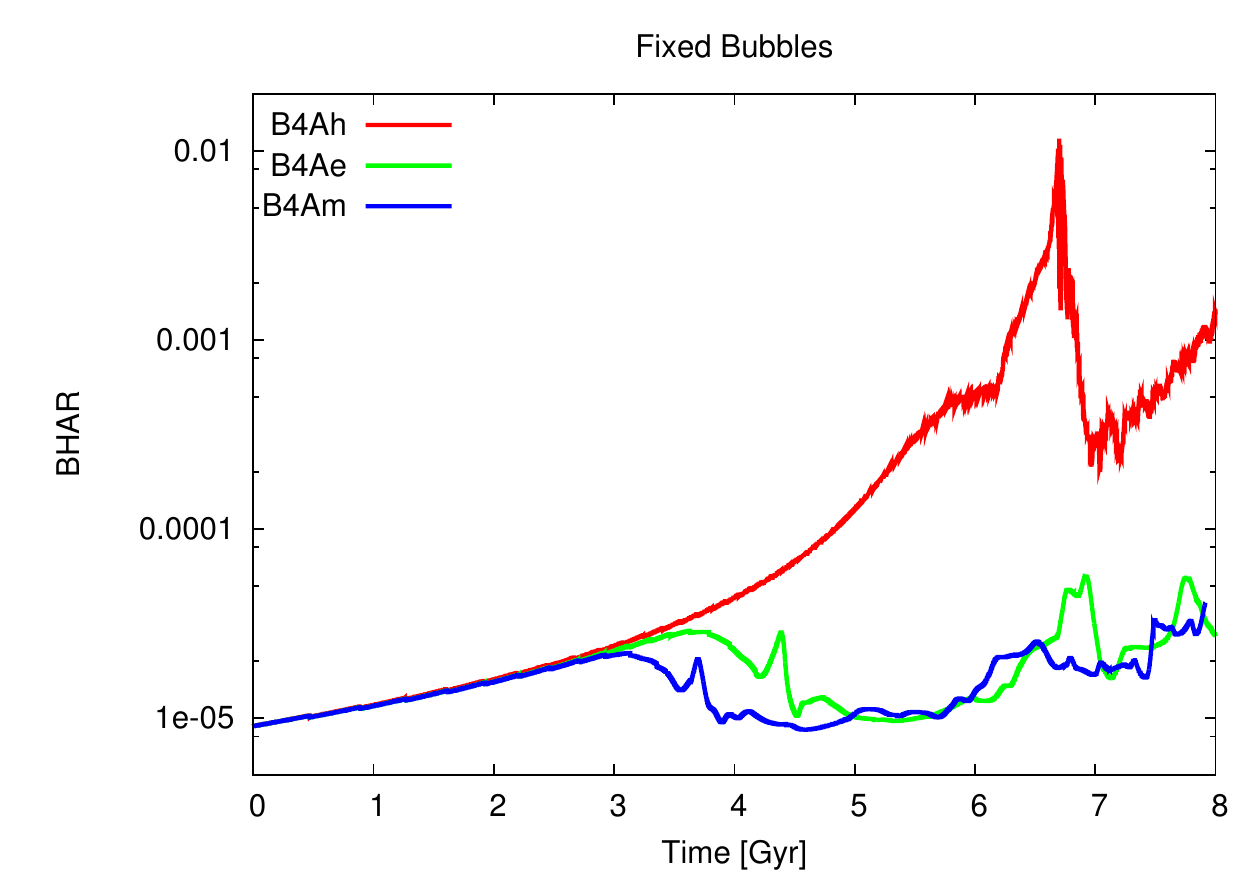}}
  \caption[Effects of magnetization on black hole accretion rate.]
           {Effects of magnetic injection on the 
            black hole accretion rate. 
            Shown is BHAR ($\dot{M}/\dot{M}_{\rm Edd}$) versus time for purely 
            hydrodynamic, equipartition, and fully magnetic modes of models 
            B1B (Bubbles), and 
            B4A (Fixed Bubbles).}
\label{fig:agn_effects_accrate_bubbles}
\end{figure*}

The effects of the injected magnetic field are much less pronounced in the
randomly-placed bubble model (B1B), as highlighted by bubble size as a function
of time, as shown in~\figref{\ref{fig:agn_effects_bubsize}} (note that the
other bubble model examined here, B4A, uses a fixed bubble radius, and thus is
not shown). The bubbles in all cases start out very large, with radii nearly
$200$~kpc. Over $4$~Gyr they grow smaller and more frequent as the time between
injection events grows shorter and the core density increases (which we will
discuss in detail later).  Eventually the hydrodynamic bubbles begin to exhibit
periodic behavior, but the magnetic fields become too complex to follow
efficiently at this resolution.  This also means that the accretion rate is
much less affected, as shown in
~\figref{\ref{fig:agn_effects_accrate_bubbles}}. Since the bubbles are large
and off-centered, the magnetic fields are not as effective at driving away gas
from the core, and hence the accretion histories of the three modes (purely
hydrodynamic, equipartition, and purely magnetic) are nearly identical (for a
discussion of the differences in accretion rate with and without feedback, see
our companion paper ~\citealt{Yang2011}). Additionally, there is almost no
difference between the equipartition and fully magnetic modes.  The
hydrodynamic injections are eventually able to reach the two-mode feedback
threshold.

When we force the bubbles to be centered on the black hole, as with model B4A,
we achieve similar suppression of the accretion rate as for the jets, as we see
in~\figref{\ref{fig:agn_effects_accrate_bubbles}}.  These differences take
$\sim 3$~Gyr to manifest, however, since the bubbles are larger than the
injection region and hence not as efficient as the jets at driving away gas
(this is also noticeable when comparing the purely hydrodynamic modes, where
the jets have an accretion rate roughly an order of magnitude smaller than the
bubbles). Once again we also see the importance of merely the presence of
magnetic injection over the relative strength of those fields: both the
equipartition and fully magnetic modes exhibit nearly the same behavior, which
after $3$~Gyr is characterized by small (factor of $\sim 2$) changes to the
accretion rate with a cyclic period of roughly $3$~Gyr, whereas the
hydrodynamic injections see much larger variations in the accretion rate over
much smaller timescales.

\begin{figure} \centering
  {\includegraphics[width=\columnwidth]{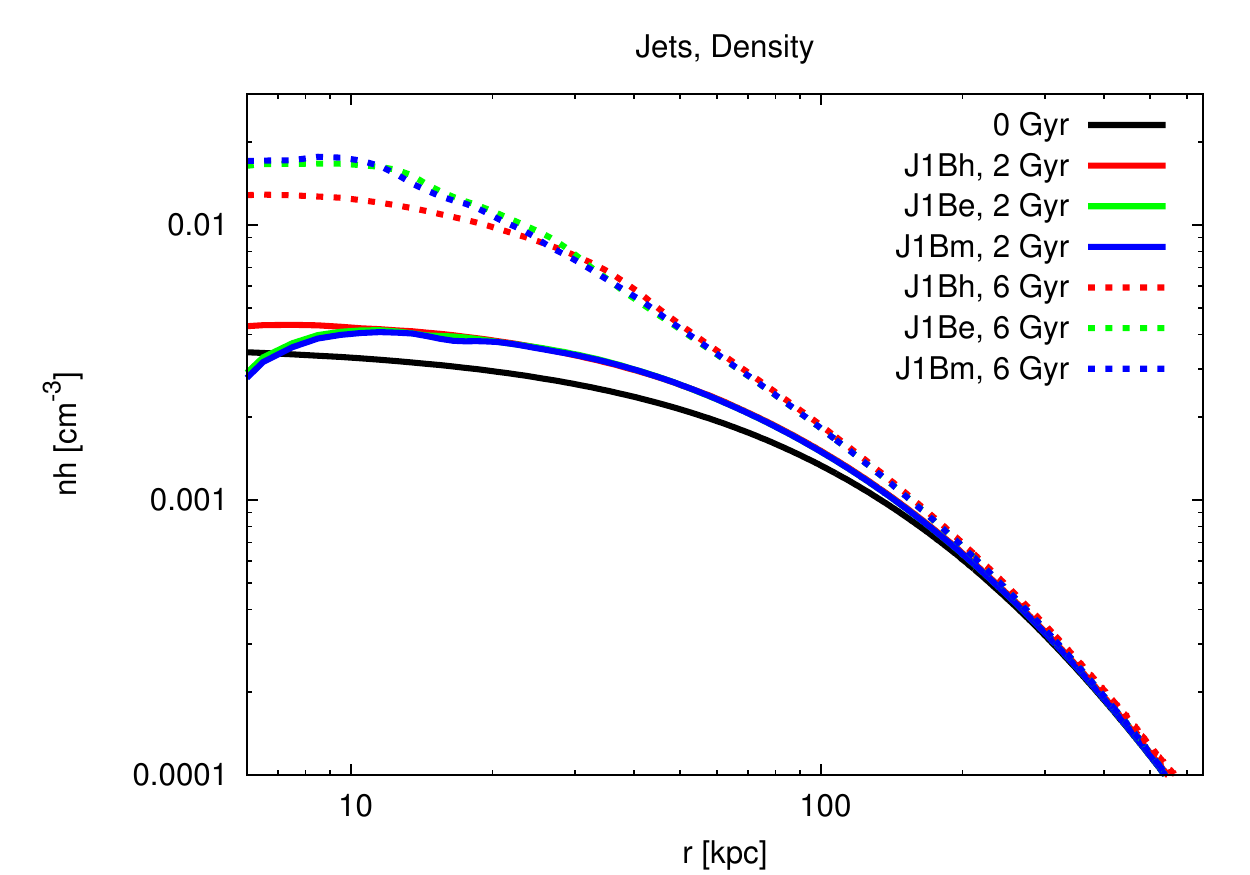}}
  {\includegraphics[width=\columnwidth]{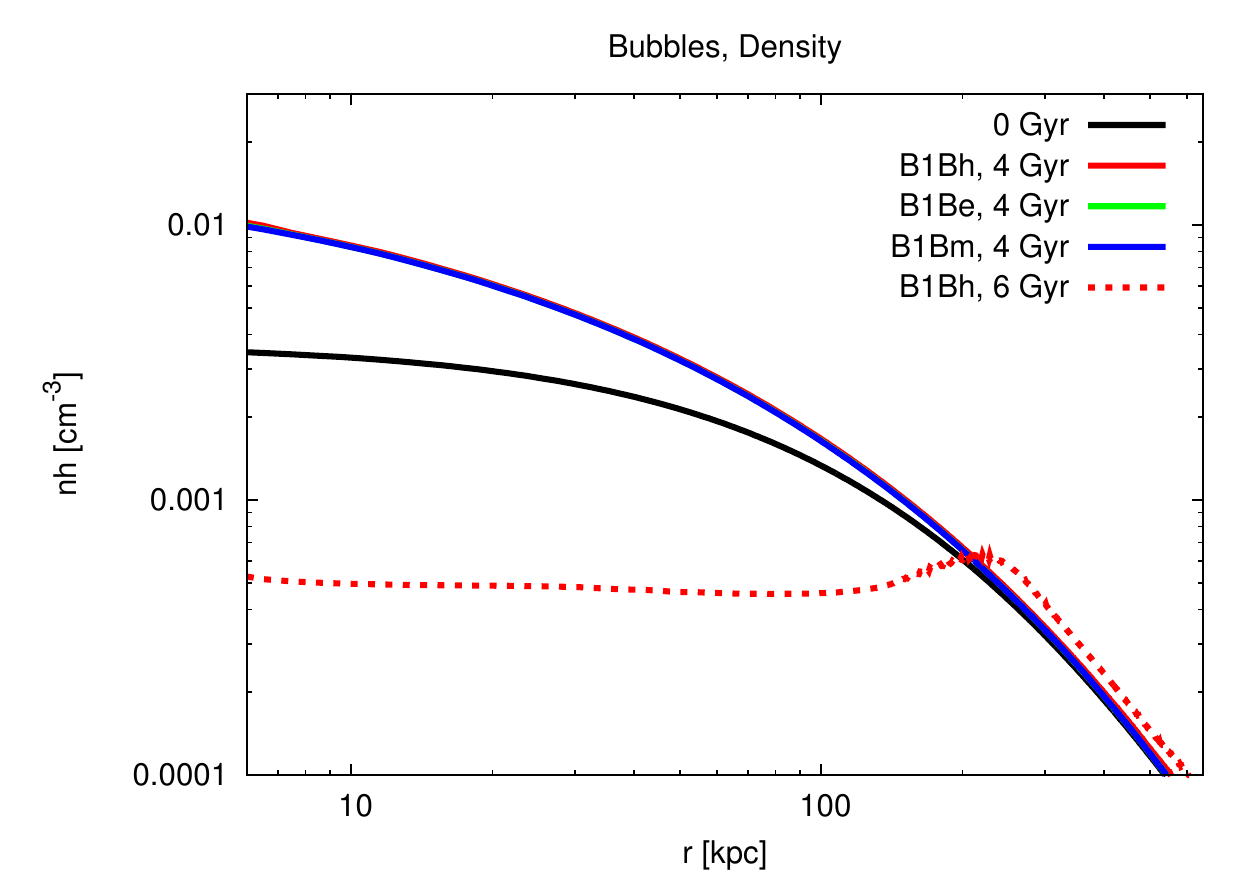}}
  {\includegraphics[width=\columnwidth]{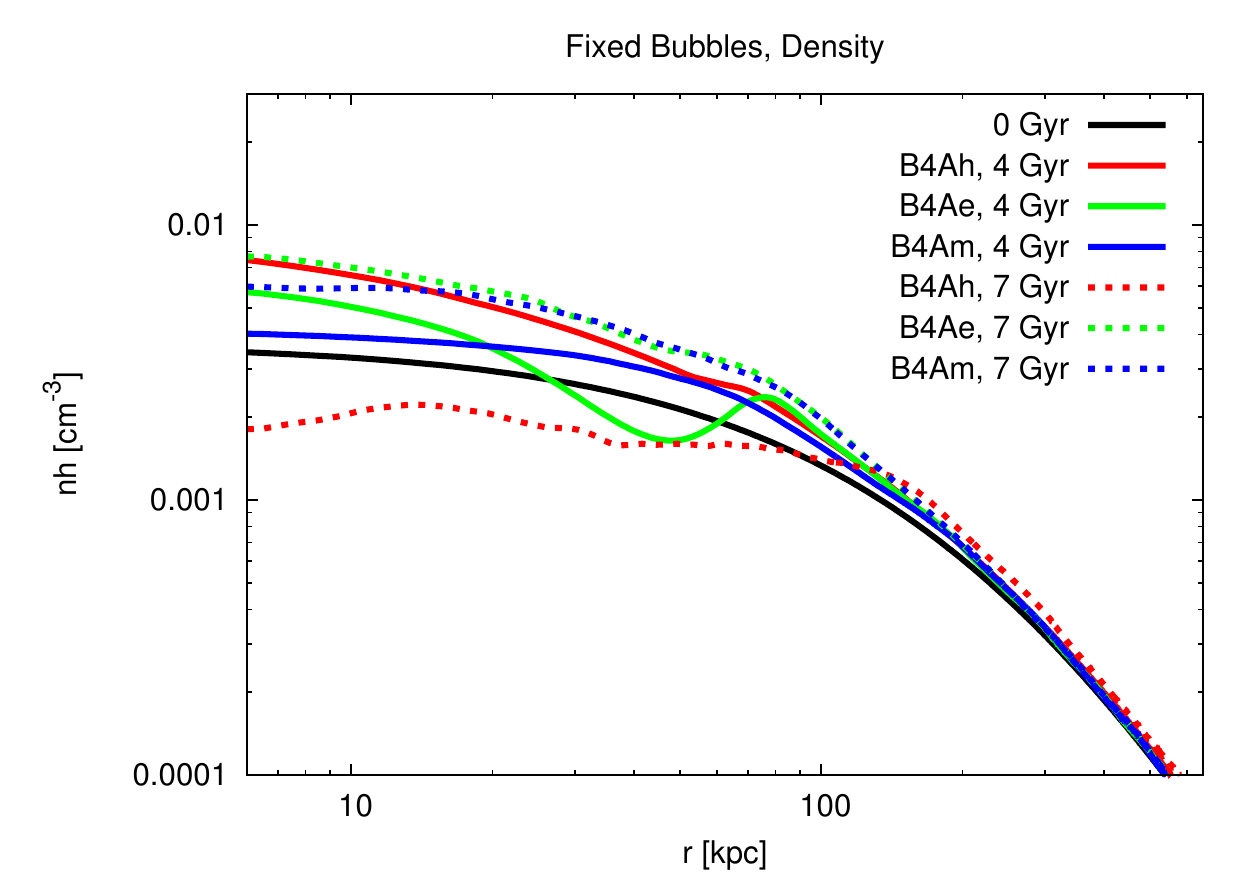}}
  \caption[Effects of magnetic injection on cluster density.]
           {Effects of magnetic injection on the 
            cluster density profile.}
\label{fig:agn_effects_profile_dens}
\end{figure}

We show radial profiles of the density in
~\figref{\ref{fig:agn_effects_profile_dens}}. These are constructed from
volume-averaged quantities in shells of thickness $1$~kpc.  For the jets we see
that at $2$~Gyr the magnetized injections are more effective at driving gas
away from the core, but outside of the central regions of the cluster there are
very few differences.  However, after $6$~Gyr the accreting gas pushes through
the magnetic fields and the magnetized outflows end up with a slightly higher
central density, though once again the injection modes are indistinguishable
past $20$~kpc. Even though the density is higher in the magnetized runs, the
sound speed is also higher, resulting in the reduced accretion rate discussed
above.

The randomly-placed bubbles, as expected, also exhibit no differences between
magnetized and unmagnetized profiles at $4$~Gyr, whereas the fixed bubbles show
remarkable differences out to a radius of $100$~kpc.  After this, the purely
hydrodynamic bubbles significantly heat the core, which drives gas into the
outer regions of the cluster atmosphere.  At $4$~Gyr the magnetized outflows
have prevented large amounts of material from collapsing into the core and the
equipartition mode has formed a low-density shell from $30$ to $70$~kpc. The
purely hydrodynamic mode has a much smaller low-density shell at $70$~kpc.  For
the hydrodynamic case the heated outwardly-expanding gas in the injection
regions prevents further accretion of material from the rest of the cluster,
while some of the inner material within the injection region falls towards the
black hole.  In the equipartition case the complex interplay of the magnetic
fields and injected thermal energy forms a larger shell. For the fully
magnetized outflow the magnetic fields simply drive gas away from the core.  By
$7$~Gyr, however, the profiles have reversed: the intense activity of the
hydrodynamic bubbles has driven gas away from the core (note that this profile
is taken from a point in time which is at the trough in a cyclic pattern, as
seen in~\figref{\ref{fig:agn_effects_accrate_bubbles}}) while the more sedate
magnetized injections see relatively little change in the density profiles.

\begin{figure}
  \centering
  {\includegraphics[width=\columnwidth]{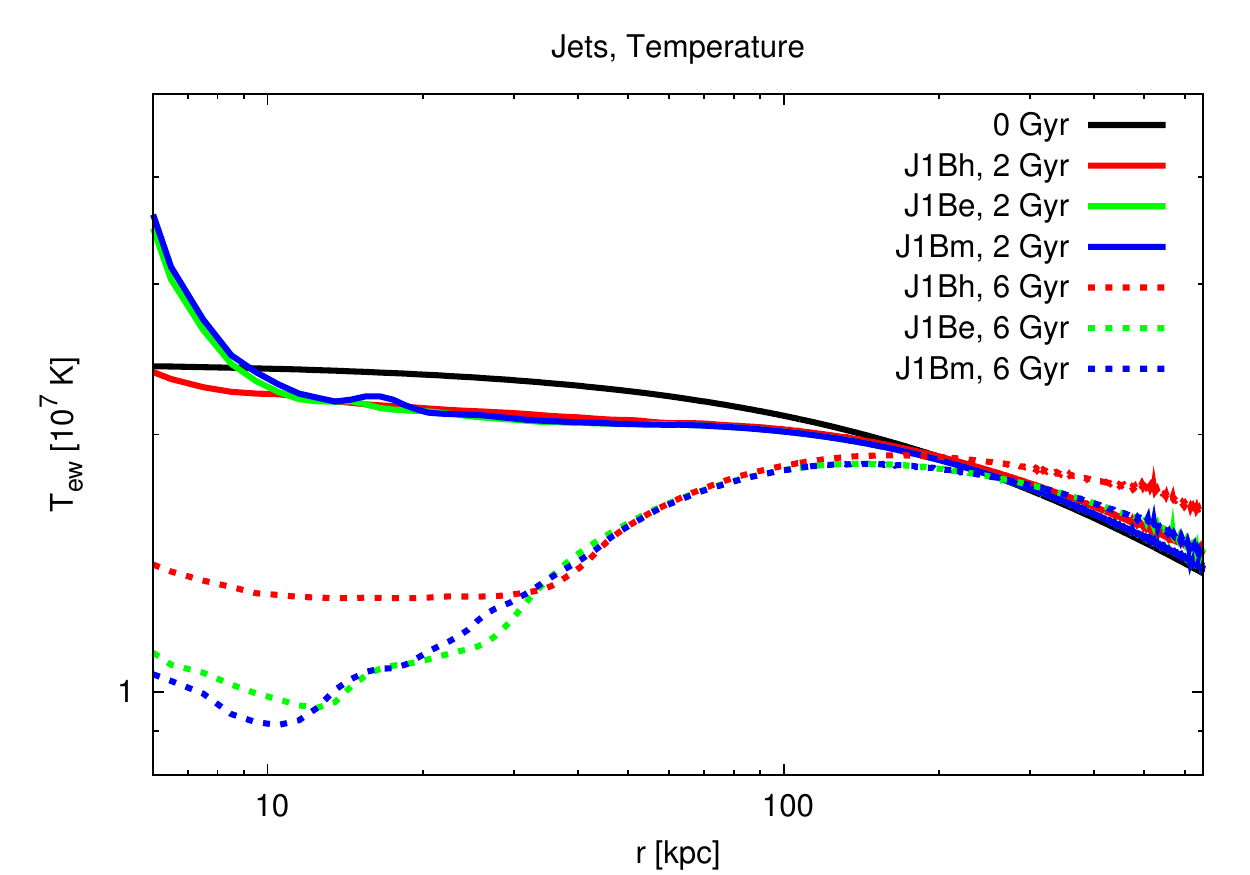}}
  {\includegraphics[width=\columnwidth]{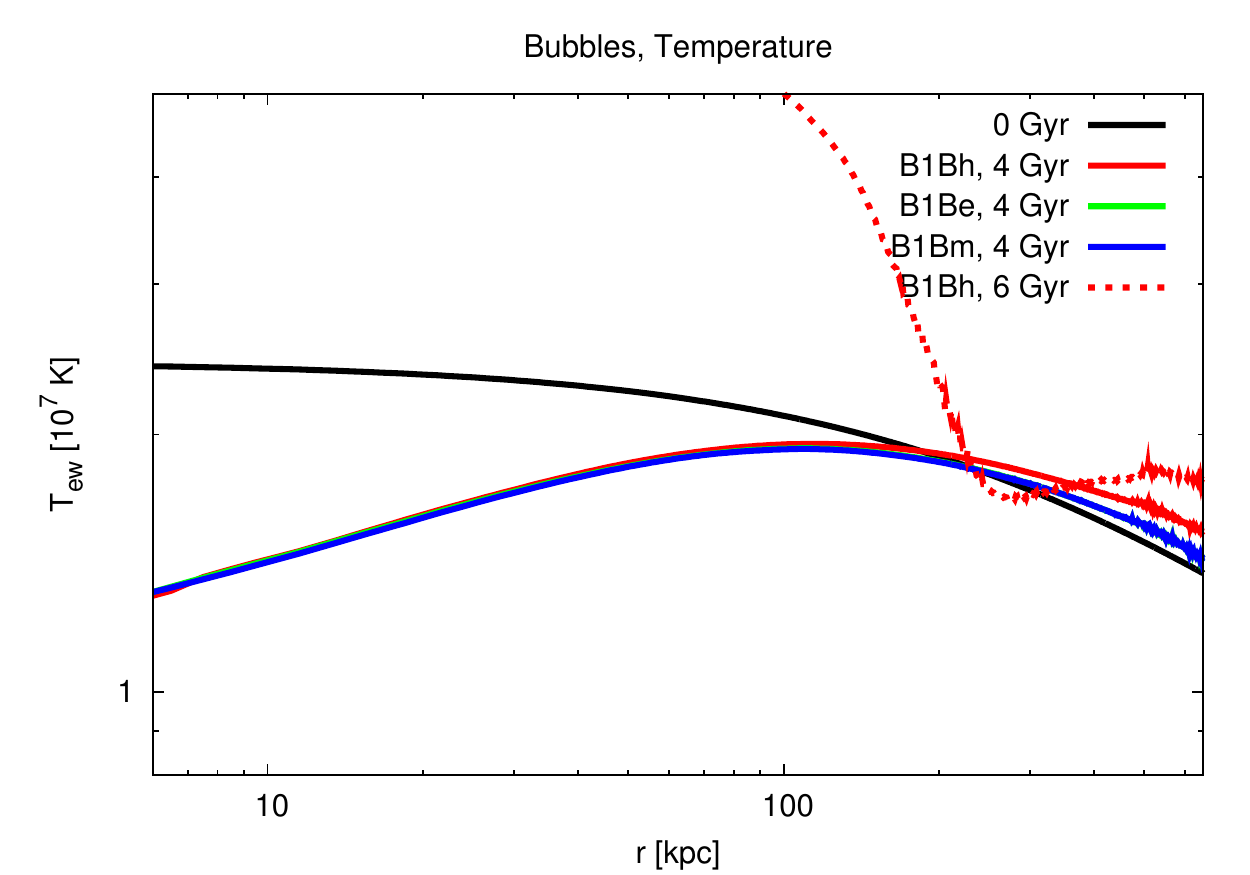}}
  {\includegraphics[width=\columnwidth]{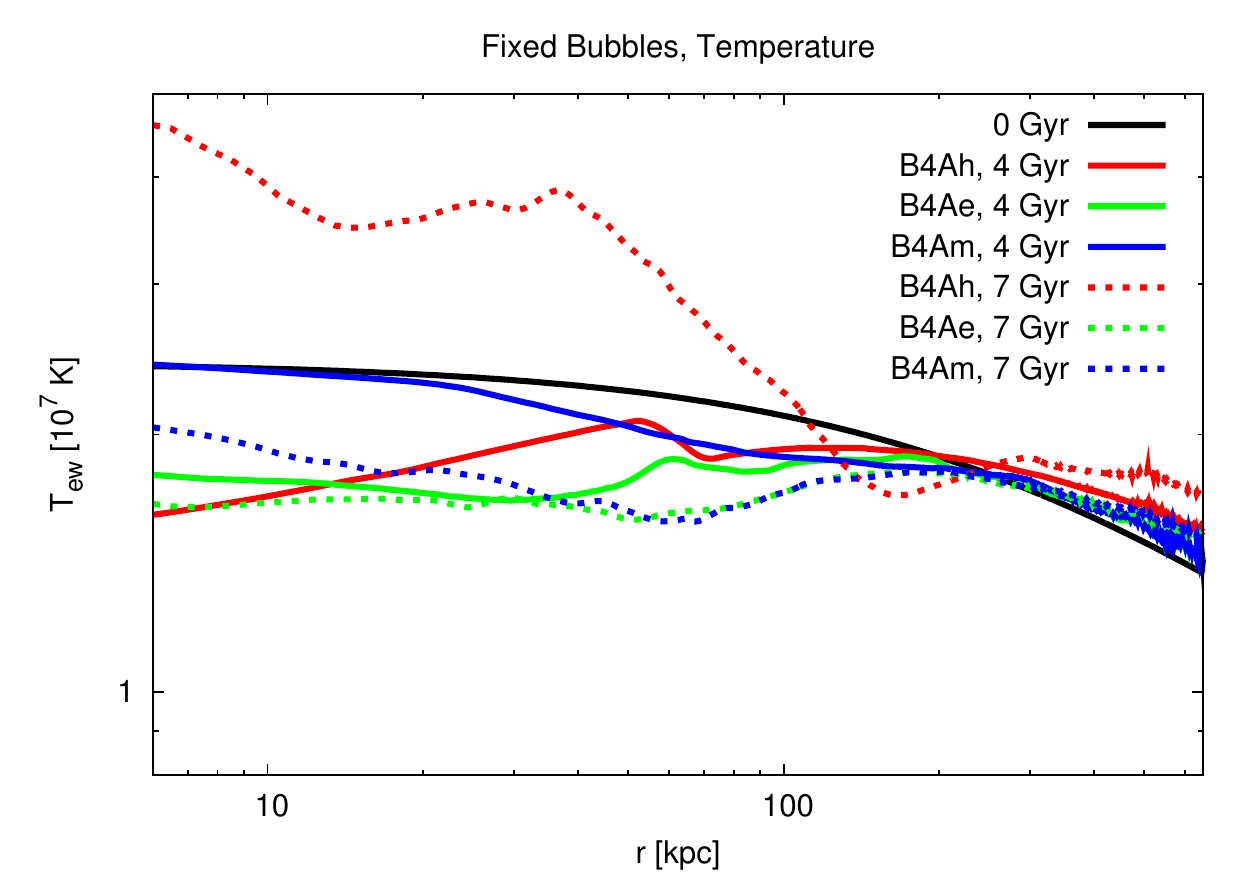}}
  \caption[Effects of magnetic injection on cluster temperature.]
           {Effects of magnetic injection on the 
            cluster temperature profile. Shown is 
            emission-weighted temperature.}
\label{fig:agn_effects_profile_temp}
\end{figure}

These conditions are also reflected by the emission-weighted temperature, as we
show in~\figref{\ref{fig:agn_effects_profile_temp}}.  Note that we did not
observe significant differences in the pressure profile and hence do not show
them here.  For the jets, the magnetic fields, which are much more effective at
removing gas from the central core, heat up the gas to $3.5 \times 10^7$~K
after $2$~Gyr of evolution. Eventually the gas is able to cool, and in the
magnetized case the gas reaches $10^7$~K after $6$~Gyr, about $30$\% lower than
in the purely hydrodynamic case. This is also the case with the fixed bubbles,
where initially the magnetized outflows maintain higher temperatures in the
core relative to the thermal outflows, but after $7$~Gyr the magnetic bubbles
change the inner temperature by at most $20$\%, whereas the hydrodynamic
bubbles heat the gas considerably.

\section{The growth of magnetic fields}
\label{sec:agn_growth}

To examine in detail the growth of magnetic fields, we will take only our
fiducial cases J1B for the jets and B1B for the bubbles as well as the
intermediate fixed bubble case of B4A. These three cases have the same peak
resolution and accretion properties and 
 - other than the form of the injection - these runs differ only 
in that the bubble runs use a larger centrally-refined region. For 
this section we only study fully magnetic injection, with no portion 
of the feedback energy in thermal or kinetic modes. 

\begin{figure}
  \centering {\includegraphics[width=\columnwidth]{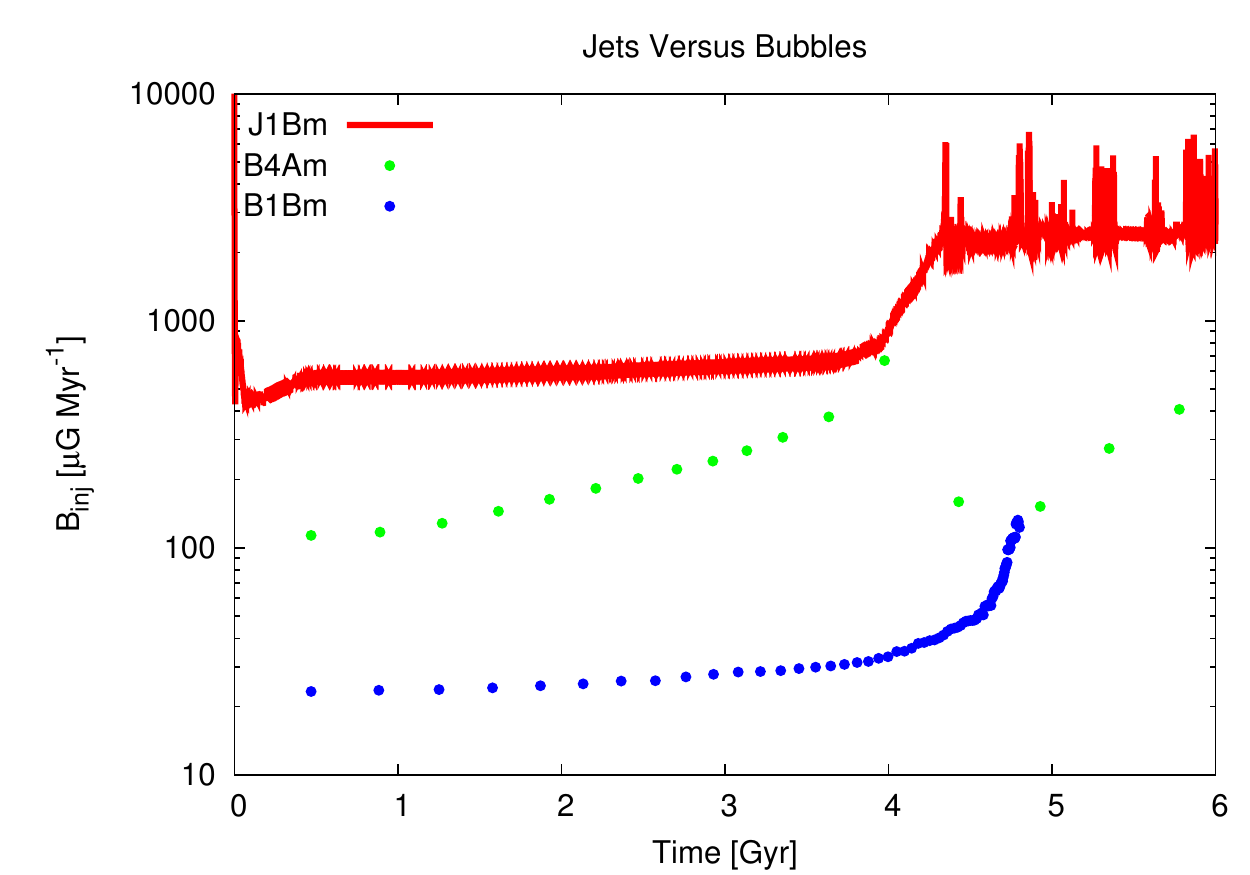}}
  \caption[Rate of injected magnetic field strength.]
           {Rate of injected magnetic field strength. Shown is
            $B_{\rm inj}$ versus time for models J1B, B4A, and B1B. 
            The continuous injection rate for J1B is shown as a line while 
            the points represent 
            the discrete injection events for models B4A and B1B.}
\label{fig:agn_growth_maginj}
\end{figure}

We first examine the rate of magnetic field injection, $B_{\rm inj}$,
in~\figref{\ref{fig:agn_growth_maginj}}. The jets (model J1B) inject magnetic
fields continuously at a rate $\sim 600 \mgmyr$ for approximately $4$~Gyr.
Once the core begins to significantly cool down the accretion rate jumps and
subsequently the jets become more powerful, seeding magnetic fields at an
average rate of $1100 \mgmyr$. Periodic behavior also sets in with spikes of up
to $8000 \mgmyr$ occurring every $\sim 500$~Myr. Since the feedback in this
case takes place in the volume which is used to measure the accretion rate, the
injected magnetic field rate is highly sensitive to small perturbations in the
gas properties of this region, and thus is very noisy. Also, the total feedback
magnetic energy is distributed over a much smaller volume than the bubbles,
hence the higher magnetic field values.

Both bubble models require $\sim 400$~Myr before the first injection event
takes place.  While the fixed bubbles (model B4A) inject fields about $5$~times
stronger than the randomly-placed bubbles (model B1B), these injections take
place within a fixed radius of $40$~kpc, while the randomly-placed bubbles have
radii of $\sim 200$~kpc initially. Thus while the injected energies are
identical for the initial injection and roughly equal for the subsequent
$4$~Gyr, the randomly-placed bubbles distribute this energy over a large
volume, and hence the magnetic field strength within that volume is lower. The
fixed bubbles reach a peak injection rate of $700 \mgmyr$ before strong
outflows prevent further accretion at $4$~Gyr, which is roughly the same time
when the jets begin their strong feedback phase. The randomly-placed bubbles,
which take longer to drive gas away from the core, only reach a peak injection
rate of $100 \mgmyr$ before the complex field topology prevents further
efficient calculations.  

\begin{figure}
  \centering {\includegraphics[width=\columnwidth]{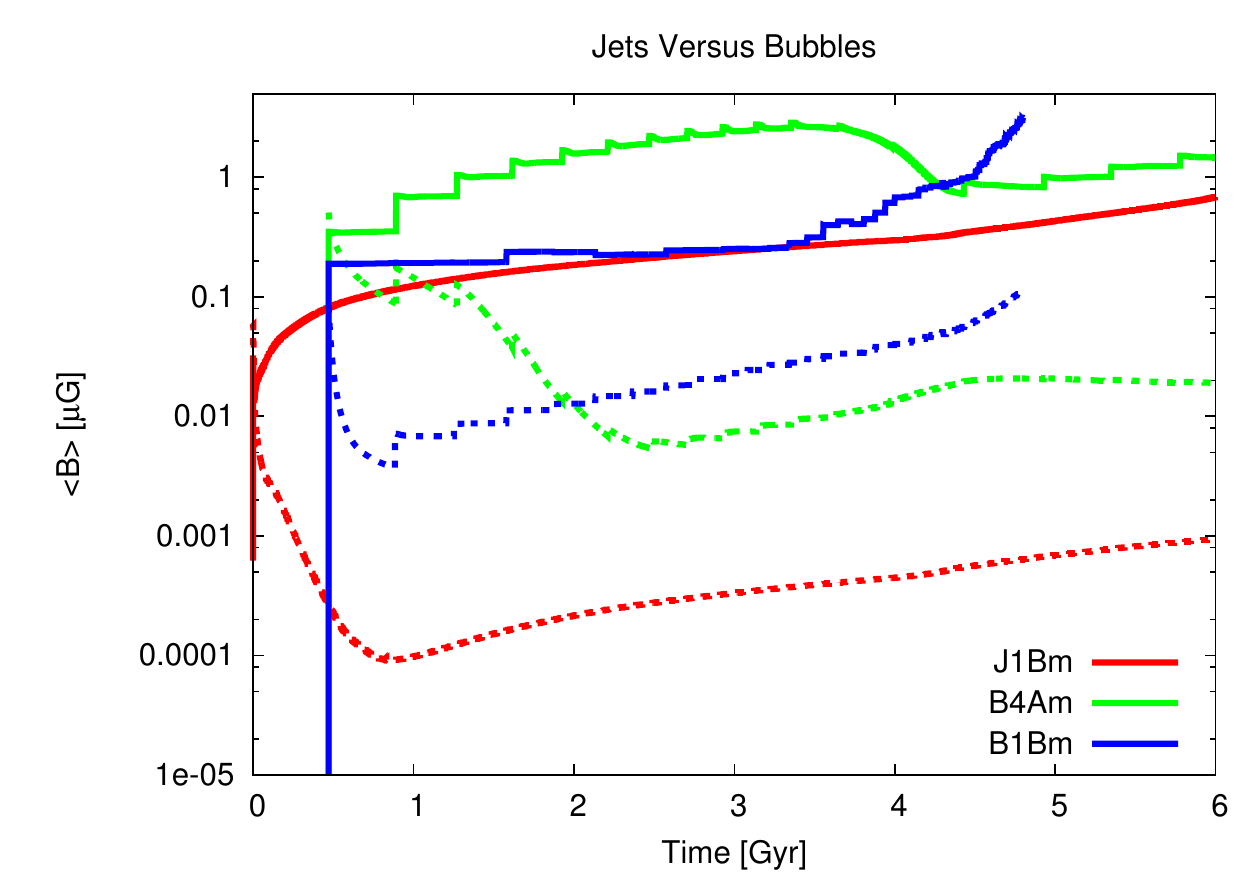}}
  \caption[Average magnetic field strength.]
           {Average magnetic field strength with jets and bubbles. 
           Solid lines are density-weighted 
           average fields within $R_{\rm core}$ and dotted lines are 
           volume-weighted fields within $R_{\rm 200}$.}
\label{fig:agn_growth_aveb}
\end{figure}

We show in~\figref{\ref{fig:agn_growth_aveb}} the average magnetic field
strength within two radii, $R=R_{\rm core} \equiv 0.15 R_{\rm 500}$ and
$R=R_{\rm 200}$. We take a density-weighted average in the core and a
volume-weighted average for the entire cluster. The fixed bubbles are most
efficient at strongly magnetizing the core, reaching $1 \mg$ after just
$1$~Gyr, which is an order of magnitude stronger than the jets and
randomly-placed bubbles, and a peak value of over $2 \mg$ averaged inside the
core after $4$~Gyr. After that time the reduced accretion rate prevents further
strengthening of the field in the core.  At that time strong outflows prevent
further strong feedback events and the magnetic fields in the core push
outwards into the cluster medium without being replenished. The jets, however,
continuously operate and are always able to magnetize the core. Thus while
initially the jets provide the weakest magnetic fields in the core they
eventually build up enough strength to provide roughly the same magnetization
as the fixed bubbles after $6$~Gyr. Randomly-placed bubbles are the least
efficient at magnetizing the core, since the bubbles are large and off-center.
However, they do begin to produce strong fields after $4$~Gyr.   

Jets and fixed bubbles exhibit similar patterns when averaging fields over the
entire cluster. Here the initially-injected fields quickly disperse throughout
the intracluster medium but eventually weaken by escaping beyond $R_{\rm 200}$
via adiabatic expansion and through numerical diffusion.  However, fields from
the core eventually make their way to the outer regions of the cluster and
after $1$-$2$~Gyr the average fields begin to rise again. The bubbles maintain
cluster-wide fields over an order of magnitude stronger than the jets, mainly
because the bubbles inject their fields farther out from the core, allowing
them to more easily propagate into the cluster. The large bubbles of the
randomly-placed bubble model eventually provide magnetic fields throughout the
cluster as strong as the fixed bubbles.

\begin{figure}
  \centering {\includegraphics[width=\columnwidth]{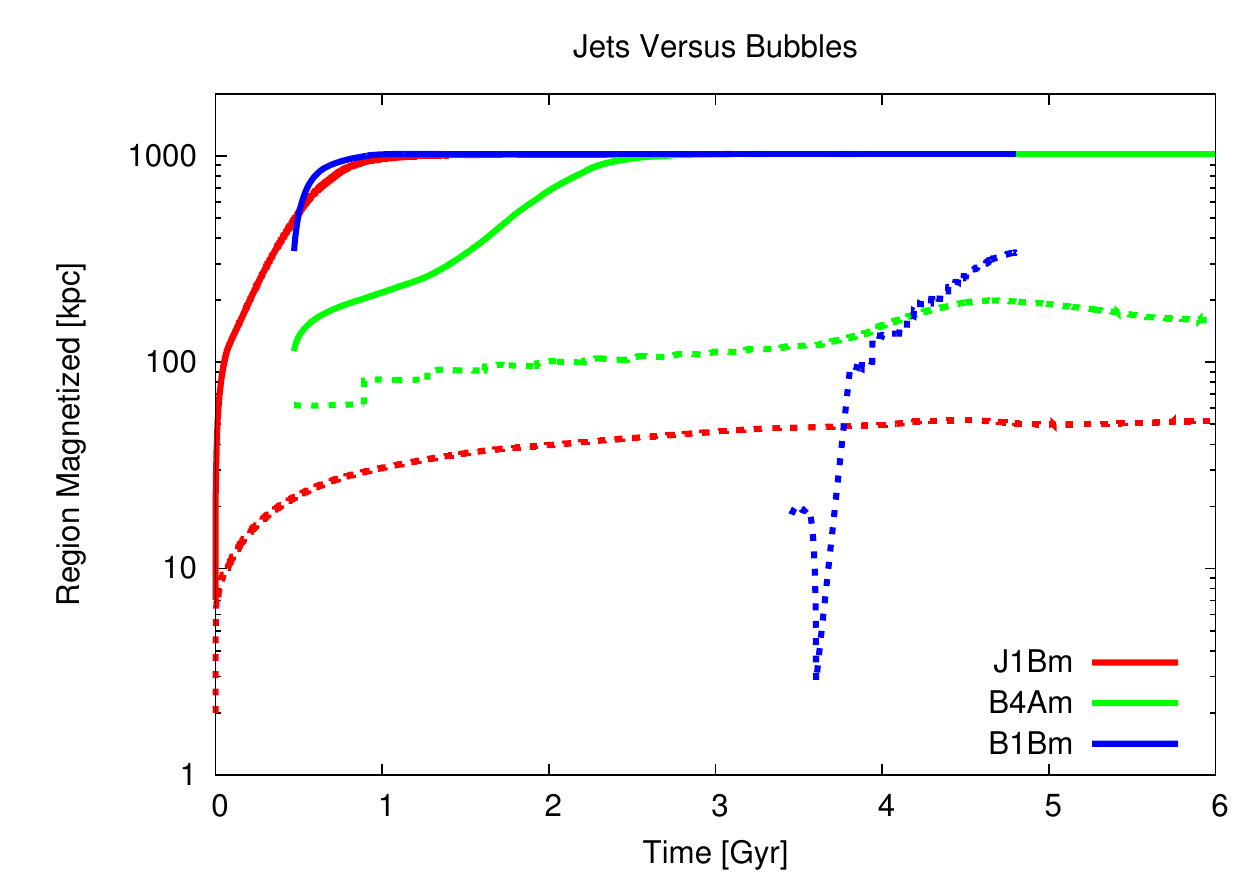}}
  \caption[Magnetized volume.]
           {Volume magnetized due to jets and bubbles. Shown is the cube root 
            of the total volume containing fields of strength at least 
            $10^{-12}$~G (solid lines) and $10^{-6}$~G (dotted lines).}
\label{fig:agn_growth_magvol}
\end{figure}

Next we can study the growth rate of magnetic fields within the cluster volume
in~\figref{\ref{fig:agn_growth_magvol}}. Here we show the total volume
encompassed by fields of strength at least $10^{-12}$~G and $10^{-6}$~G. While
all models are able to weakly magnetize the entire cluster, they differ
substantially in the time taken to do so.  The jets (model J1B) continuously
inject strong fields in the core which push out weaker fields into the cluster,
saturating the cluster within $1$~Gyr. The randomly-placed bubbles (model B1B)
saturate the cluster almost immediately since the bubbles are very large and
centered away from the core.  The fixed bubbles (model B4A) take the longest,
$2$~Gyr, since these injected fields are centered at the core but are much
weaker than the fields injected by the jets and hence less efficient at pushing
weaker fields throughout the cluster. 

Strong fields permeate a much smaller volume, with jets only able to provide
$\sim \mg$~fields within the inner $50$~kpc. The fixed bubbles, with radii of
$40$~kpc, are naturally able to strongly magnetize almost the entire core out
to a radius of $100$~kpc and maintain this magnetization for five billion
years. The randomly-placed bubbles cannot produce significant magnetic fields
until multiple bubbles overlap so that their magnetic contributions can add
together, which does not occur until $3.5$~Gyr of evolution. These fields
quickly disperse but recover when the bubble feedback becomes strong enough to
continuously produce strong fields. 

\begin{figure}
  \centering {\includegraphics[width=\columnwidth]{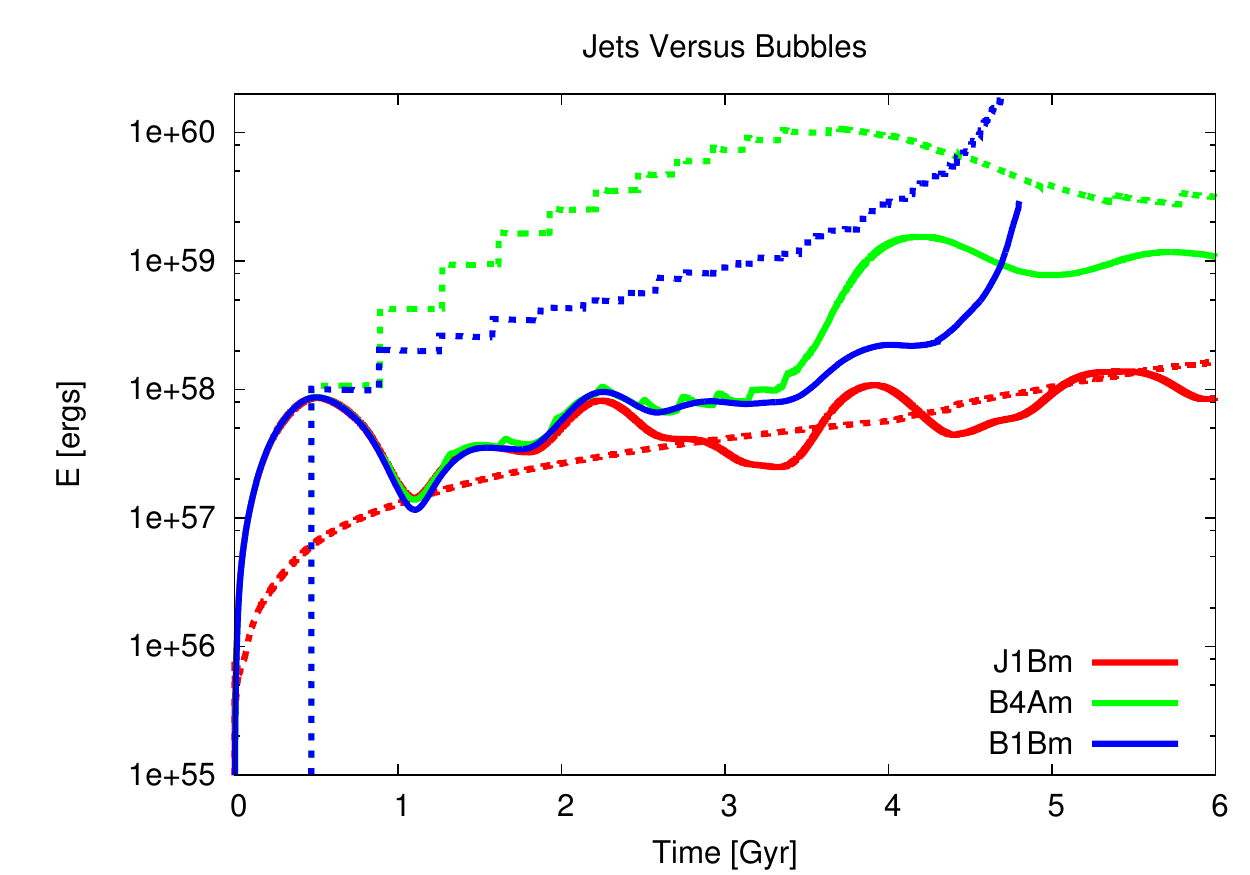}}
  \caption[Total cluster kinetic and magnetic energy.]
           {Total cluster kinetic and magnetic energy versus time. 
           Solid lines are kinetic and dotted lines are magnetic energy.}
\label{fig:agn_growth_ener}
\end{figure}

\figref{\ref{fig:agn_growth_ener}} shows the total kinetic and magnetic
energies within the entire cluster volume as a function of time. The evolution
of the magnetic energy follows the same evolution as the injection rate.  The
kinetic energy follows a strongly periodic behavior as gas cools and accretes
onto the black hole and is driven outwards by outflows.  The kinetic energy for
all cases also follows the magnetic energy, since magnetic tension in the
injected fields drives additional gas motions.  Since the fields of the jet
model are centrally located, they are less likely to drive large-scale
complicated gas motions, whereas the bubbles with large, diffuse, tangled
magnetic fields generate complicated gas motions. The turbulence due to these
bubbles produces kinetic energies an order of magnitude larger than that due to
the jets.

\begin{figure}
  \centering 
  \centering {\includegraphics[width=\columnwidth]{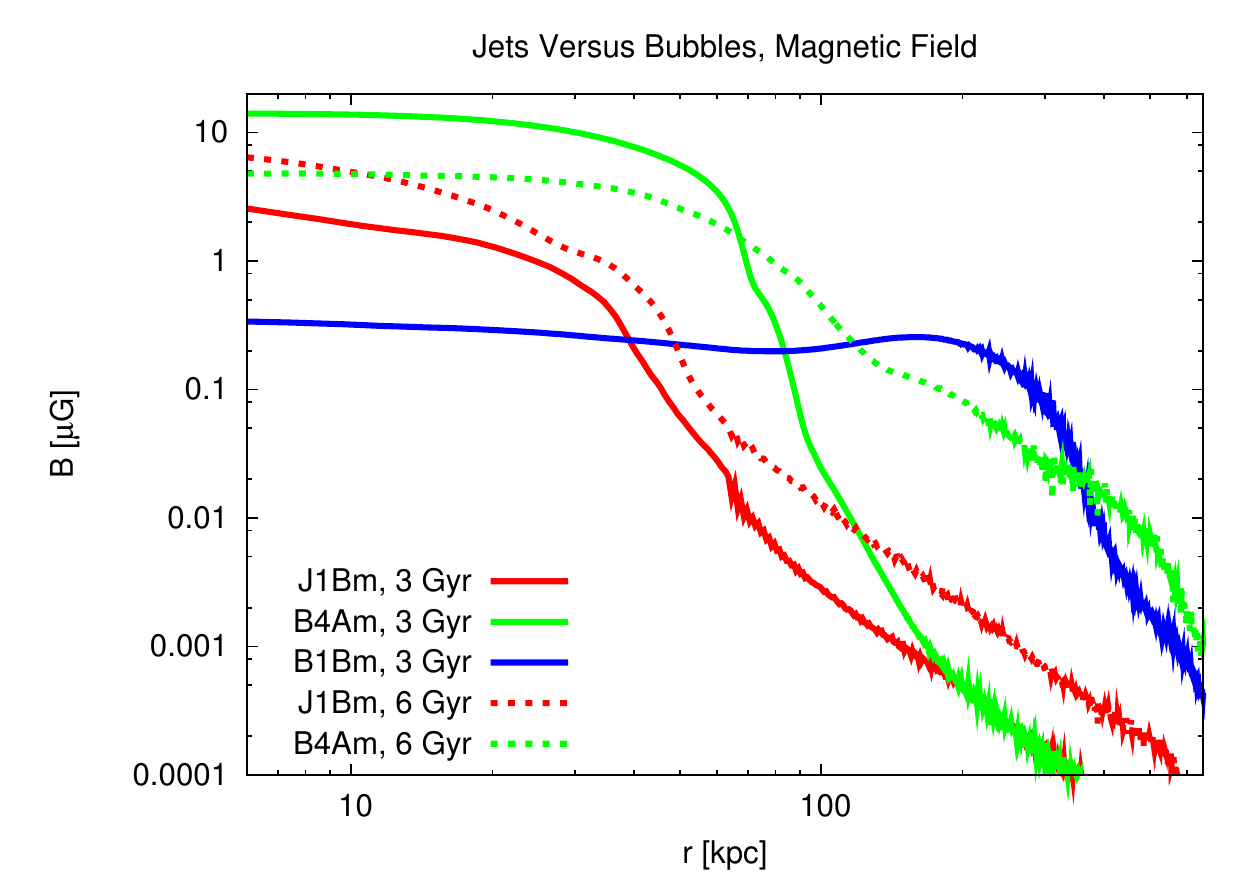}}
  \caption[Magnetic field radial profiles.]
           {Volume-weighted magnetic field as a function of radius.}
\label{fig:agn_growth_profile_mag}
\end{figure}

Finally we show radial profiles of the magnetic field
in~\figref{\ref{fig:agn_growth_profile_mag}} at $3$~and $6$~Gyr. We construct
these profiles from volume-weighted averaging of the field in $1$~kpc shells.
At $3$~Gyr, the fixed bubbles maintain the strongest fields ($10$~times
stronger than the jets and $30$~times stronger than the randomly-placed
bubbles) within the central $100$~kpc but at large radii these fields diminish
rapidly. Both bubble models have the same profile shape: an inner plateau where
the fields are injected and a steep reduction in field strength beyond that
radius. The randomly-placed bubbles, which place weaker magnetic fields within
a larger region, maintains this plateau to almost $300$~kpc, which is sensible
because the bubbles involved have typical radii of $150$~kpc. The jets produce
a gradually declining field within the inner $50$~kpc and the strongest fields
in the innermost core, as expected. All the models produce only weak fields ($<
10^{-12}$~G) beyond a radius of $600$~kpc.

At $6$~Gyr, both the jets and fixed bubbles (the randomly-placed bubbles only
evolved to $4$~Gyr) exhibit opposite trends. The jets continuously amplify the
fields in the core, boosting the field strength in the inner $50$~kpc by a
factor of four while continuing to strengthen the magnetic fields at larger
radii. The fixed bubbles, however, experience a drop in accretion rate at $\sim
4$~Gyr and hence their ability to magnetize the core after that diminishes.
Thus the fields from the inner regions of the cluster just propagate outwards,
magnetizing the remainder of the cluster while reducing the magnetization of
the core. After $6$~Gyr the jets eventually produce the strongest magnetic
fields in the core, but the fixed bubbles generate a more highly-magnetized
cluster overall.

\section{Magnetic field topology}
\label{sec:agn_topology}

\begin{figure*}
  \centering 
  {\includegraphics[scale=0.15]{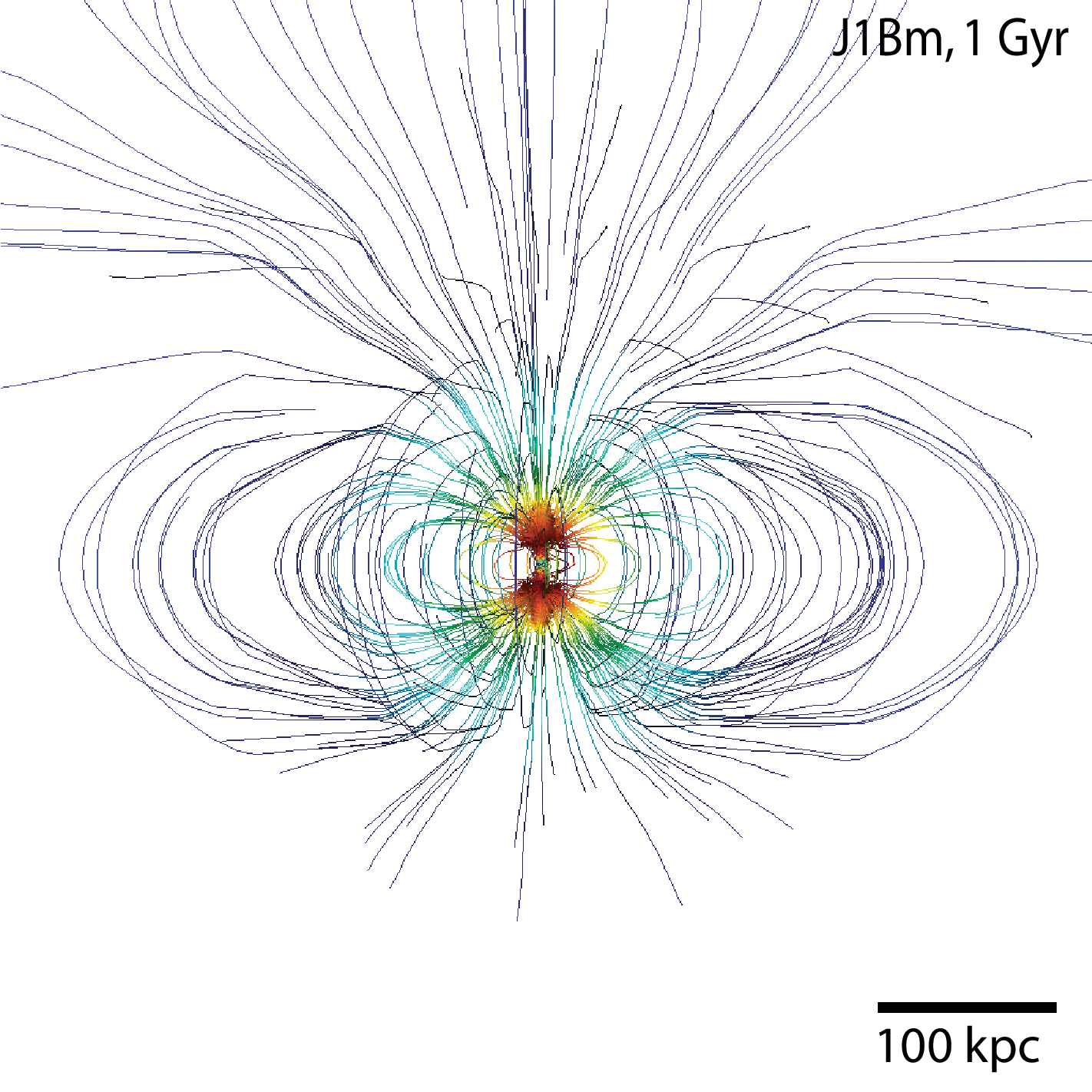}}
  {\includegraphics[scale=0.15]{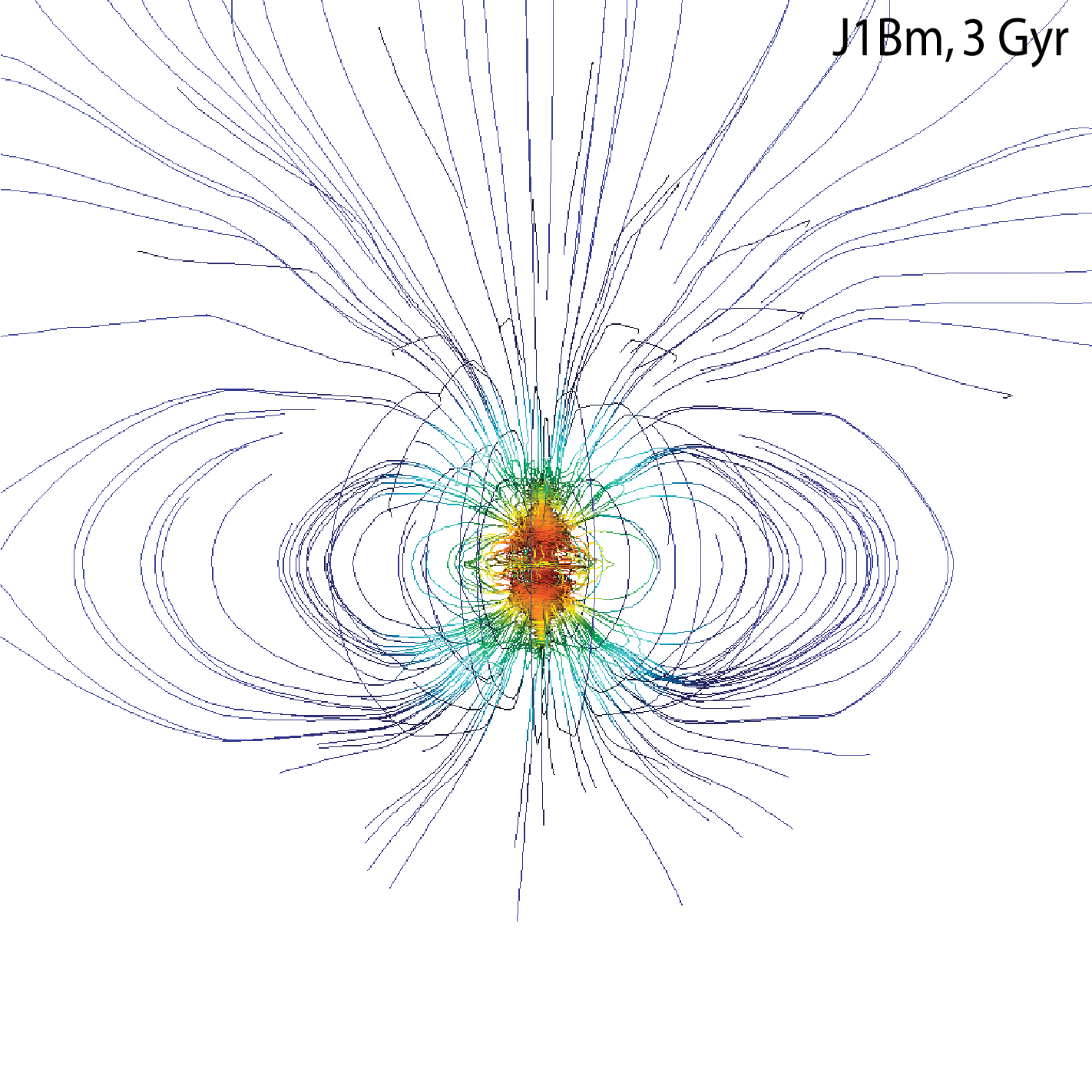}}
  {\includegraphics[scale=0.15]{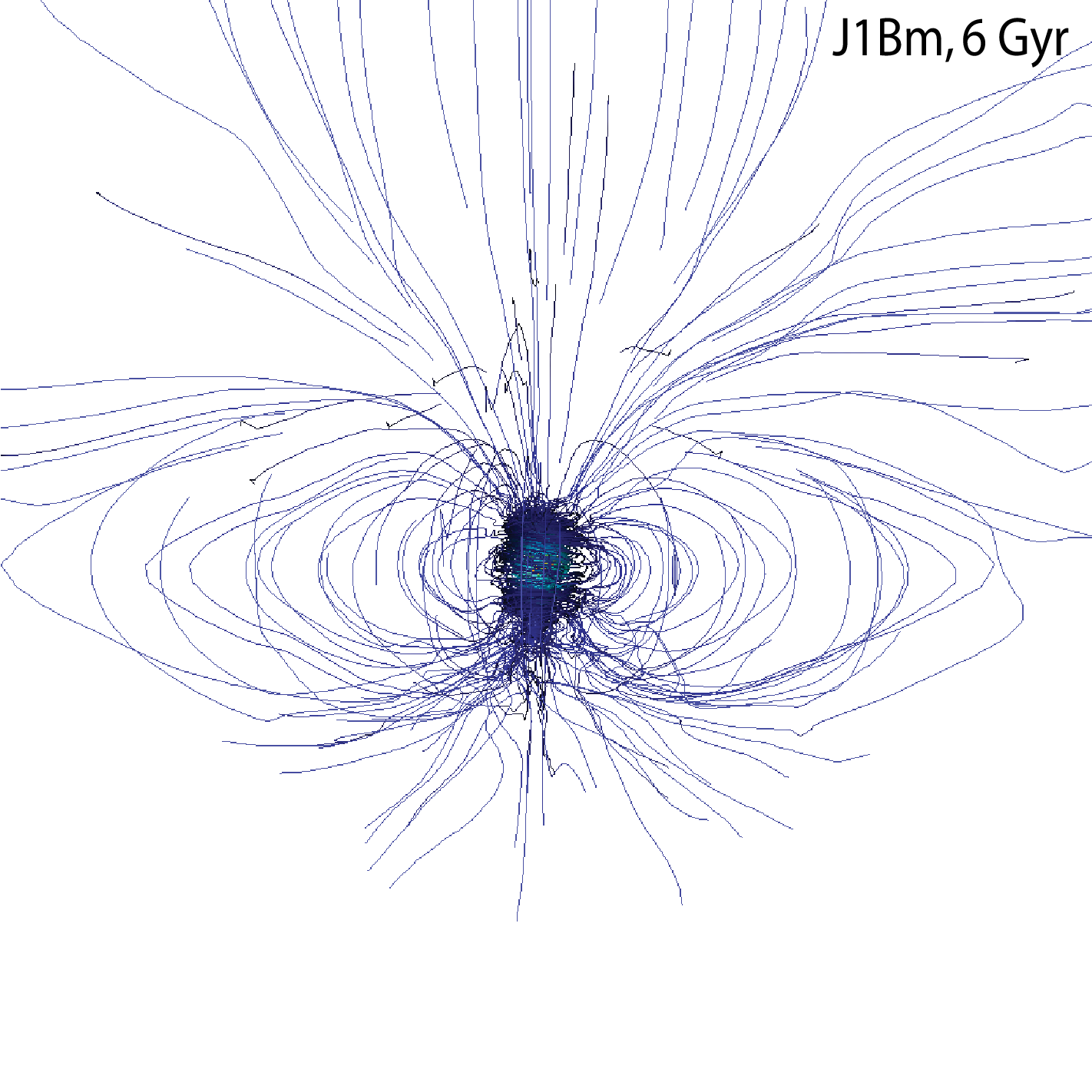}}
  
  {\includegraphics[scale=0.15]{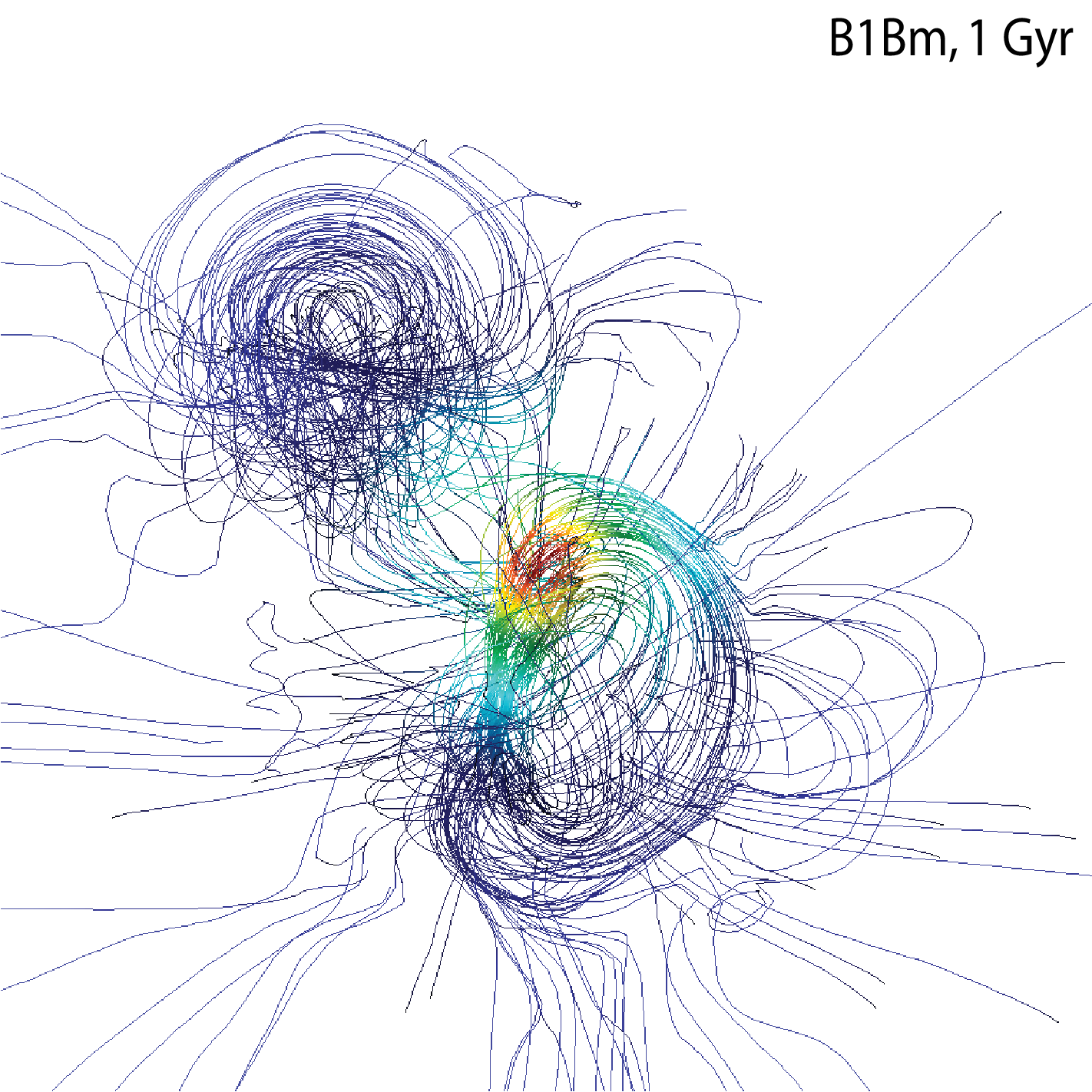}}
  {\includegraphics[scale=0.15]{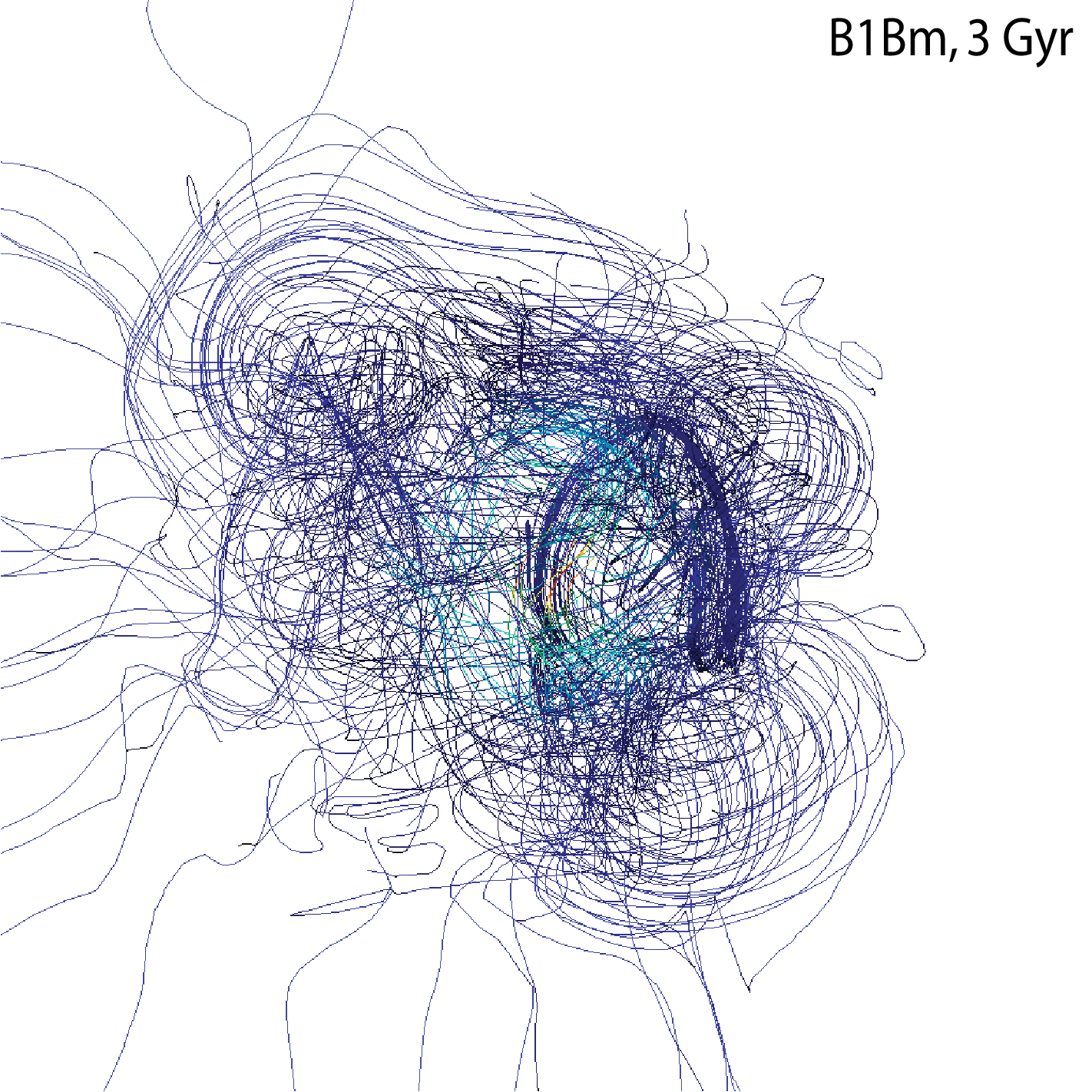}}
  {\includegraphics[scale=0.15]{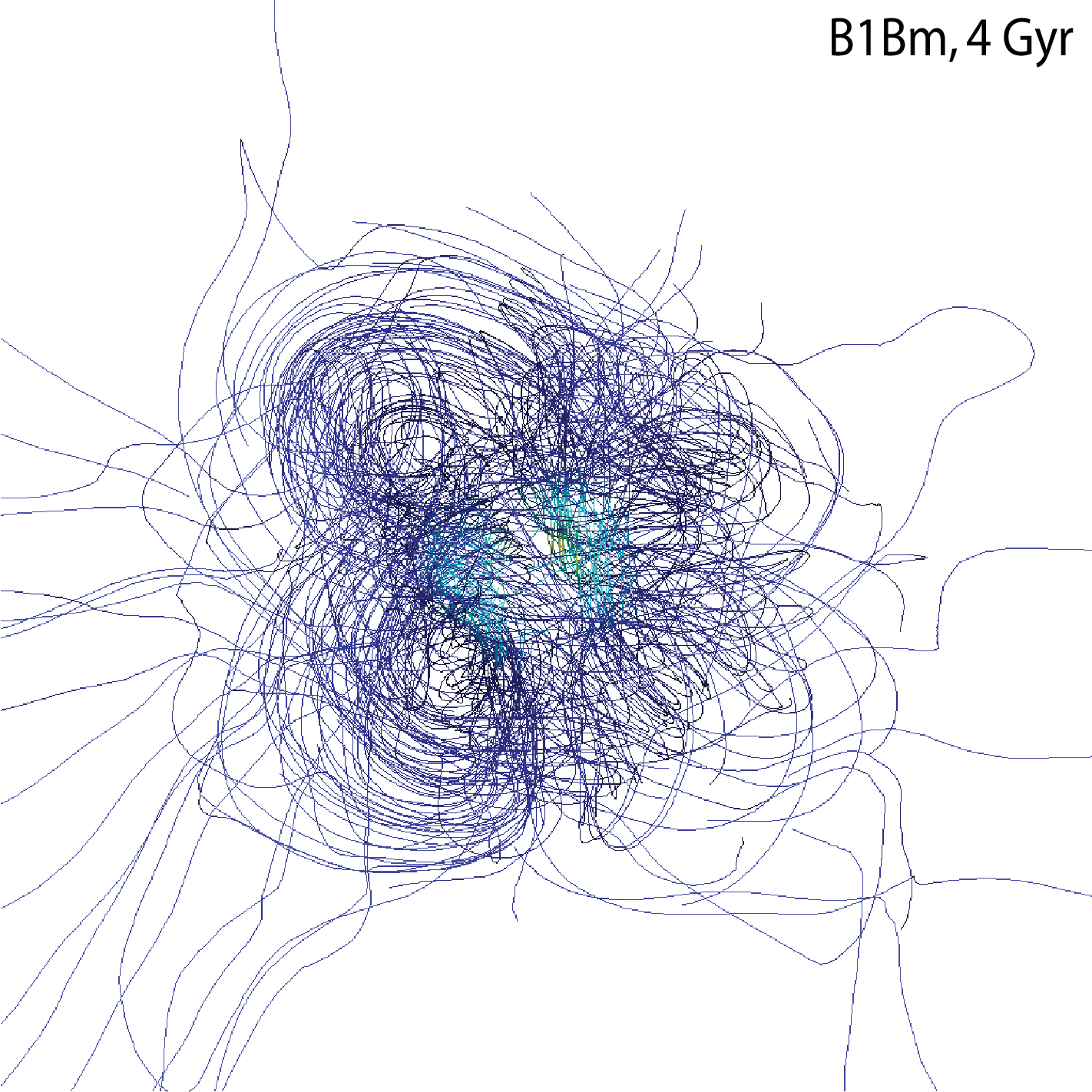}}
  
  {\includegraphics[scale=0.15]{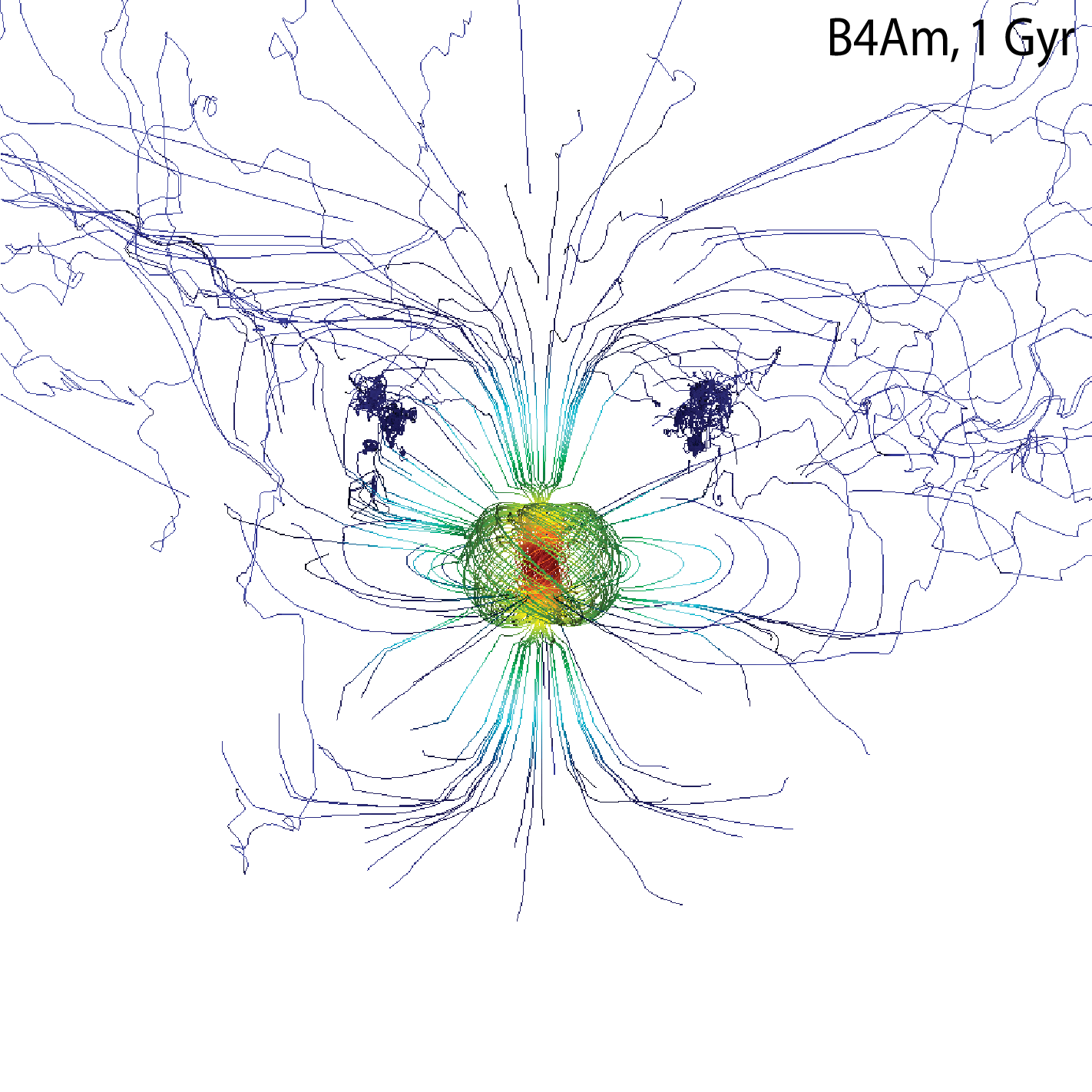}}
  {\includegraphics[scale=0.15]{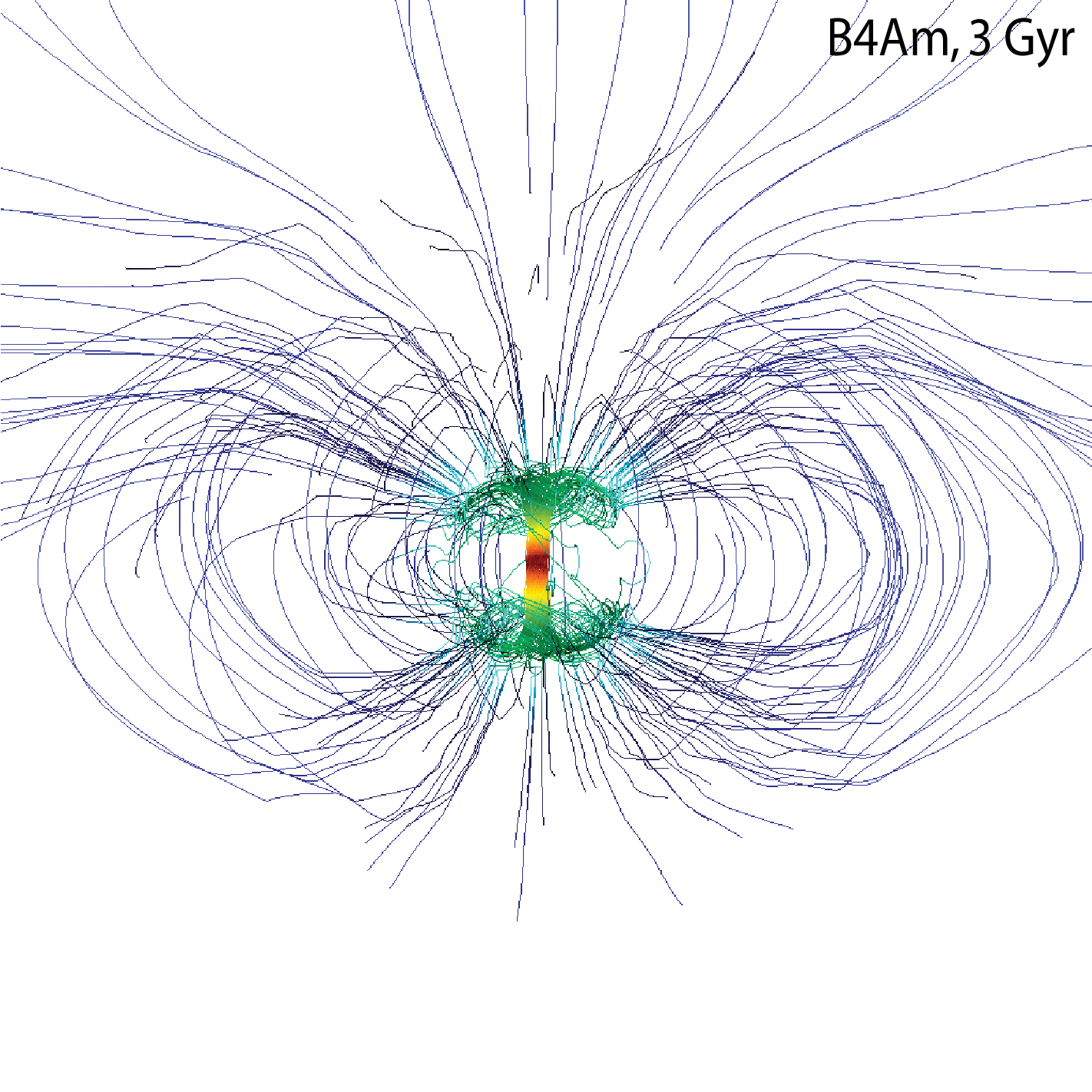}}
  {\includegraphics[scale=0.15]{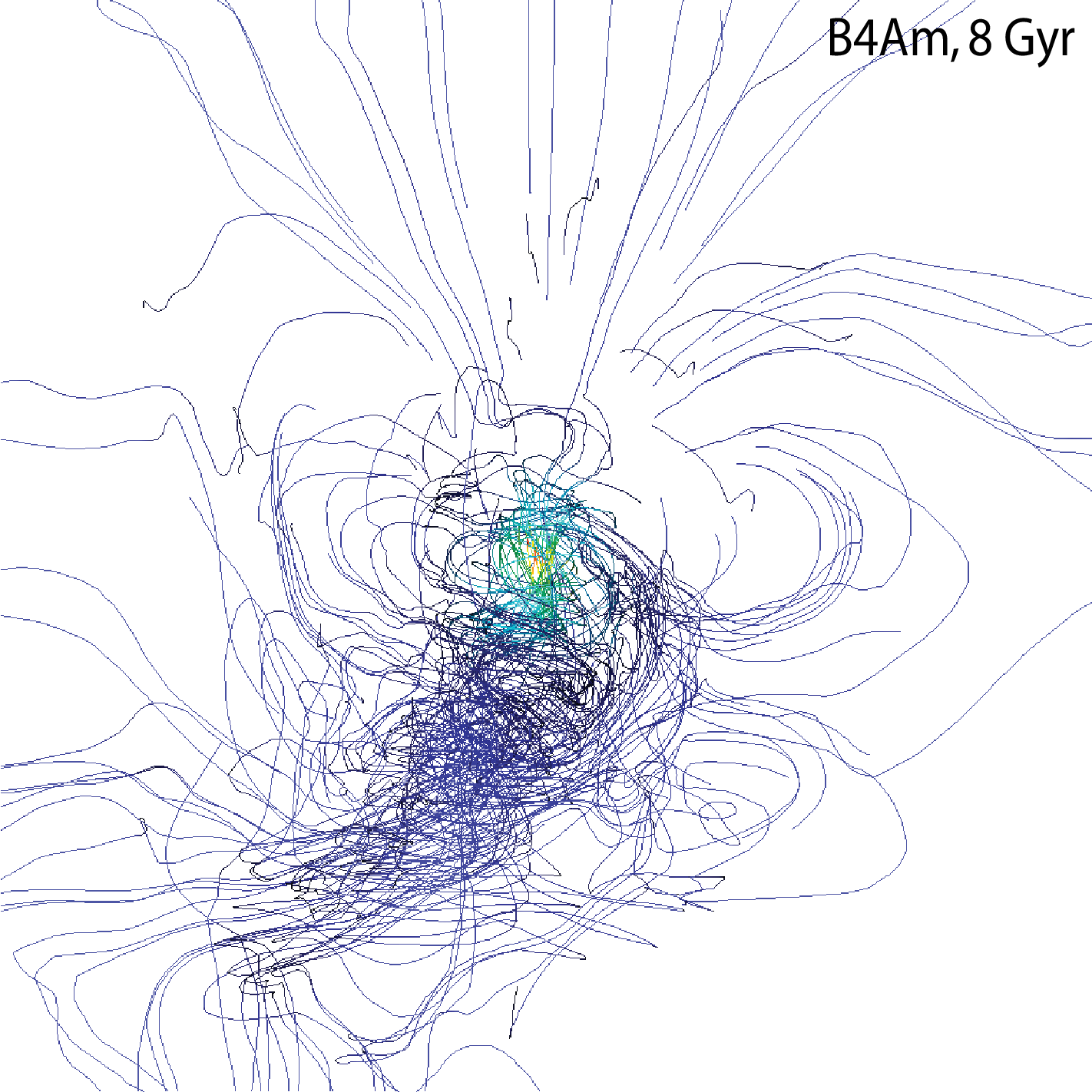}}

 %
 %
  
 %
 %
  \caption[Evolution of magnetic field lines.]
           {Evolution of magnetic field lines for the fiducial models. 
            Shown are models J1Bm (top), B1Bm (middle), and B4Am (bottom).
            From left to right, model J1Bm is shown at times $1$, $3$, and 
            $6$~Gyr, model B1Bm is shown at times $1$, $3$, and $4$~Gyr, and 
            model B4Am is shown at times $1$, $3$, and $8$~Gyr.
            All field lines are constructed using $100$ tracer particles 
            uniformly distributed on a sphere with radius $300$~kpc.
            The field lines are colored using the local gas density, with 
            blue colors indicating low density and red 
            colors denoting high-density regions. The view is 
            roughly $600$~kpc across.}
\label{fig:agn_topology_streams_evolution}
\end{figure*}

We now turn to the topology of the fields produced by our fiducial jet and
bubble models: J1B to represent the jets, B1B for randomly-placed bubbles, and
B4A for fixed bubbles.  Field lines demonstrate the complex morphology produced
by the AGN-based injection. We generate field lines by seeding $100$ tracer
particles uniformly on a sphere of radius $300$~kpc.
\figref{\ref{fig:agn_topology_streams_evolution}} shows the field lines for the
fiducial models at various times. Each model is shown at $1$ and $3$~Gyr and at
a time just before complex fields prevent further calculation. Thus we show the
field lines at $6$~Gyr for the jets, $4$~Gyr for the randomly-placed bubbles,
and $8$~Gyr for the fixed bubbles.

The jets initially produce a toroidal shape with a very regular structure: as
the fields are injected as a torus they push out already-present fields which
maintain their structure.  After $3$~Gyr the fields elongate along the jet axis
forming a tower with cavities opening in the plane perpendicular to the jet.
The magnetic field is able to escape quickly along the jet axis, quickly
magnetizing volumes outside the core. Finally, at $6$~Gyr, the gas is able to
cool sufficiently to begin accreting, dragging the field lines along with it as
it accretes onto the core. We thus form a dense cocoon of magnetic field lines.
These fields are only slightly tangled and maintain their regular torus-like
structure, though stronger fields have been able to propagate farther into the
cluster.

The randomly-placed bubbles exhibit complex, tangled behavior almost
immediately.  While each bubble contains a torus-like magnetic field, as
multiple bubbles form and overlap the fields entangle. The bubbles are able to
form strong fields with complex structures far from the central core. At later
times more bubbles form and overlap, increasing the average magnetic field in
the core.  Few magnetic field lines are able to escape into the outer cluster
regions, even after $4$~Gyr. However, the volume magnetized by bubbles is much
larger than that of the jets. The bubbles are also much more asymmetric than
the fields produced by the jets, since each injected bubble has a new, random
field orientation.

The fixed bubbles initially exhibit only somewhat ordered topology.  While each
bubble has the same magnetic field orientation, between bubble events some
magnetic field lines can escape and turbulence caused by the infalling gas can
twist and tangle the fields.  At $3$~Gyr, however, the bubbles appear
frequently enough to maintain a torus-like topology. Eventually the fields
collapse due to the asymmetric infill of the gas, and a large parcel of
magnetic field lines escapes along the jet axis, leading to a complex topology
containing both ordered torus-like and tangled components.

\begin{figure*}
  \centering 
  {\includegraphics[scale=0.145]{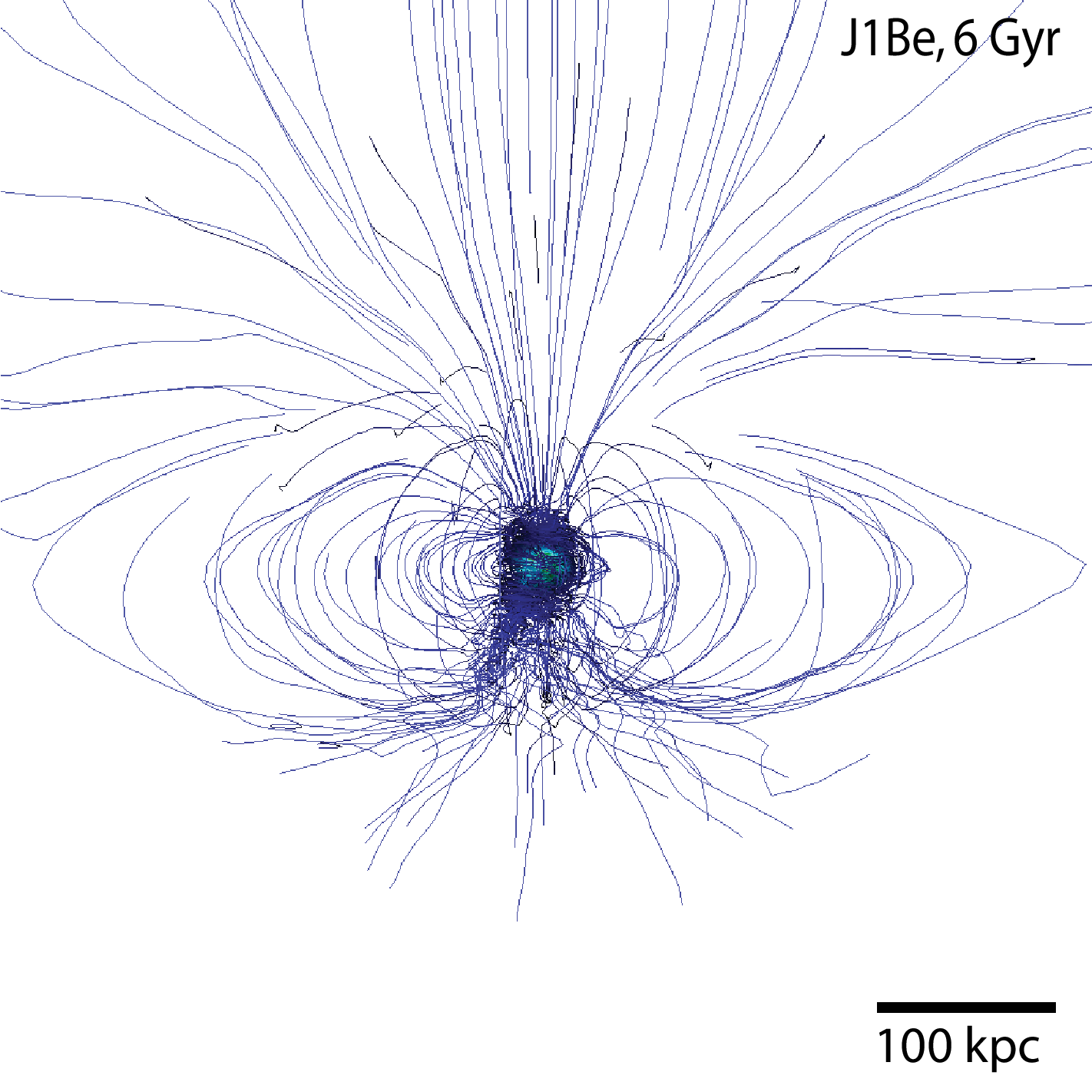}}
  {\includegraphics[scale=0.145]{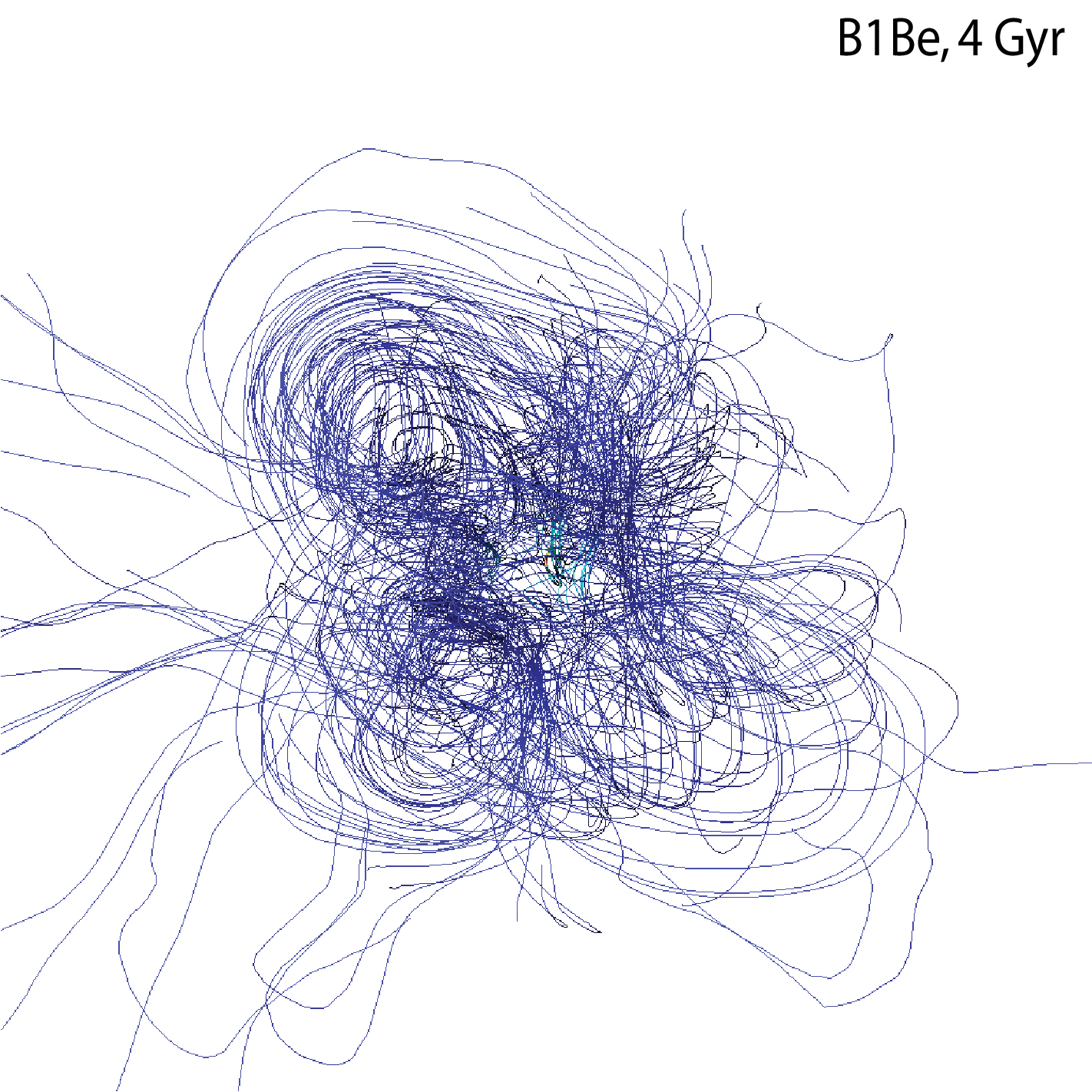}}  
  {\includegraphics[scale=0.145]{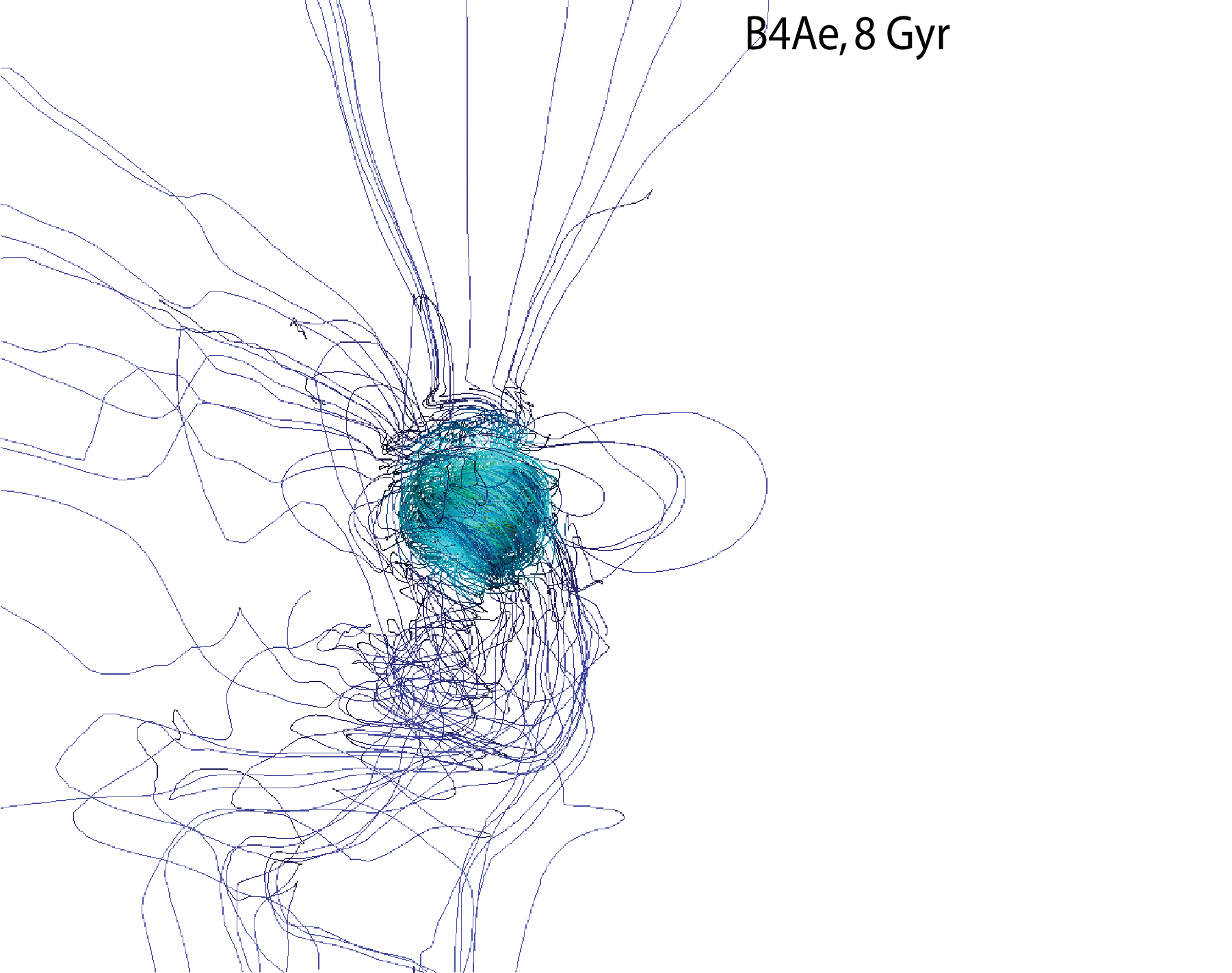}}  
  \caption[Magnetic field lines of equipartition injection]
           {Magnetic field lines from the equipartition injection mode. 
            Shown are models J1Be at $6$~Gyr (top), 
            B1Be at $4$~Gyr (middle), 
            and B4Ae at $8$~Gyr (bottom).
            Construction of the lines and coloring is identical 
            to~\figref{\ref{fig:agn_topology_streams_evolution}}.}
\label{fig:agn_topology_streams_equi}
\end{figure*}

We also show the equipartition case for each fiducial model
in~\figref{\ref{fig:agn_topology_streams_equi}} at the latest time available
for each model. For each model the magnetic fields are weaker than in the
purely magnetic case, but the larger topology remains unaffected.  The final
cocoon created by the jets is smaller since the injected fields are not as
strong. The weaker fields of both the randomly-placed and fixed bubbles produce
much more disperse field lines. Thus while we affect the overall strength of
magnetic fields we do not change the topological structure of the fields when
considering injection modes with less magnetic energy. Also, the addition of
thermal energy in the equipartition case does not affect the topology of the
fields.

\begin{figure*}
  \centering 
  {\includegraphics[scale=0.28]{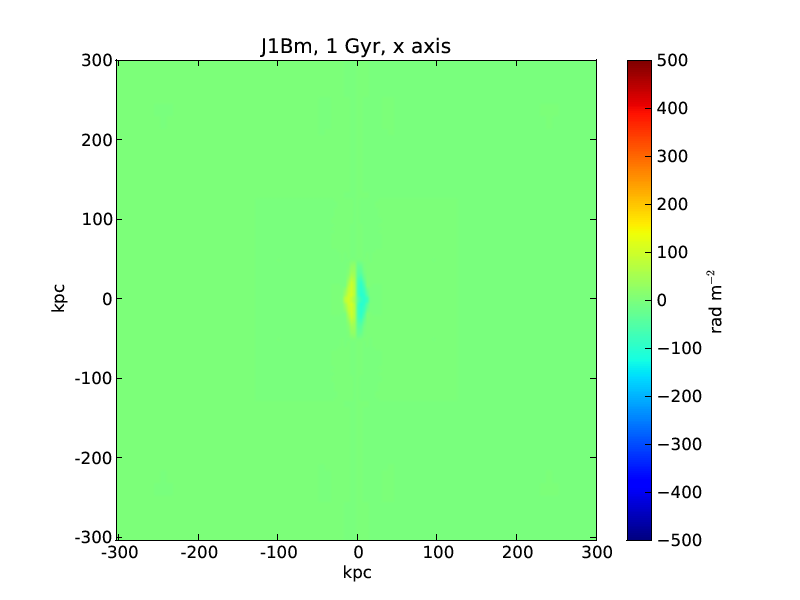}} 
  {\includegraphics[scale=0.28]{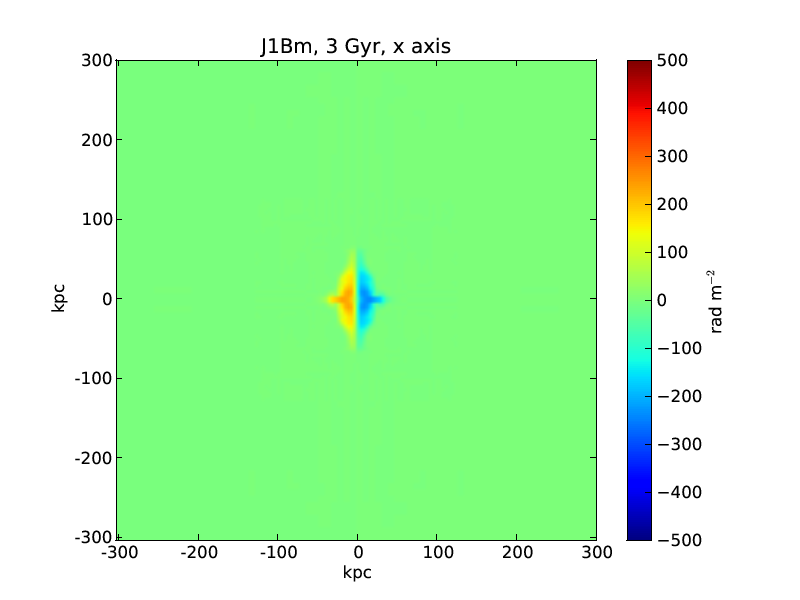}} 
  {\includegraphics[scale=0.28]{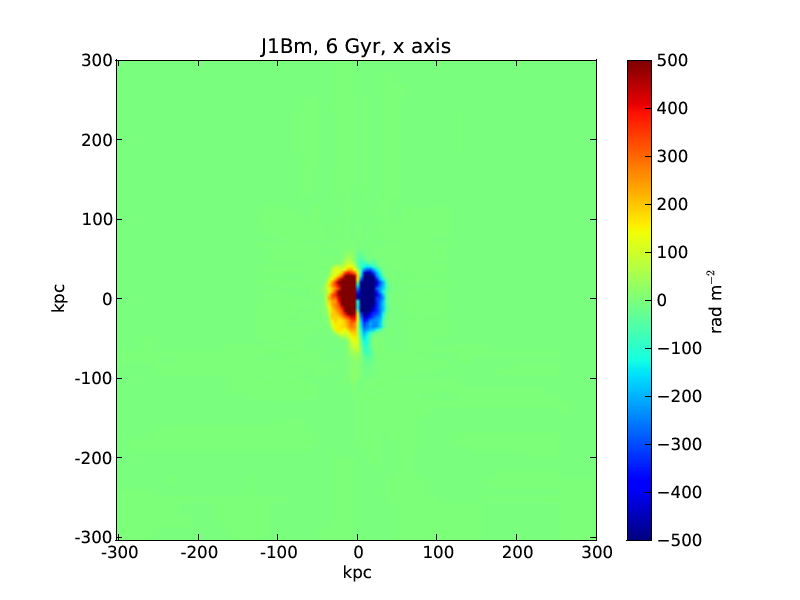}}
  {\includegraphics[scale=0.28]{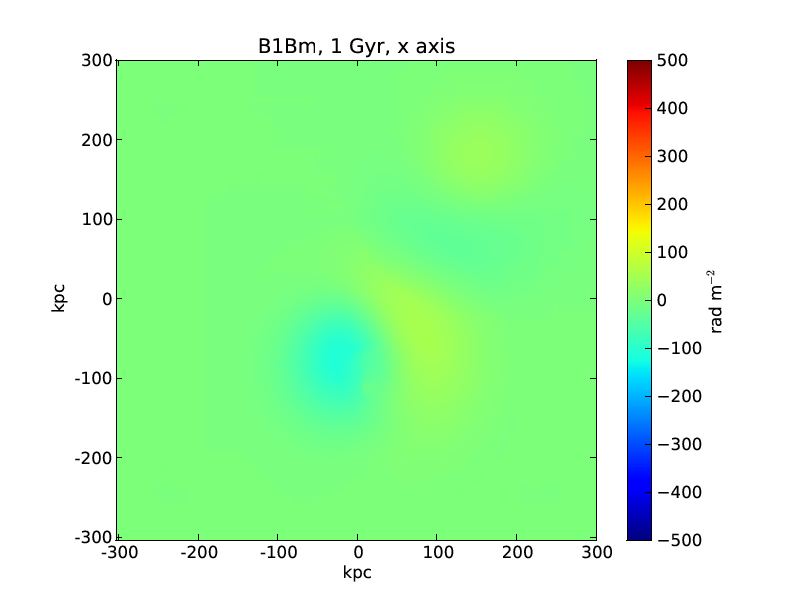}}
  {\includegraphics[scale=0.28]{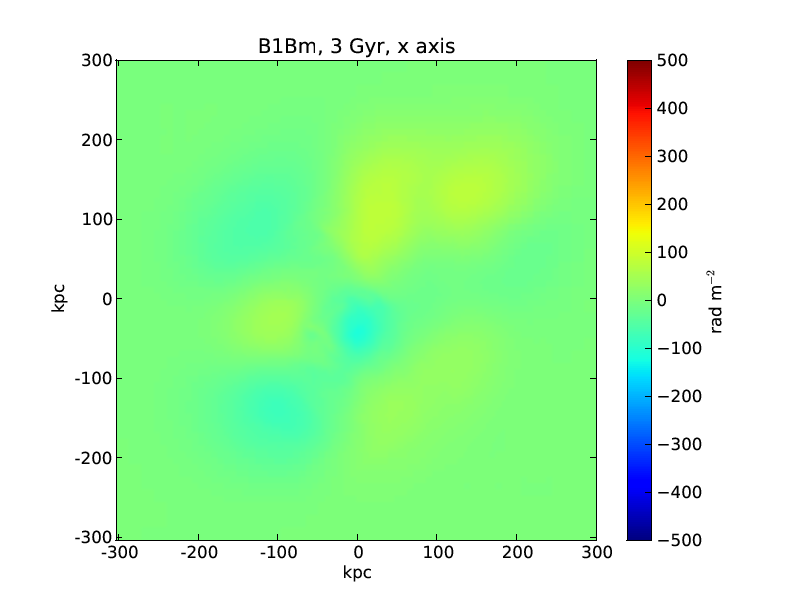}}   
  {\includegraphics[scale=0.28]{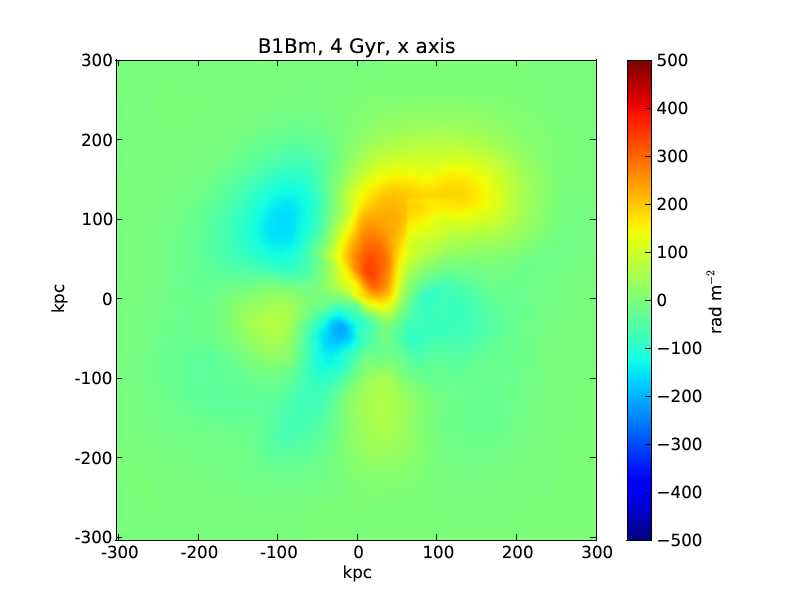}}
  {\includegraphics[scale=0.28]{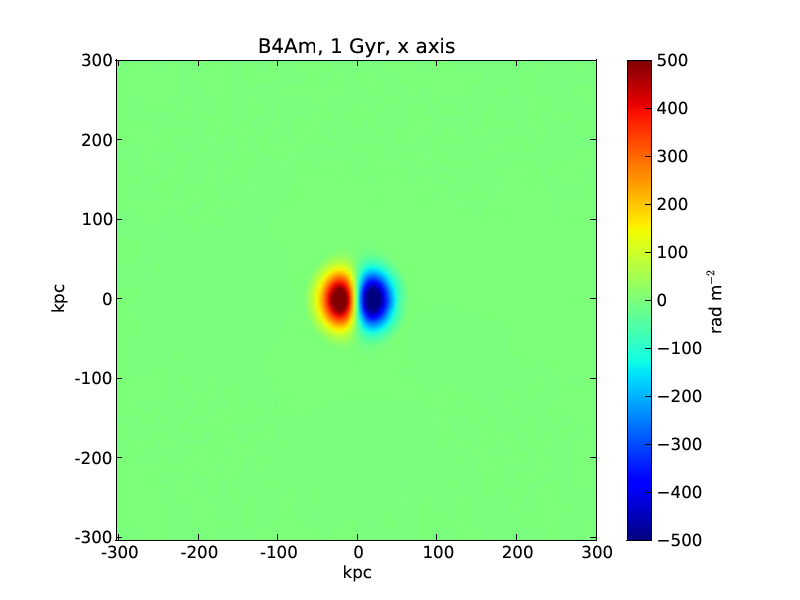}}
  {\includegraphics[scale=0.28]{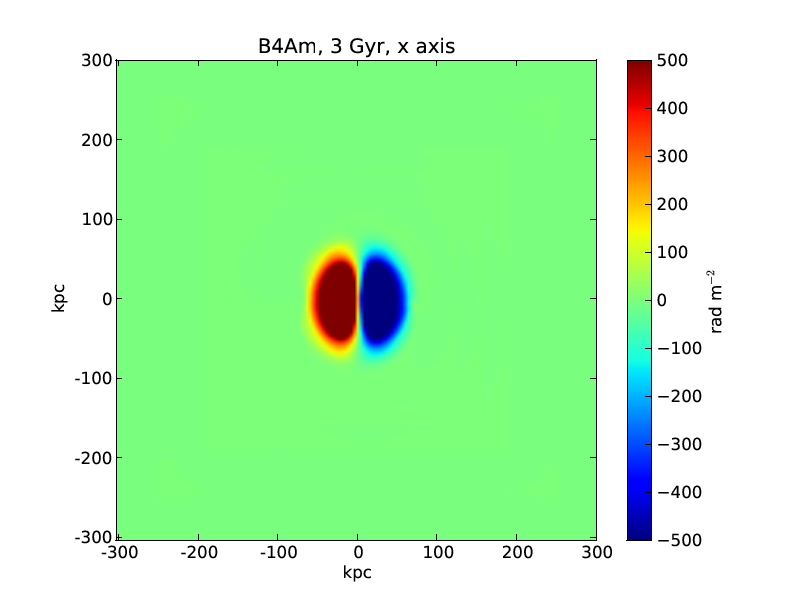}}
  {\includegraphics[scale=0.28]{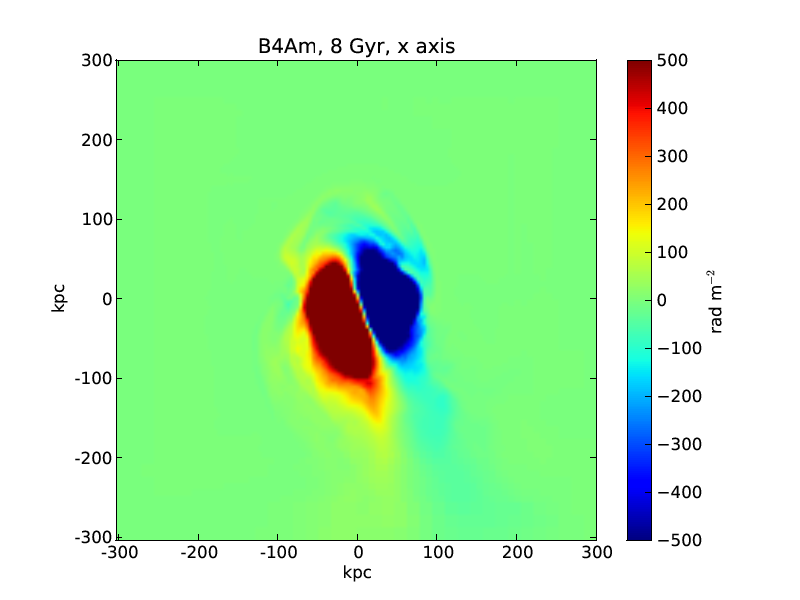}}
  \caption[Evolution of the rotation measure.]
           {Evolution of the rotation measure for the fiducial models. 
            Shown are models J1Bm (top), B1Bm (middle), and B4Am (bottom).
            From left to right, model J1Bm is shown at times $1$, $3$, and 
            $6$~Gyr, model B1Bm is shown at times $1$, $3$, and $4$~Gyr, and 
            model B4Am is shown at times $1$, $3$, and $8$~Gyr.
            All rotation measures are taken along the $x$-axis and have a 
            uniform resolution of $4$~kpc. Note that the color scale is chosen to
            enhance the regions outside the core. 
            The true minima and maxima for each model are listed 
            in Table~\ref{tab:agn_rm}.}
\label{fig:agn_topology_rm_evolution}
\end{figure*}

Next in~\figref{\ref{fig:agn_topology_rm_evolution}} we see
the evolution of the 
 magnetic field in terms of the rotation measure (RM) taken along the 
 $x$-axis. The rotation measure is defined as
\begin{equation}
  {\rm RM} = 812 \int_0^L n_e {\bf B} \cdot d{\bf l} 
             ~{\rm radians}~{\rm m}^{-2},
\label{eq:agn_rm}
\end{equation}
where ${\bf B}$ is the magnetic field 
in $\mu$G, $n_e$ is the electron density, and ${\bf l}$ 
is the direction of light propagation in \kpc 
(positive values indicate fields pointed towards the observer). 
The maps shown have a 
spatial resolution of $4$~kpc.
In our plots, we have truncated the color scale to indicate RM values between -500 and 500 rad m$^{-2}$ so that we may readily compare the structures of different 
injection models. The true minima and maxima for each model are listed 
in Table~\ref{tab:agn_rm}.

\begin{table}
  \centering
  \caption{Extreme rotation measure values.}
\begin{tabular}{ccccc}
 \hline
 \hline
 Model & Time (Gyr) & Axis & Min & Max  \\
 \hline
 fJ1Bm & 6 & x & -2132.38 & 2050.55 \\
fJ1Bm & 6 & y & -2588.3 & 1847.2 \\
fJ1Bm & 6 & z & -437.649 & 4016.64 \\
fJ1Be & 6 & x & -1547.93 & 1695.04 \\
 fB1Bm & 4 & x & -209.177 & 338.916 \\
fB1Bm & 4 & y & -244.671 & 275.929 \\
fB1Bm & 4 & z & -561.889 & 115.311 \\
fB1Be & 4 & x & -150.565 & 444.04 \\
 fB4Am & 8 & x & -2400.3 & 3101.37 \\
fB4Am & 8 & y & -3483.68 & 1906.73 \\
fB4Am & 8 & z & -1227.99 & 7642.16 \\
fB4Ae & 8 & x & -2448.87 & 3706.81 \\
 \hline
 \end{tabular}
\label{tab:agn_rm}
\end{table}

The jets at $1$~Gyr show high rotation measure values with very little spatial
extent: only $\sim 10$~kpc perpendicular to the jet and $\sim 50$~kpc along the
jet axis. The RM exhibits the handedness given to the injected magnetic fields,
with a value of $150~{\rm rad}~{\rm m}^{-2}$ to the left of the jet and
$-150~{\rm rad}~{\rm m}^{-2}$ to the right.  At $4$~Gyr the strong RM region in
the center is slightly larger with peak values about three times stronger.  By
$6$~Gyr the strong feedback episodes have dramatically increased the peak RM
value to over $2500~{\rm rad}~{\rm m}^{-2}$ and the collapsing gas has shrunk
the volume enclosed by strong magnetic fields.  Also, turbulent gas motions
have begun to introduce some asymmetry in the RM map; however, the overall
structure remains intact.

The magnetic fields produced by the randomly-placed bubbles result in
large-scale ($\sim 400$~kpc) weak RM structures within $1$~Gyr. However, the RM
here is much weaker - reaching a value of no more than $400~{\rm rad}~{\rm
m}^{-2}$ by $4$~Gyr - than those generated by the jets since the magnetic
fields are distributed over a much larger volume.  As more bubbles form, the
peak RM values increase and distribute over a large volume, eventually forming
$\sim 100$~kpc structures surrounding the central core.  The fixed bubbles show
a similar handedness as the jets since once again each injection is given the
same axis.  However, the RM here reaches a peak value much stronger than both
the randomly-placed bubbles and the jets. Additionally, strong RM values are
distributed over a much larger volume, reaching $\sim 100$~kpc into the cluster
atmosphere.

\begin{figure*}
  \centering 
  {\includegraphics[scale=0.28]{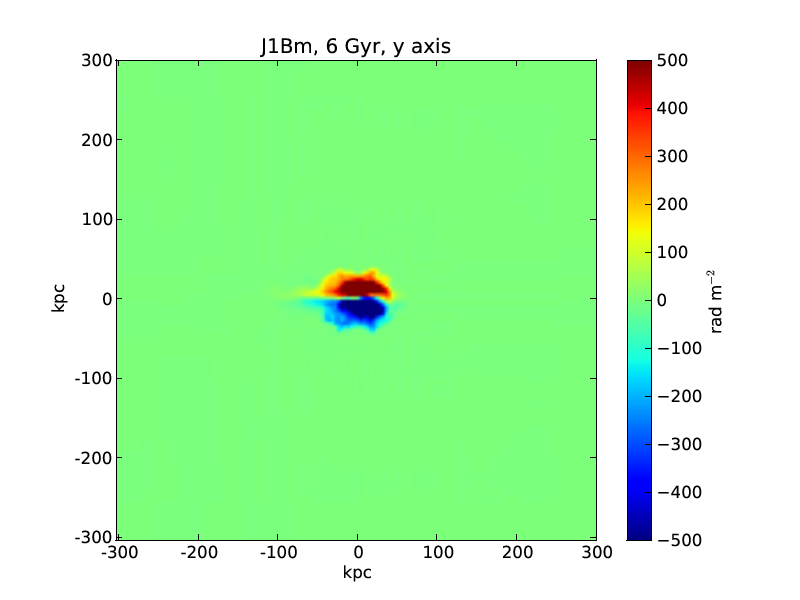}} 
  {\includegraphics[scale=0.28]{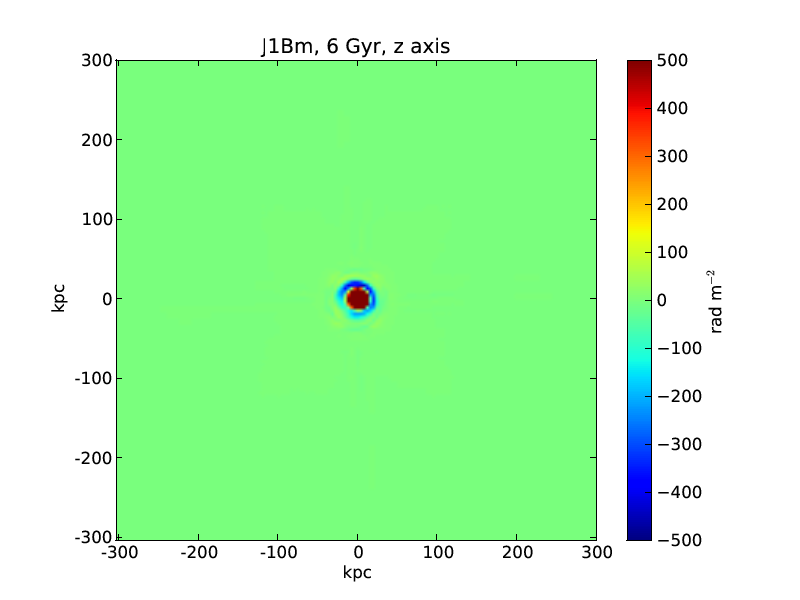}}\\ 
  {\includegraphics[scale=0.28]{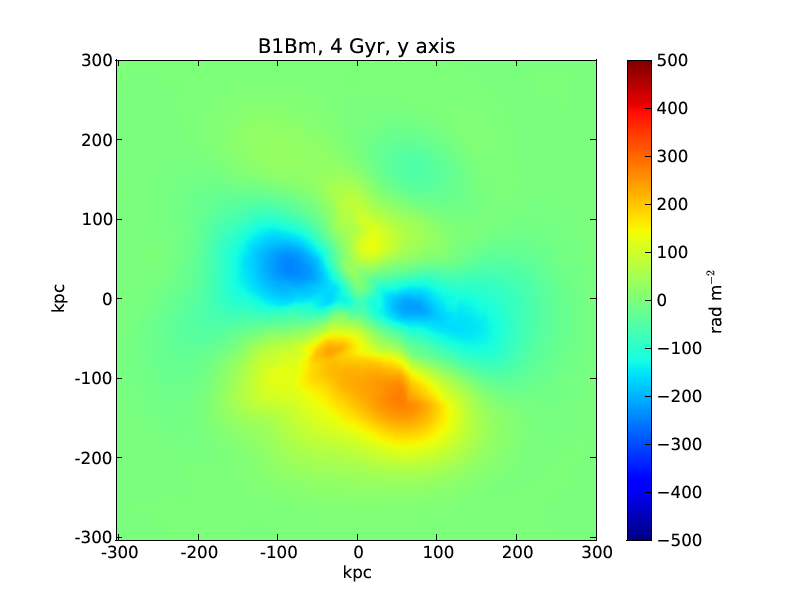}} 
  {\includegraphics[scale=0.28]{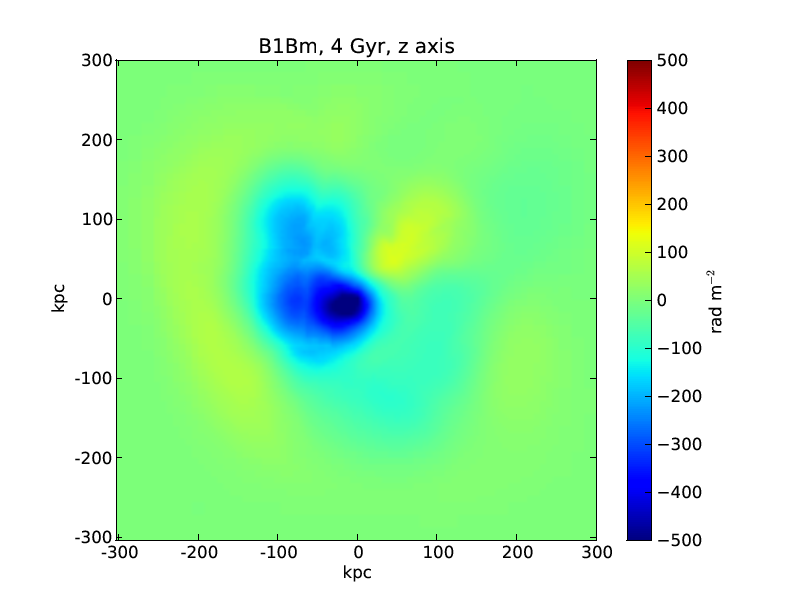}} \\
  {\includegraphics[scale=0.28]{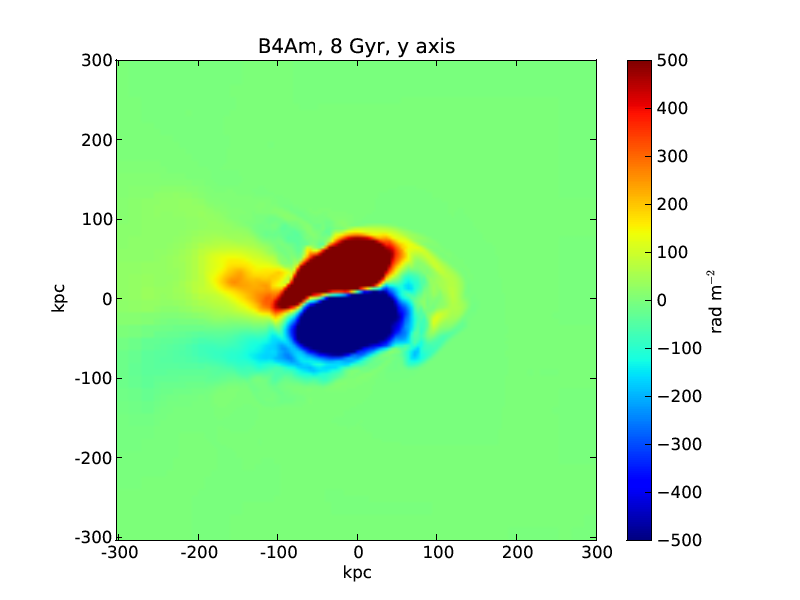}} 
  {\includegraphics[scale=0.28]{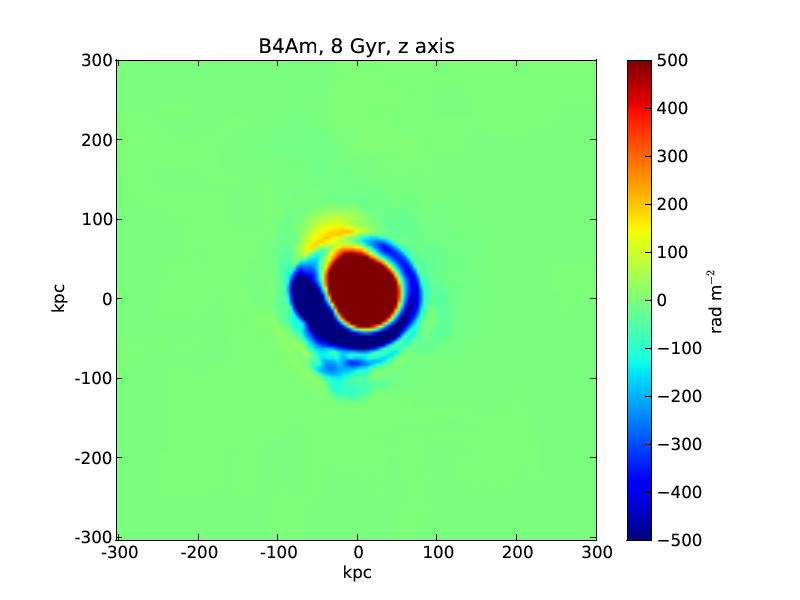}}
  \caption[Directional dependence of the rotation measure.]
           {Rotation measure taken along the $y$-axis (left) 
           and $z$-axis (right). 
            Shown are models J1Bm at $6$~Gyr (top), 
            B1Bm at $4$~Gyr (middle), 
            and B4Am at $8$~Gyr (bottom).
            All rotation measures have a 
            uniform resolution of $4$~kpc.
            Note that the color scale is chosen to
            enhance the regions outside the core. 
            The true minima and maxima for each model are listed 
            in Table~\ref{tab:agn_rm}.}
\label{fig:agn_topology_rm_dir}
\end{figure*}

We additionally view the latest times available for each model in the $y$ and
$z$ directions in~\figref{\ref{fig:agn_topology_rm_dir}}.  The
highly-directional jets show identical RM distributions in both the $x$- and
$y$-directions, but the rotation measure along the jet axis shows little
spatial variation and a very high value within the central $10$~kpc.  The
randomly-placed bubbles exhibit nearly identical morphology along the other
axes, as expected by the random orientations given to each injected bubble. The
fixed bubbles have similar features to the jets, with large RM values along the
$z$-axis, but with some discernible rotation measure extending out to $\sim
100$~kpc, much farther than for the smaller jets.

\begin{figure*}
  \centering 
  {\includegraphics[scale=0.28]{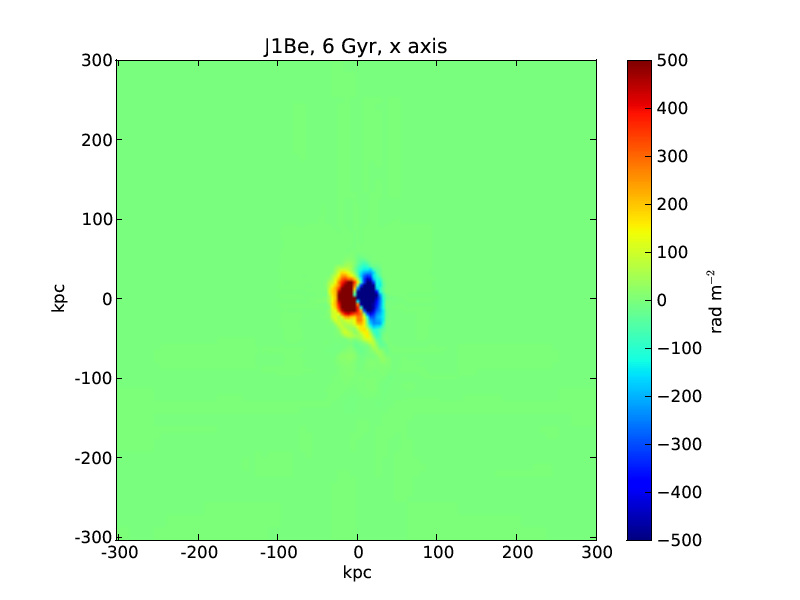}} 
  {\includegraphics[scale=0.28]{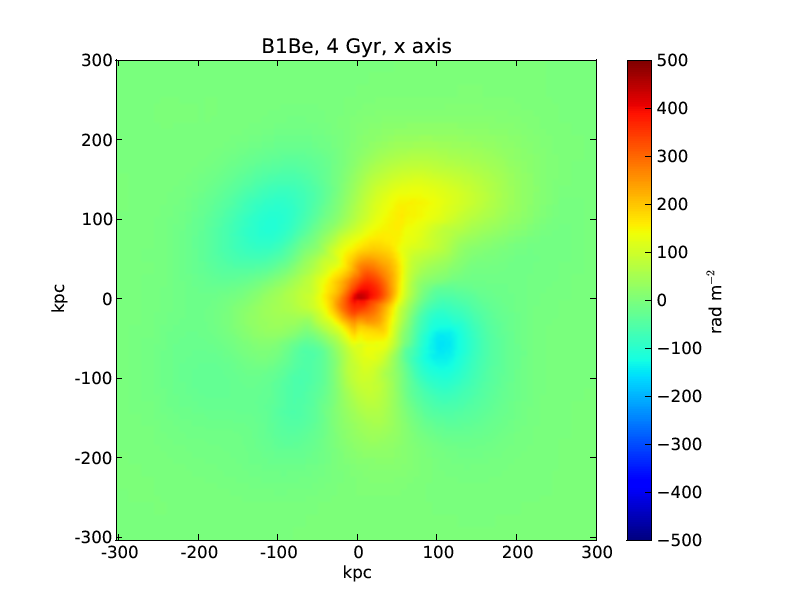}} 
  {\includegraphics[scale=0.28]{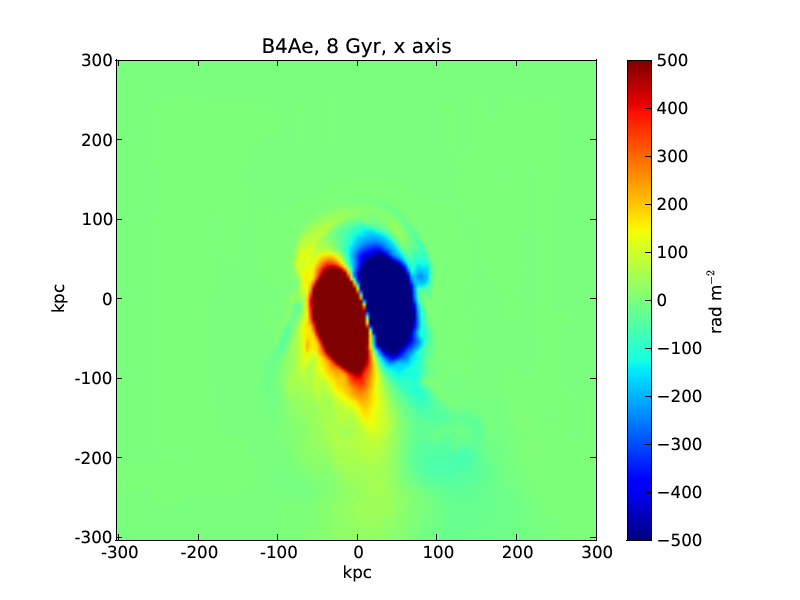}}
  \caption[Rotation measure of equipartition injection]
           {Rotation measure from the equipartition injection mode. 
            Shown are models J1Be at $6$~Gyr (left), 
            B1Be at $4$~Gyr (middle), 
            and B4Ae at $8$~Gyr (right).
            All rotation measures are taken along the $x$-axis and have a 
            uniform resolution of $4$~kpc.
            Note that the color scale is chosen to
            enhance the regions outside the core. 
            The true minima and maxima for each model are listed 
            in Table~\ref{tab:agn_rm}.}
\label{fig:agn_topology_rm_equi}
\end{figure*}

Finally, we can compare equipartition modes to fully magnetized injection
with~\figref{\ref{fig:agn_topology_rm_equi}}. Once again we see little
difference in the overall structure for the jets and both bubble models. The
values of the peak rotation measures are only slightly less in the
equipartition cases than they are in their fully magnetic counterparts.
Additionally, the randomly-placed bubbles show a much more smooth and uniform
distribution with equipartition injection.  The jets and fixed bubbles maintain
their spatial extent. Thus we once again see that merely the presence of
magnetic injection has significant effects, and that the relative proportion of
magnetic to thermal energy is less important.

\section{Model parameter survey}
\label{sec:agn_survey}

\subsection{Jets}

Our jet model parameter survey includes variations in the underlying peak grid
resolution, the size of the jet, the value of the accretion strength parameter
$\alpha$, and the size of the accretion and depletion regions. For the plots in
this section, we will label the model groups as follows: models J1A-J1C are
labeled as ``Jet Resolution'', models J2A-J2C are the ``Jet Size'' group, ``Jet
Accretion Strength'' will refer to models J3A-J3C, and finally models J4A-J4C
will be referenced by ``Jet Depletion and Accretion''.

\begin{figure*}
  \centering  
  {\includegraphics[width=\columnwidth]{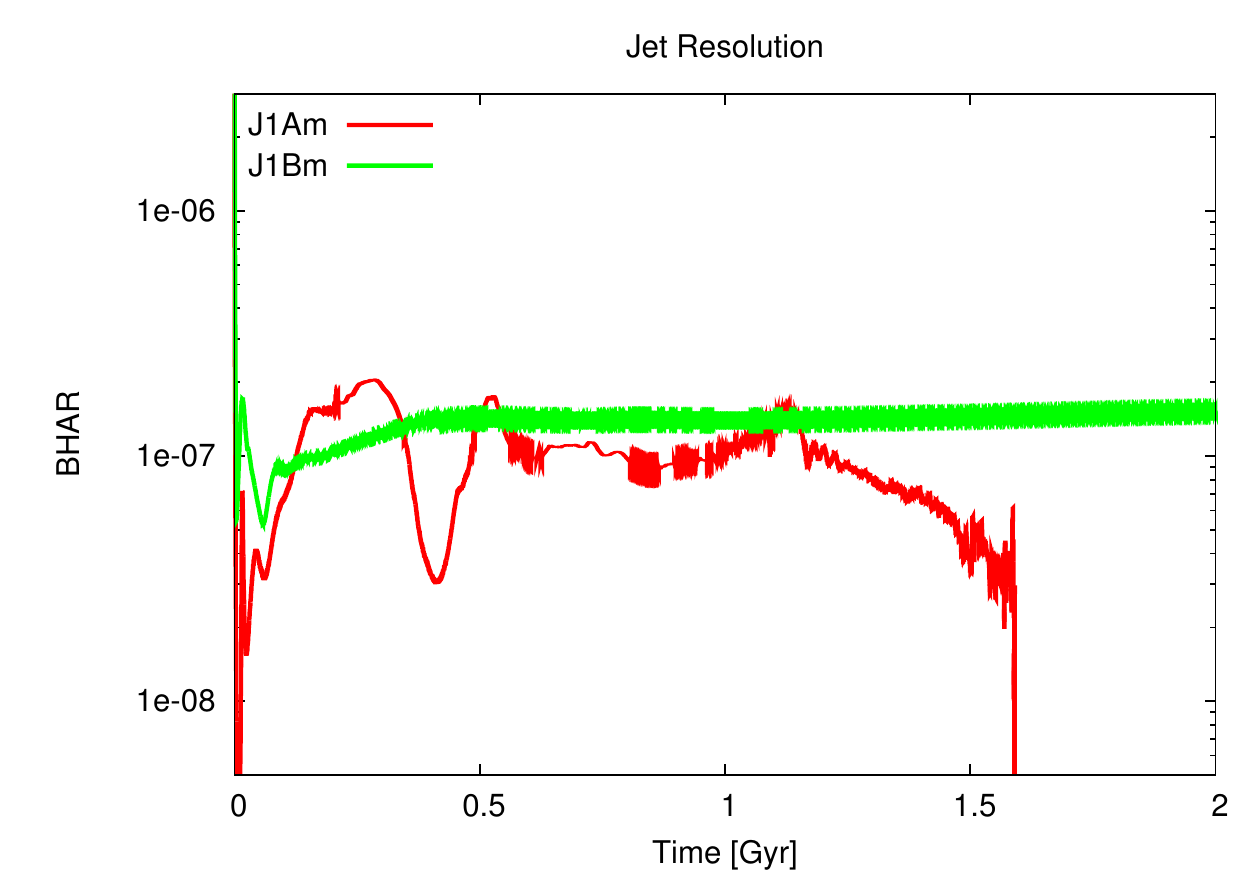}}
  {\includegraphics[width=\columnwidth]{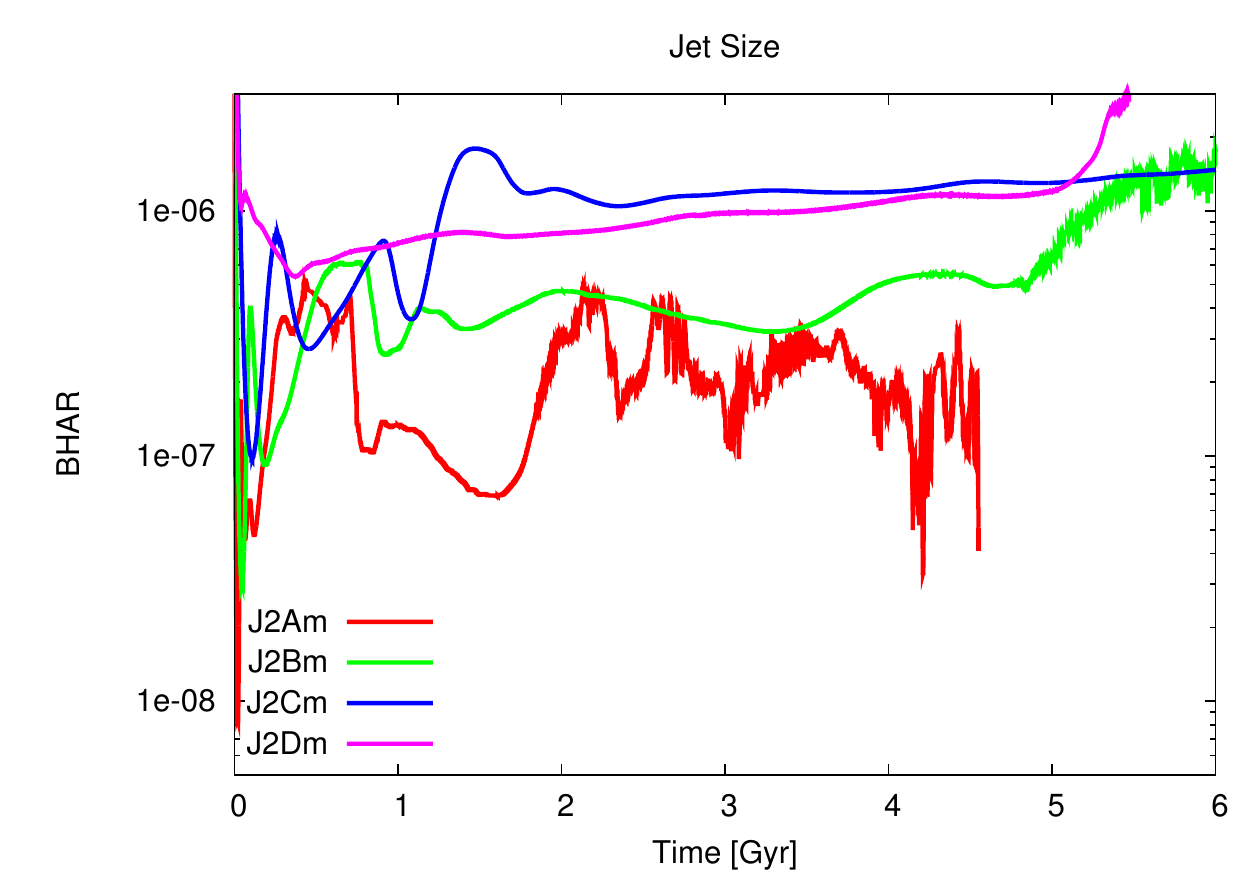}} 
  {\includegraphics[width=\columnwidth]{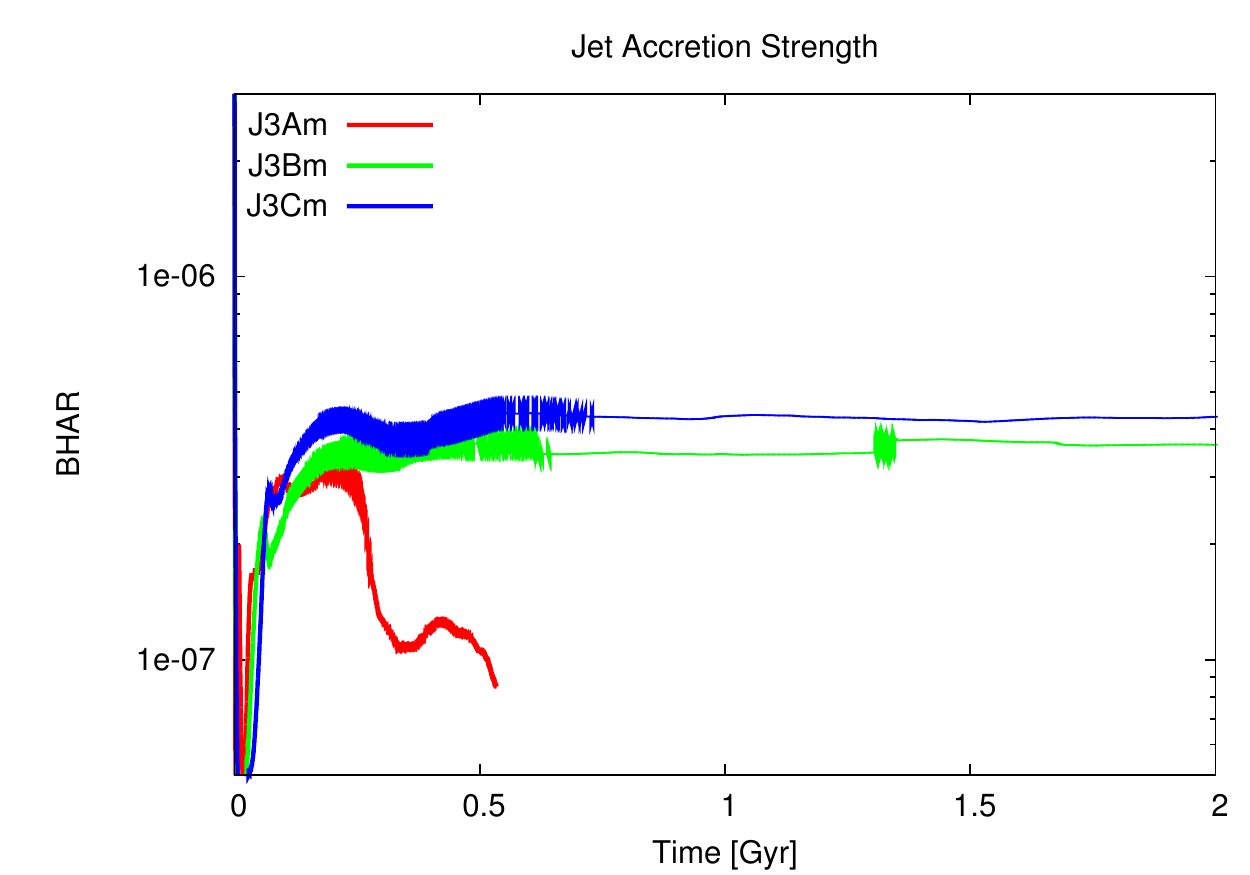}}  {\includegraphics[width=\columnwidth]{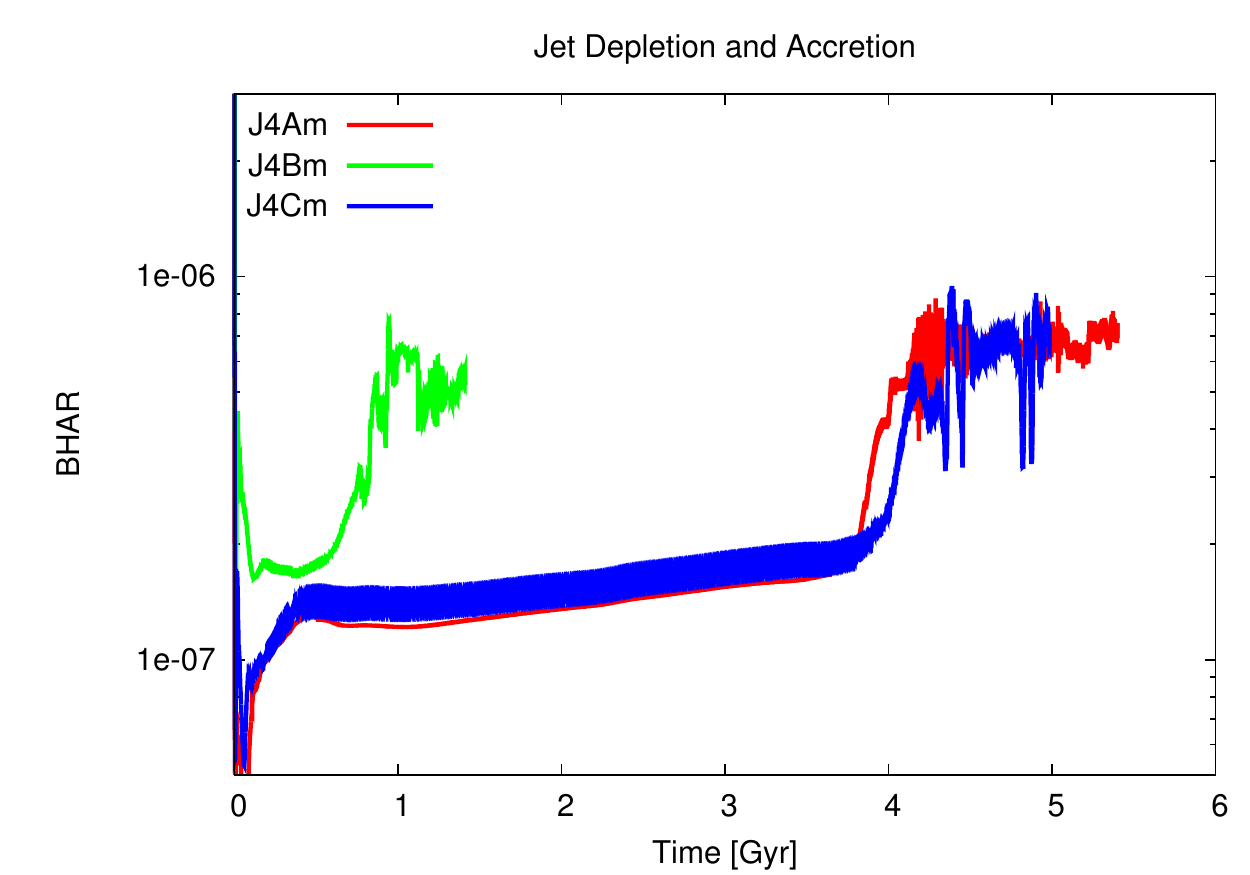}}
  \caption[Jet model survey - black hole accretion rate.]
           {Affecting the black hole accretion rate by varying the 
           jet model parameters. 
            Shown is BHAR ($\dot{M}/\dot{M}_{\rm Edd}$) versus time.}
\label{fig:agn_survey_jet_accrate}
\end{figure*}

These parameter changes significantly affect the accretion history of the SMBH,
as shown in~\figref{\ref{fig:agn_survey_jet_accrate}}.  With higher resolution
(model J1A) the accretion history is much more variable, but the average
accretion rate is roughly equal to the fiducial case. In this case the strong
injected fields make continuation of the simulation very difficult past
$1.5$~Gyr, and at this point the combination of turbulent gas motions and
strong fields make the accretion rate highly variable. When we vary the jet
size we get more diverse behavior and longer-lasting jets. In model J2A, where
we double the size of the jets, we maintain the same average accretion strength
as in the fiducial case but add the periodic behavior seen with the
well-resolved jets of model J1A. As we increase the jet size while fixing
$\Delta x= 1.0$~kpc in models J2B and J2C we see a consistent increase in the
average accretion rate.  As the jet energy gets deposited in a larger volume
the magnetic fields become less effective at keeping gas away from the central
black hole.  When we decrease the relative size of the jet in model J2D, the
smaller jet becomes more effective at removing gas, thereby lowering the
accretion rate relative to the J2C model, which has the same resolution but a
larger jet.

When we vary the accretion strength parameter $\alpha$ (models J3A-C), we see
similar evolution patterns in the accretion rate, though the rates are
uniformly higher than the fiducial case. Model J3A, which both lowers the
resolution to $\Delta x = 0.5$~kpc and raises $\alpha$ to $100$, produces
incredibly strong fields within $0.5$~Gyr, preventing further simulation.
However, prior to this point there still appears to be some periodicity with
features similar to model J1A, which has the same resolution but $\alpha=1$.
Interestingly, setting $\alpha=100$ does not produce an accretion rate that is
$100$~times higher. Instead, the stronger outflows are more effective at
pushing away gas from the core, lowering the overall accretion rate to just a
factor of five larger than the fiducial case. Similarly, with $\alpha=300$ in
model J3C we do not see a corresponding tripling of the accretion rate. 

Varying the depletion and accretion radii can have some significant effects,
especially if these values are larger than the injection region. Model J4A sets
$R_{\rm dep} = R_{\rm acc} = \Delta x$~(i.e., one zone), and we do not see
significant differences between this and the fiducial case. This is also the
case when we remove depletion altogether, as in model J4C. Since the accretion
rate is incredibly small for the jets in the first place, removing gas has no
significant effects. However, when we set the accretion and depletion radii to
$4$~zones, as in model J4B, we see an initially larger accretion rate followed
by a factor of five jump after only $1$~Gyr, rather than the $4$~Gyr it takes
in the fiducial model. In this case, we are measuring within a volume slightly
larger than the injection region itself (note that the injected magnetic fields
are strongest at the scaling radius, which is one half the injection radius)
and the gas that the magnetic fields drive away from the core gets included in
the accretion rate calculation, leading to an increase in the accretion rate.
Also the large jump in accretion rate occurs earlier since we are measuring at
a larger radius, and the cooled gas does not take as long to reach the inner
$4$~kpc as it does to reach the inner $2$~kpc, where the magnetic injection is
strongest. 

\begin{figure*}
  \centering  
  {\includegraphics[width=\columnwidth]{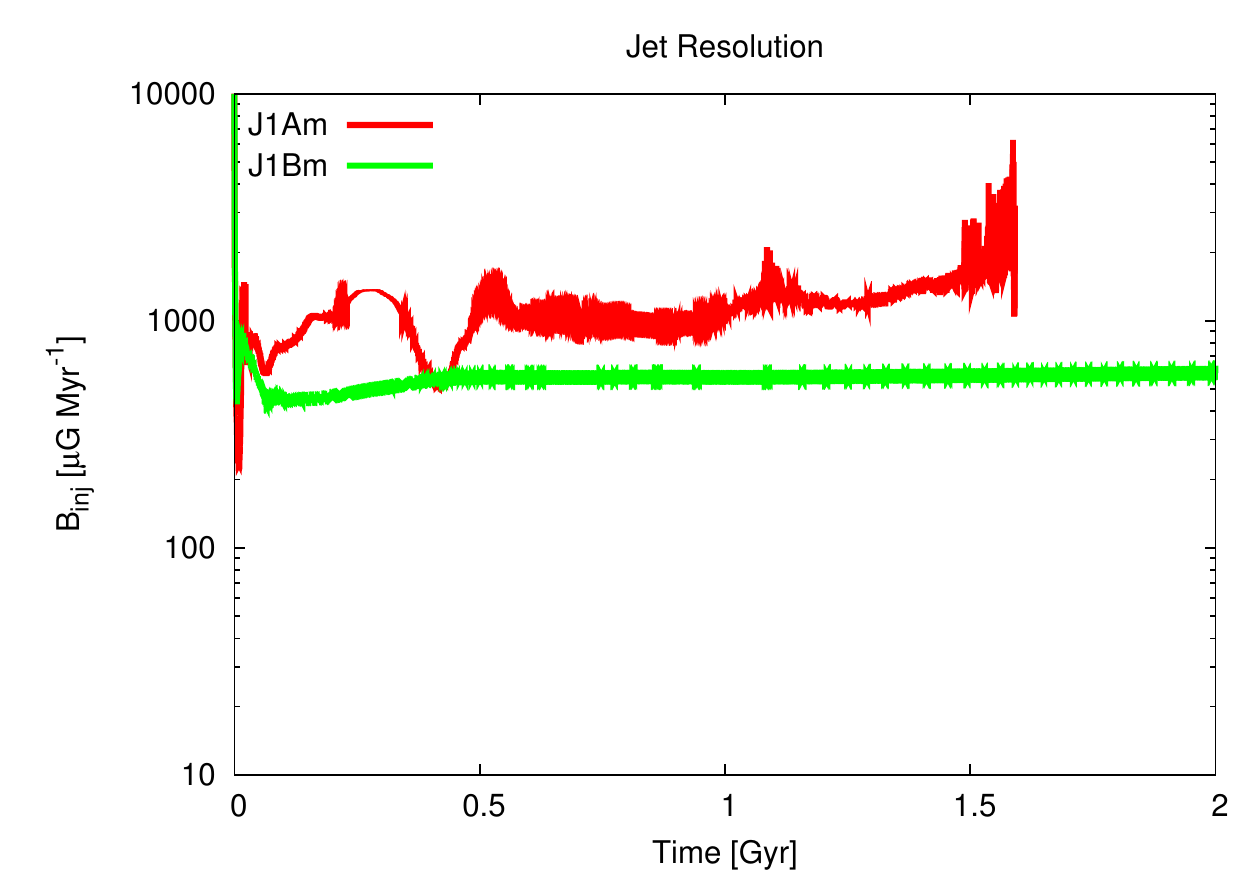}}
  {\includegraphics[width=\columnwidth]{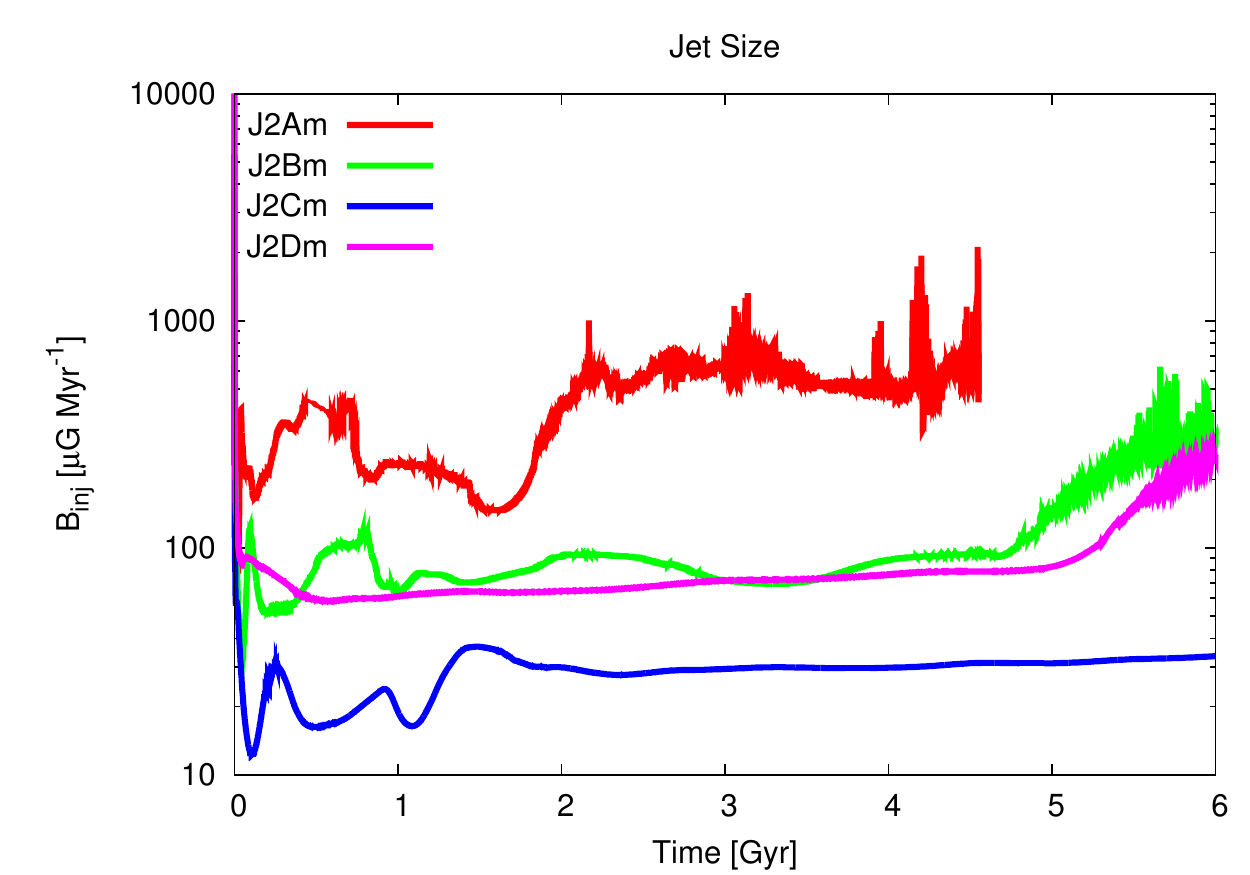}} 
  {\includegraphics[width=\columnwidth]{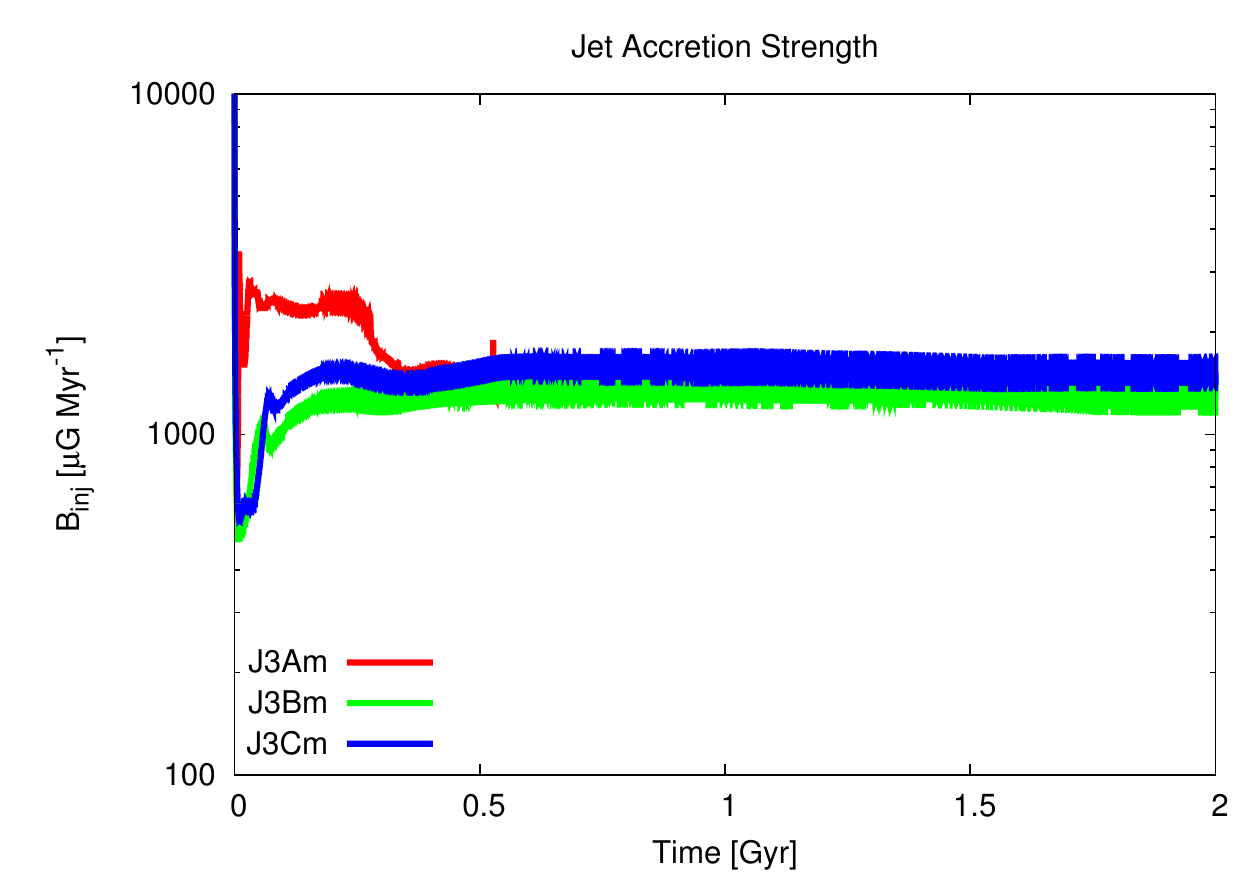}}
  {\includegraphics[width=\columnwidth]{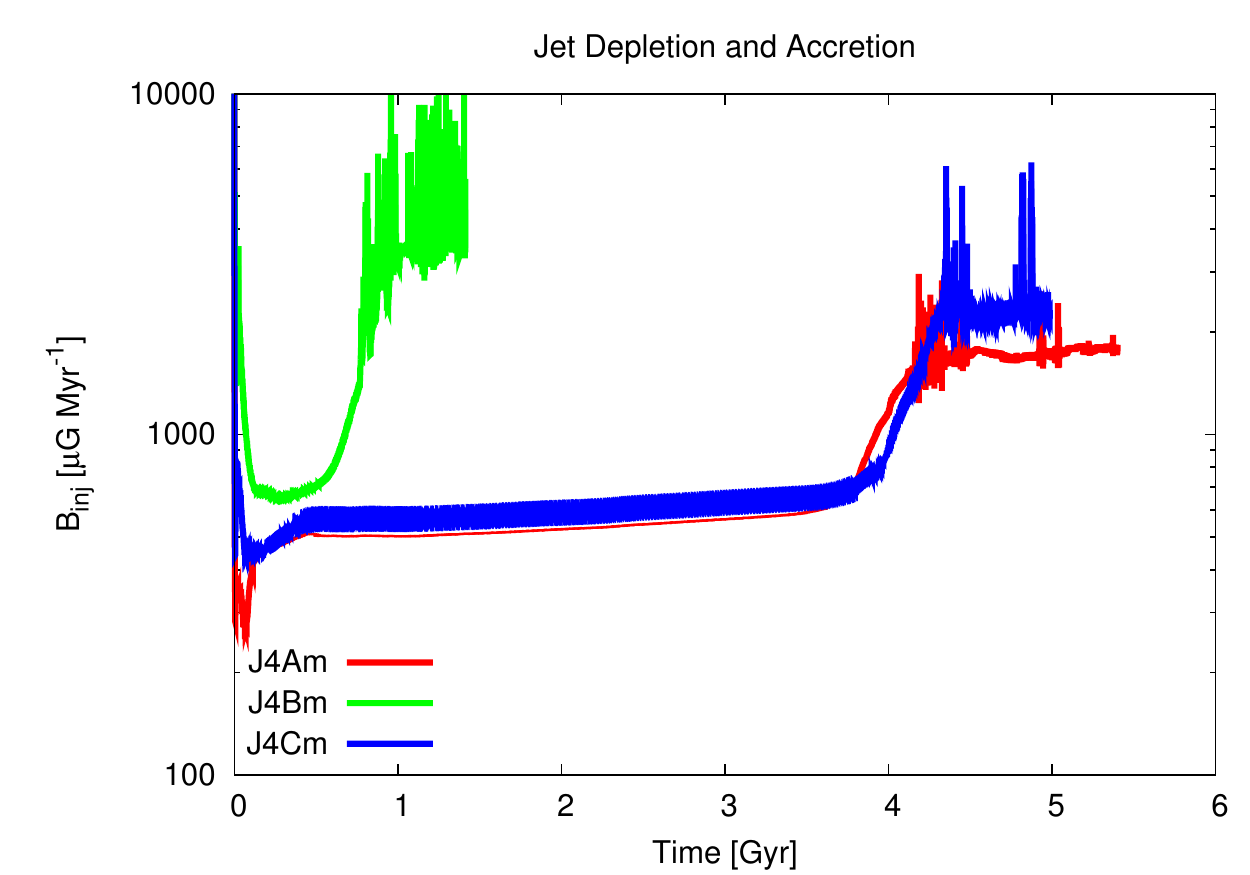}}
  \caption[Jet model survey - rate of injected magnetic field strength.]
           {Affecting the rate of injected magnetic field strength 
           by varying the jet model parameters.}
\label{fig:agn_survey_jet_maginj}
\end{figure*}

In~\figref{\ref{fig:agn_survey_jet_maginj}} we see how varying the jet
parameters affects the rate of injected magnetic field strength.  For model
J1A, despite the small changes to the accretion rate, the injected magnetic
field rate is almost a factor of two higher. This is mainly due to a severely
decreased time step in the J1A run: although the instantaneous accretion rate,
and hence the injected magnetic energy, may be lower, the total injected
magnetic field over long periods of time is larger in the higher-resolution
run. When we vary the jet size, we find that model J2A, despite having the
lowest accretion rate, has the highest rate of injection of magnetic fields.
This is because this model injects magnetic fields into the smallest region.
The resulting field is more strongly concentrated, leading to a higher $B$. As
we scale to ever-larger jets, the fields are distributed over larger volumes,
and hence the $B_{\rm inj}$ rate drops.  Model J2D, which has a lower
resolution but same jet size as J2B, maintains roughly the same injection rate,
consistent with the view that lower resolutions do not significantly alter the
resulting magnetic fields.

Varying the accretion strength parameter $\alpha$ has almost no effect: since
the accretion rate self-regulates with changes in $\alpha$, models J3B and J3C
are not significantly different. However, model J3C, with a larger $\alpha$ of
$300$, does produce slightly stronger magnetic fields.  When we vary the
accretion and depletion radii, we once again see only strong differences with
model J4B, which produced fields almost an order of magnitude stronger than the
fiducial case after only $1$~Gyr.

\begin{figure*}
  \centering  
  {\includegraphics[width=\columnwidth]{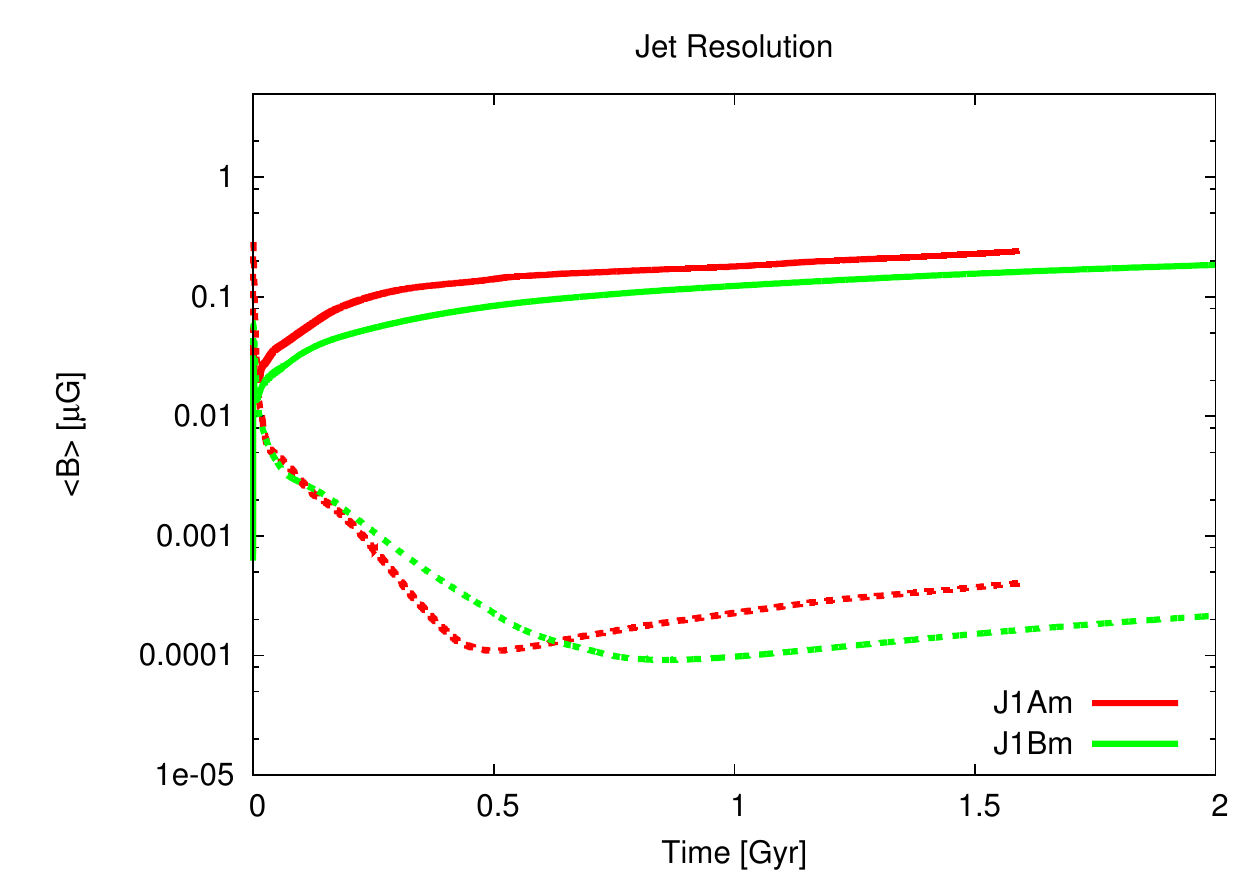}}
  {\includegraphics[width=\columnwidth]{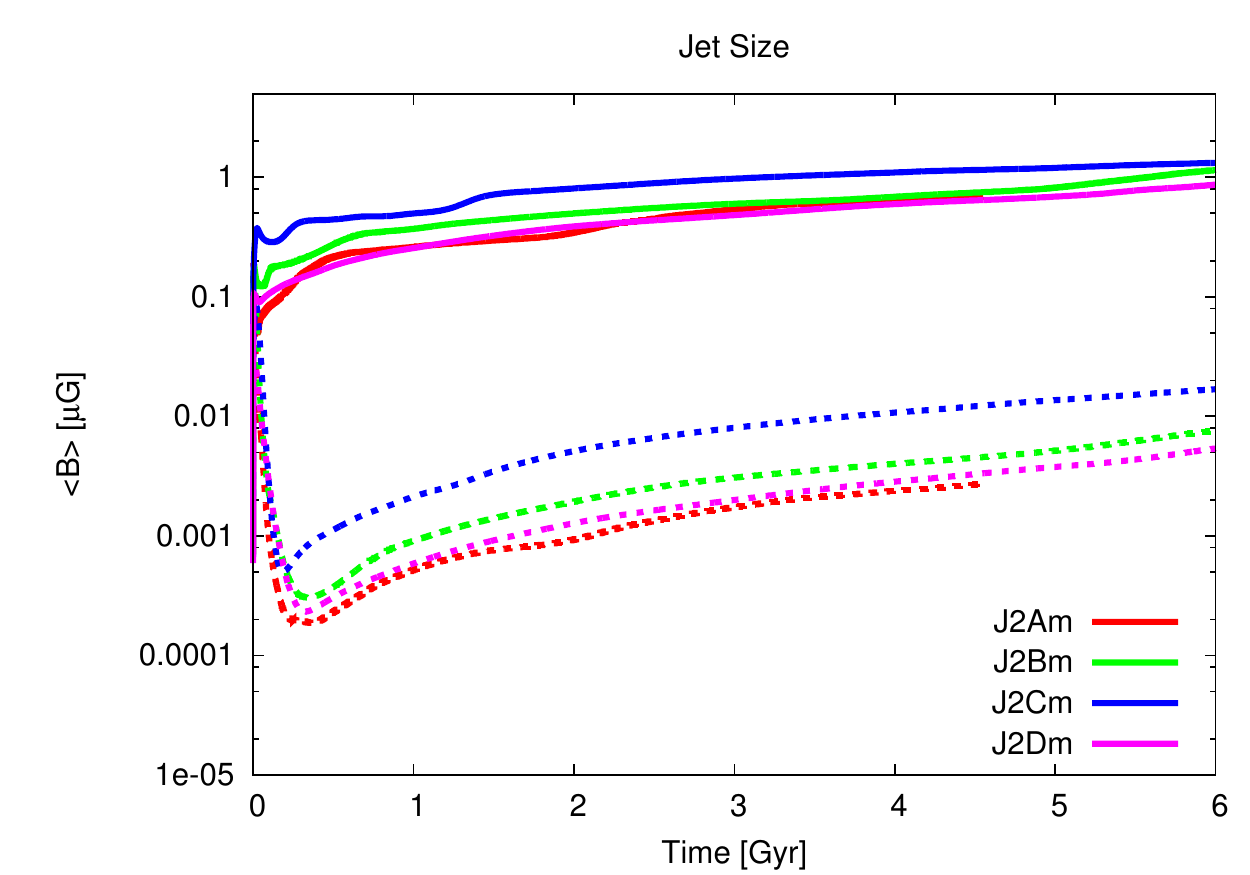}} 
  {\includegraphics[width=\columnwidth]{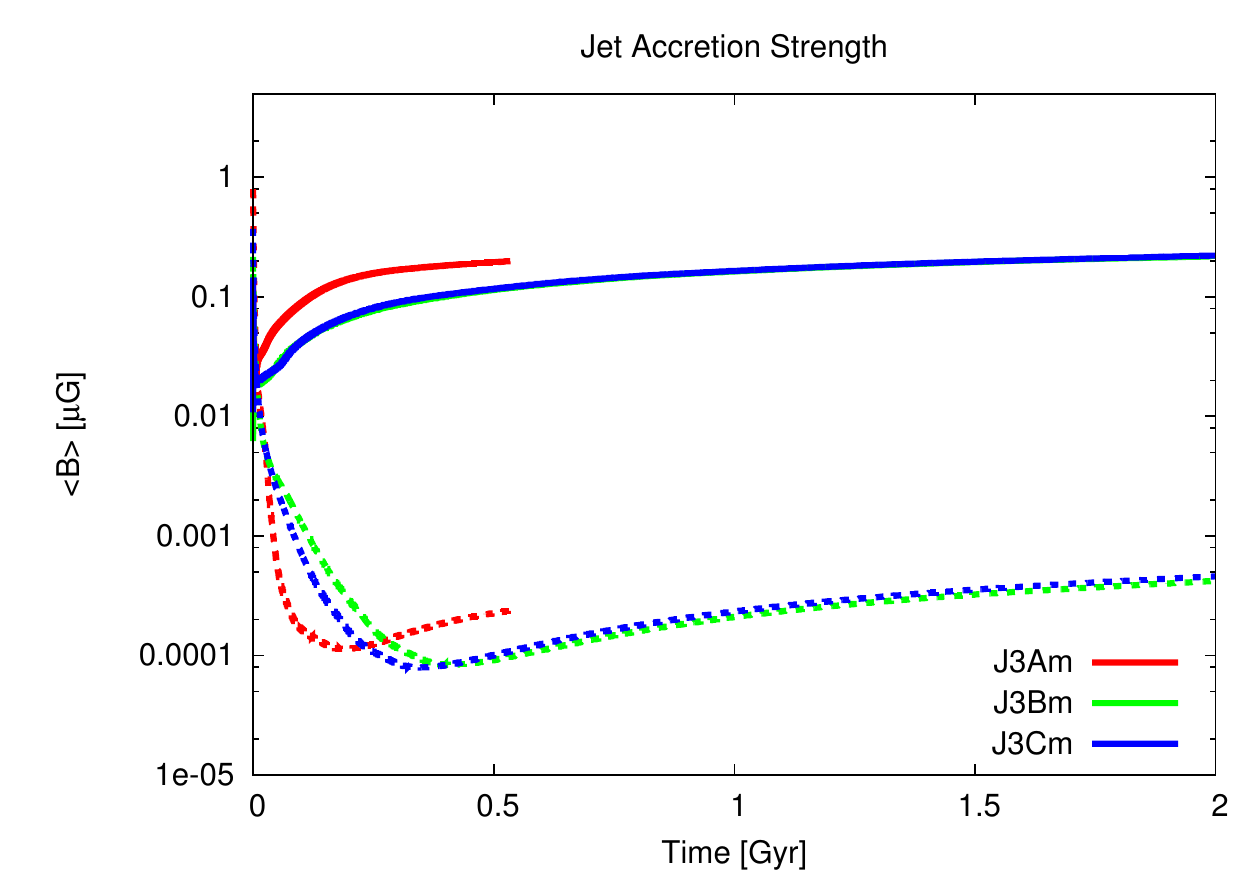}}  
  {\includegraphics[width=\columnwidth]{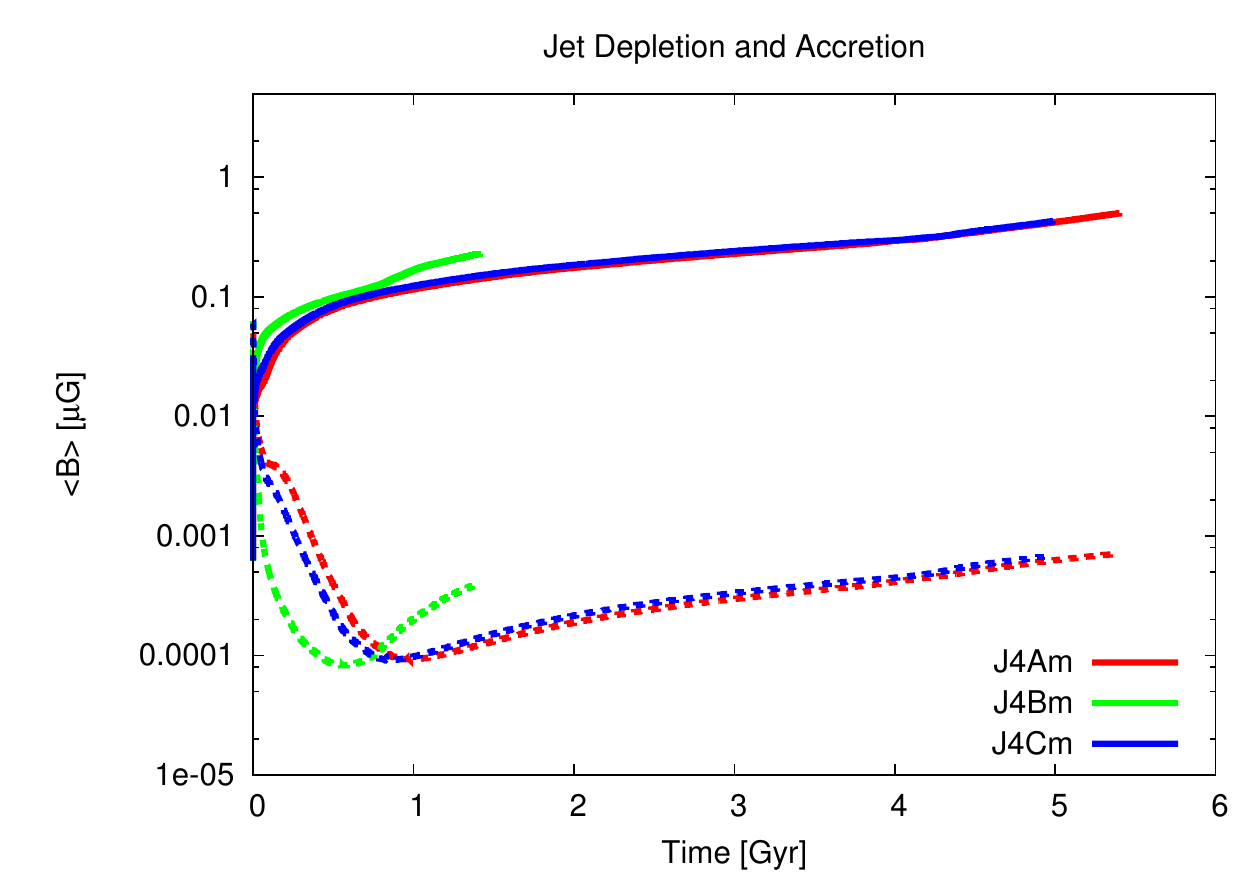}}
  \caption[Jet model survey - average magnetic field strength.]
           {Affecting the average magnetic field strength 
           by varying the jet model parameters. 
           Solid lines are density-weighted 
           average fields within $R_{\rm core}$ and dotted lines are 
           volume-weighted fields within $R_{\rm 200}$.}
\label{fig:agn_survey_jet_aveb}
\end{figure*}

\figref{\ref{fig:agn_survey_jet_aveb}} shows the changes to the average
magnetic field strength within $R_{\rm core}$ and $R_{\rm 200}$. The
higher-resolution jets (J1A) and large jets (J2C) are best at magnetizing both
the inner and outer cluster regions. As expected, the large jets are able to
distribute fields over a much larger volume than the fiducial case, although
they cannot produce fields stronger than $\sim 1 \mg$ in the core. There is
again no significant difference when increasing $\alpha$ from $100$ to $300$,
although these models produce slightly stronger fields within the core compared
to the fiducial case. Despite the enhanced accretion rate of model J4B, we do
not evolve the jet long enough to see significant differences in the average
field of the core. However, we do begin to see an increase in the overall
magnetization of the cluster, with roughly a doubling of the average
cluster-wide field.

\begin{figure*}
  \centering  
  {\includegraphics[width=\columnwidth]{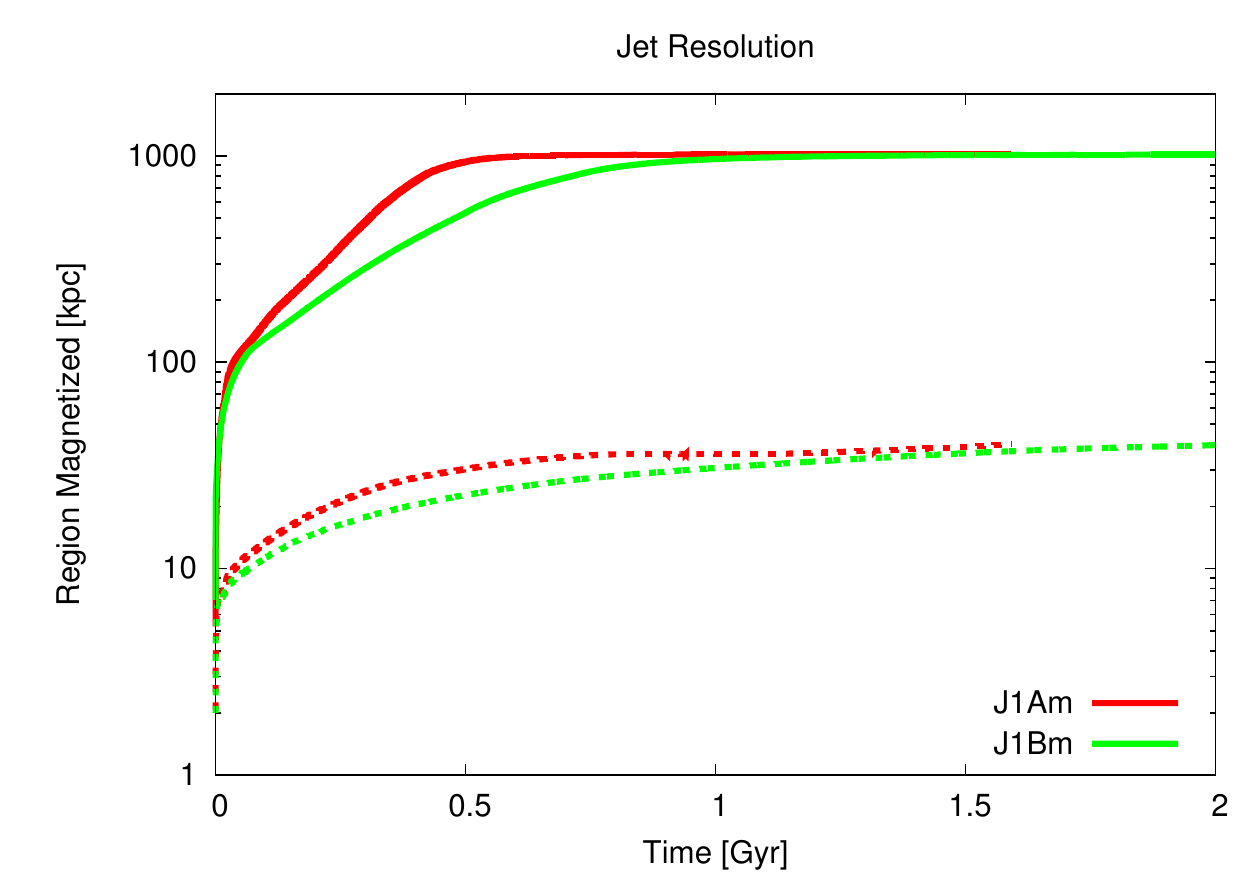}}
  {\includegraphics[width=\columnwidth]{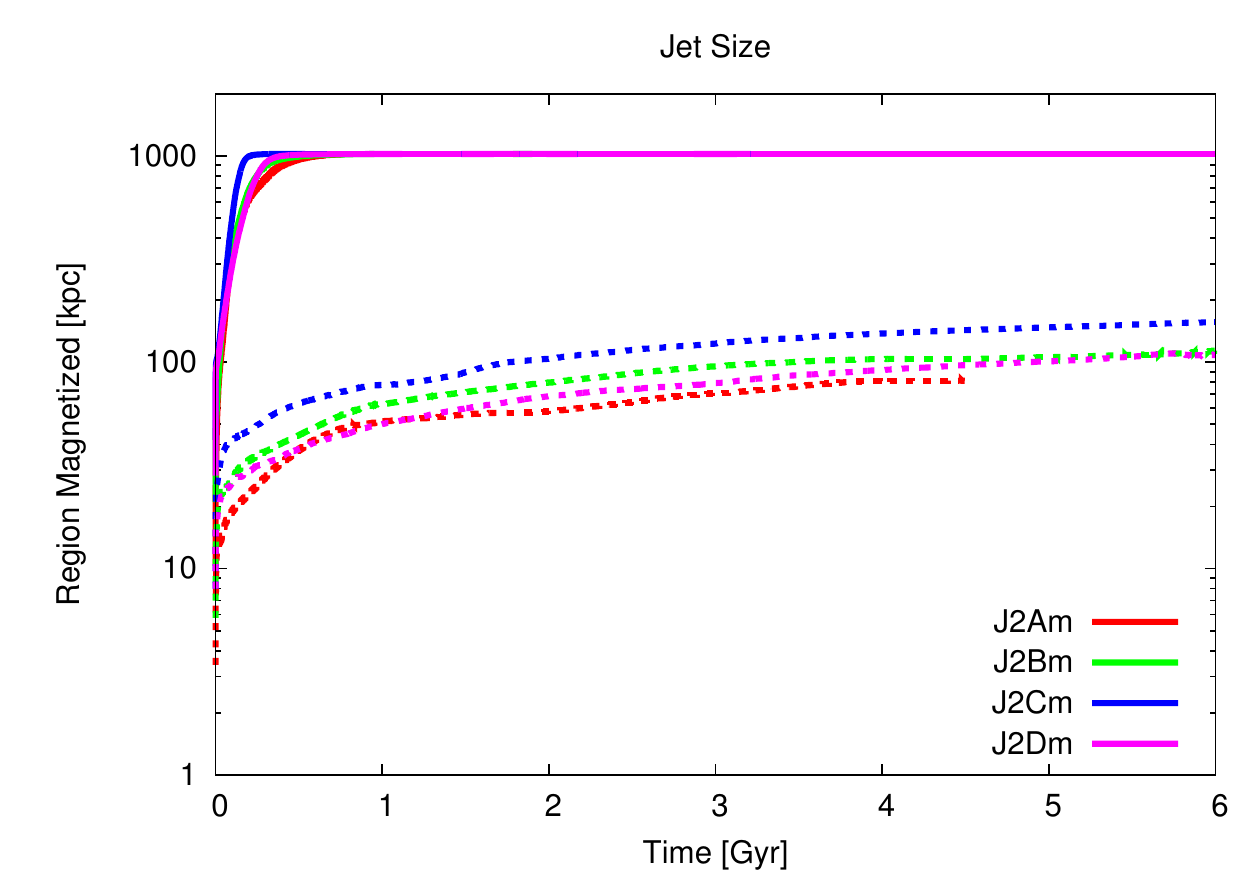}} 
  {\includegraphics[width=\columnwidth]{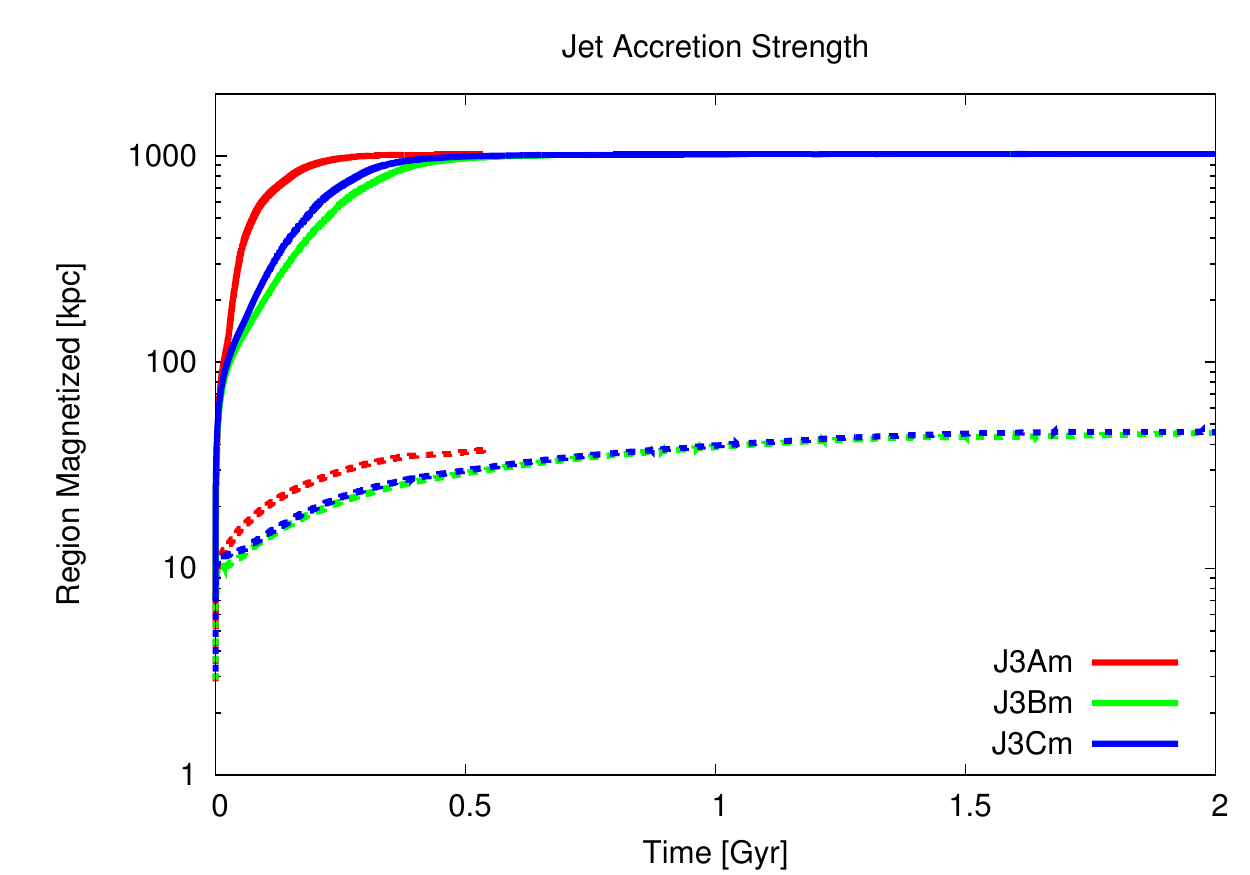}}
  {\includegraphics[width=\columnwidth]{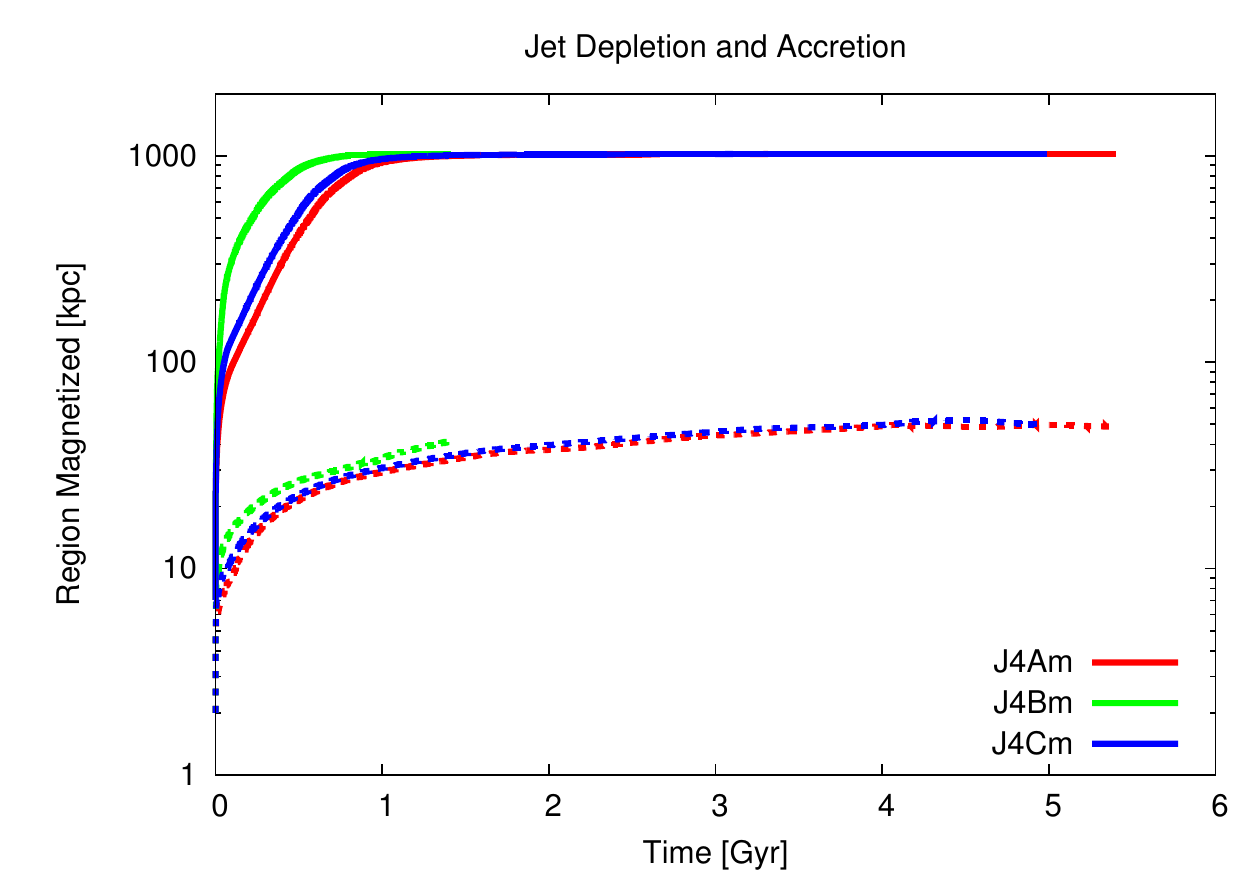}}
              \caption[Jet model survey - magnetized volume.]
           {Affecting the volume magnetized by varying the jet model 
           parameters. Shown is the cube root 
            of the total volume containing fields of strength at least 
            $10^{-12}$~G (solid lines) and $10^{-6}$~G (dotted lines).}
\label{fig:agn_survey_jet_magvol}
\end{figure*}

We see similar trends in the magnetized volume, as shown
in~\figref{\ref{fig:agn_survey_jet_magvol}}. The large jets and high-$\alpha$
jets push weak magnetic fields throughout the cluster volumes in less than
$500$~Myr, whereas the fiducial case takes almost twice as long. However, the
high-$\alpha$ models still do not push strong fields very far into the cluster
- only to $\sim 50$~kpc. On the other hand, the large jets can generate strong
fields out to $100$~kpc. Also, here the effects of higher resolution are less
pronounced. With the boosted accretion rate of model J4B, the cluster saturates
with weak magnetic fields much more quickly (less than half the time of the
fiducial case), with a slight increase in the volume of strong magnetic fields.
Once again, we do not see large differences with the smaller accretion and
depletion radii relative to the fiducial model.

\begin{figure*}
  \centering  
  {\includegraphics[width=\columnwidth]{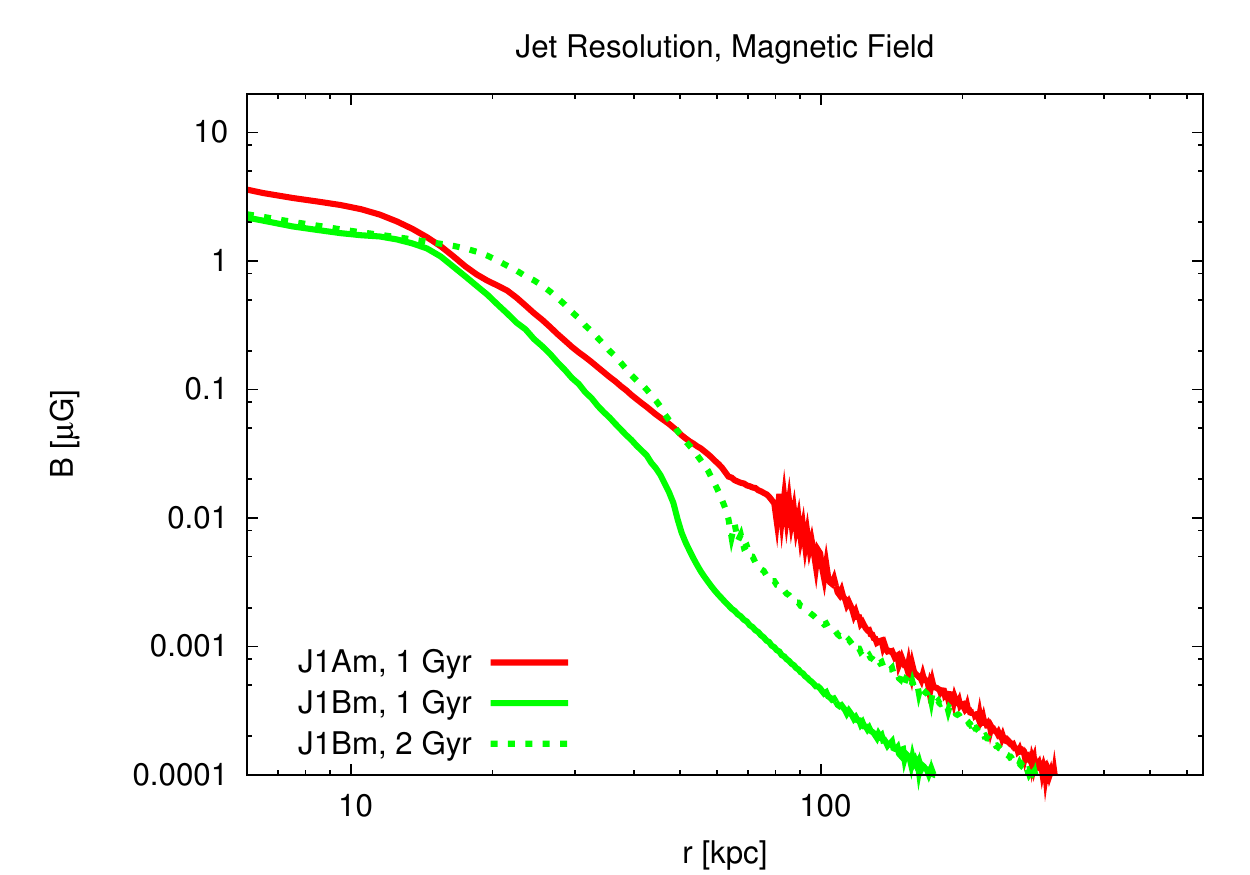}}
  {\includegraphics[width=\columnwidth]{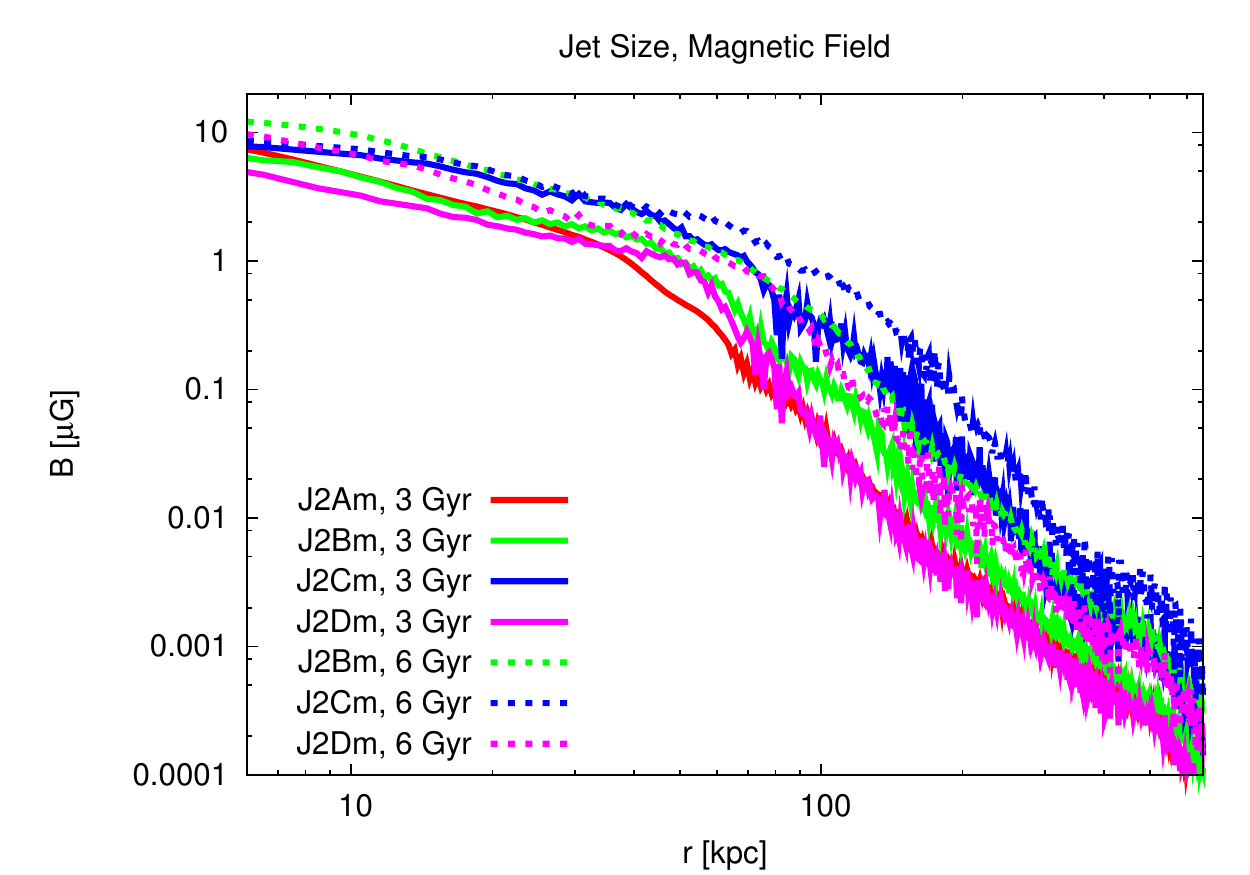}}
  {\includegraphics[width=\columnwidth]{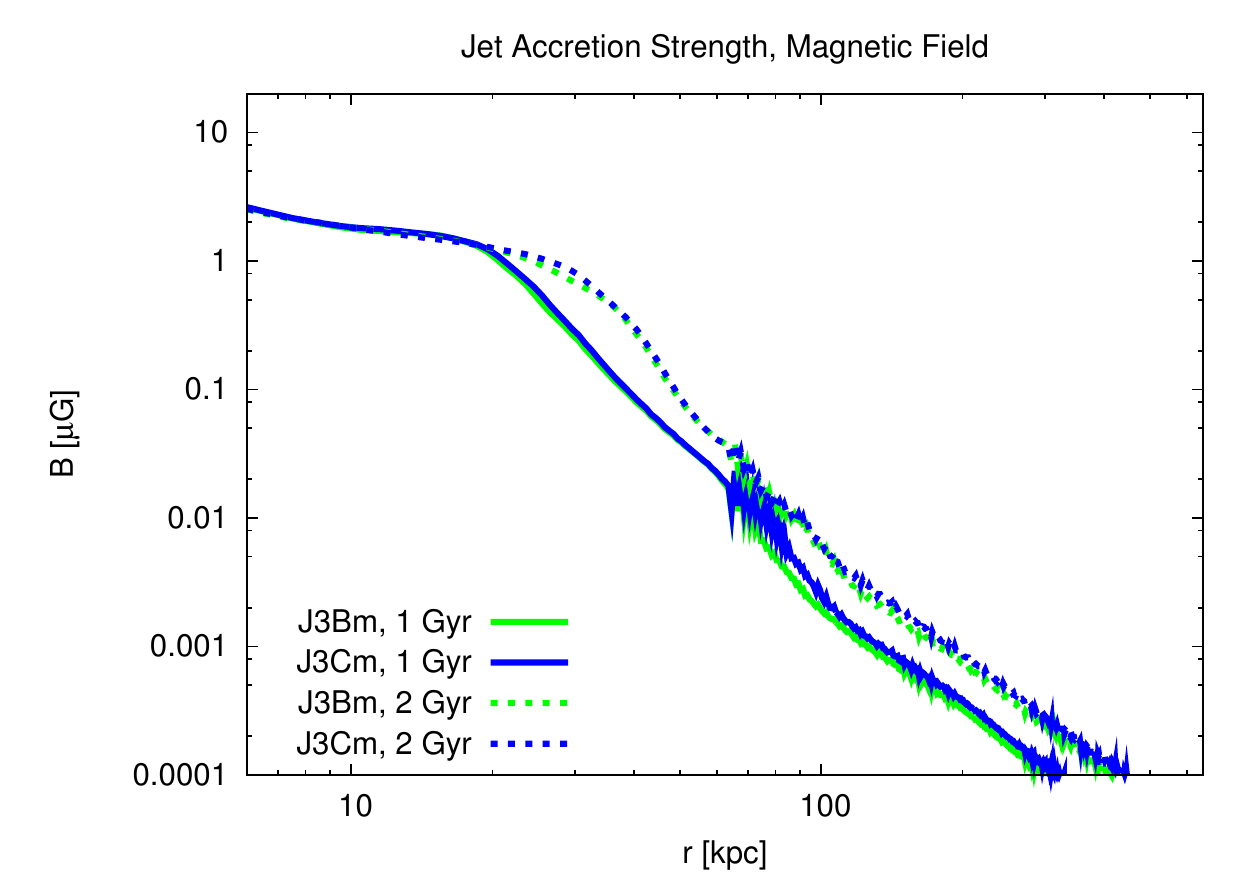}}
  {\includegraphics[width=\columnwidth]{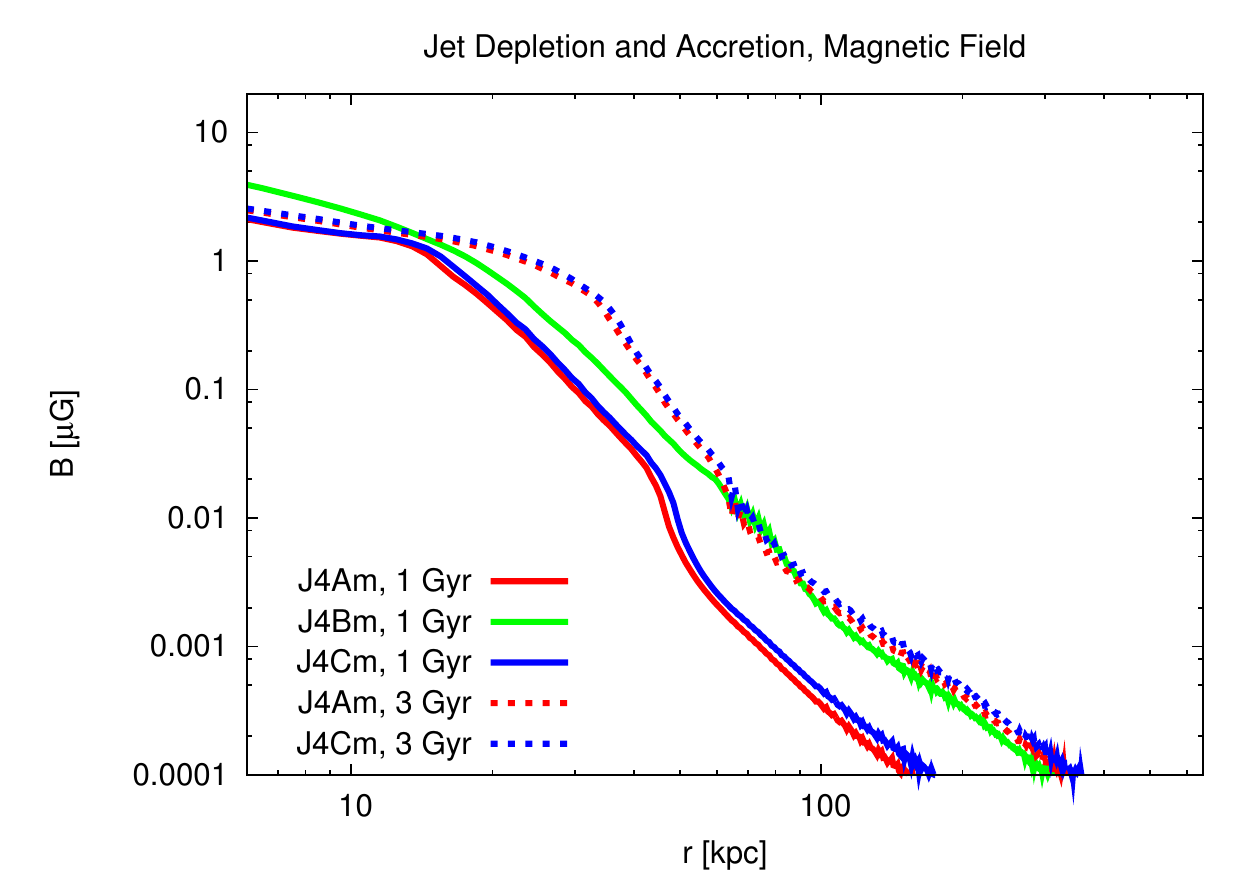}}
  \caption[Jet model survey - magnetic field radial profiles.]
           {Affecting the magnetic profiles 
            by varying the jet model parameters. 
            Shown are volume-weighted magnetic fields as a function 
           of radius.}
\label{fig:agn_survey_jet_profile_mag}
\end{figure*}

Finally, we see in~\figref{\ref{fig:agn_survey_jet_profile_mag}} the radial
profiles of the magnetic fields. We first see that model J1A, which has higher
resolution than the fiducial case, is able to provide more magnetization
throughout the cluster in $1$~Gyr than the fiducial case is able to produce in
twice that time.  All of the large jets are able to generate strong fields in
the core (over $10 \mg$ in some case) after $6$~Gyr. However, since the
magnetic fields of model J2C are distributed over such a large volume, the
strongest fields in the core come from the slightly smaller jets of model J2B.
When we increase $\alpha$ in models J3B and J3C, we see almost no differences
in the profiles at $1$~and $2$~Gyr. Also, there is very little difference
between either of these models and the fiducial case. Model J4B, with the
larger accretion and depletion radii, produces an order of magnitude stronger
field at large radii and nearly a doubling of the field in the inner cluster
after only $1$~Gyr. 

\subsection{Bubbles}    
  
By varying the peak resolution and accretion strength in the bubble models we
 also see significant differences in the accretion rate, magnetic outflow
properties, and overall magnetization of the cluster. The plots in this section
are split into two groups: ``Bubble Resolution'' will contain models B1A-B2E,
and ``Bubble Accretion Strength'' will reference models B2A, B2B, B3A, and B3B.
 
\begin{figure*}
  \centering 
  {\includegraphics[width=\columnwidth]{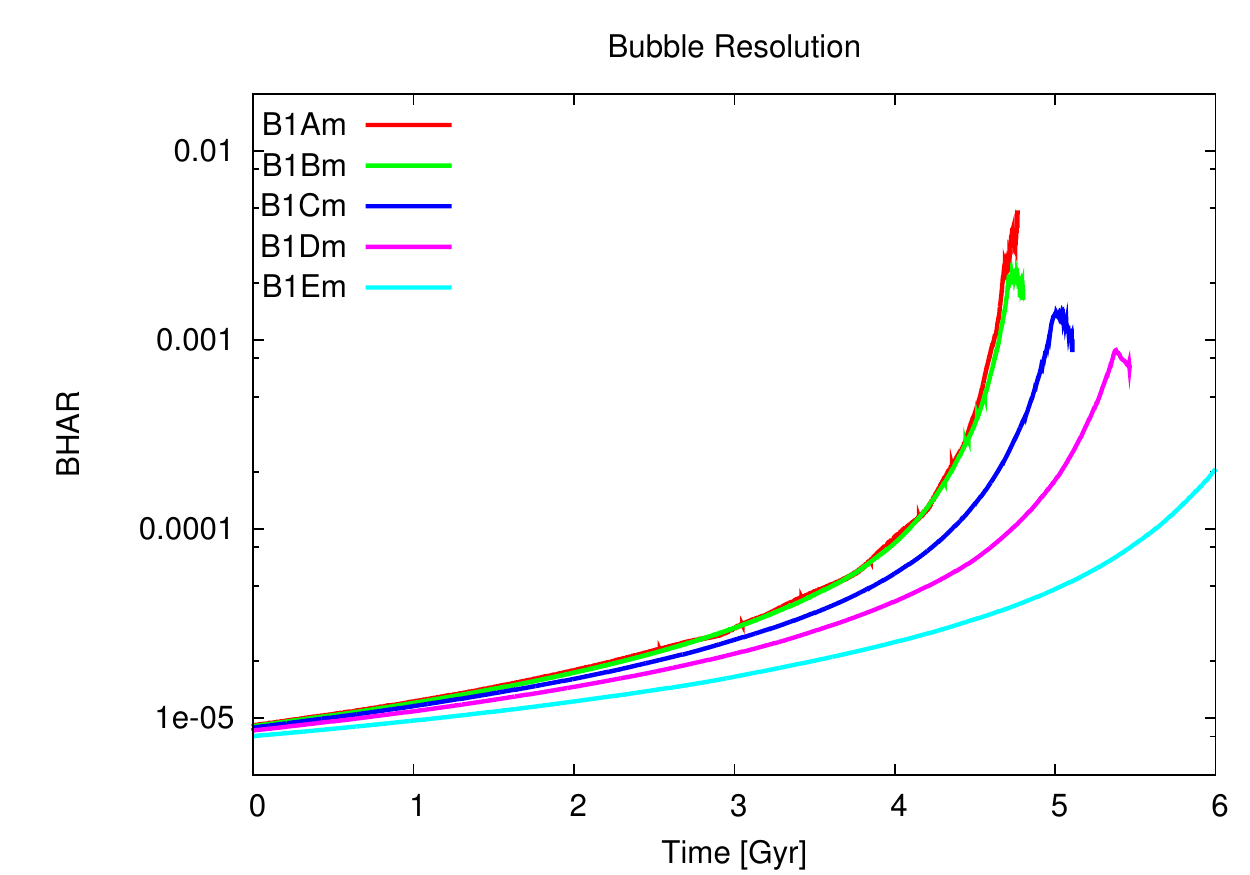}} 
  {\includegraphics[width=\columnwidth]{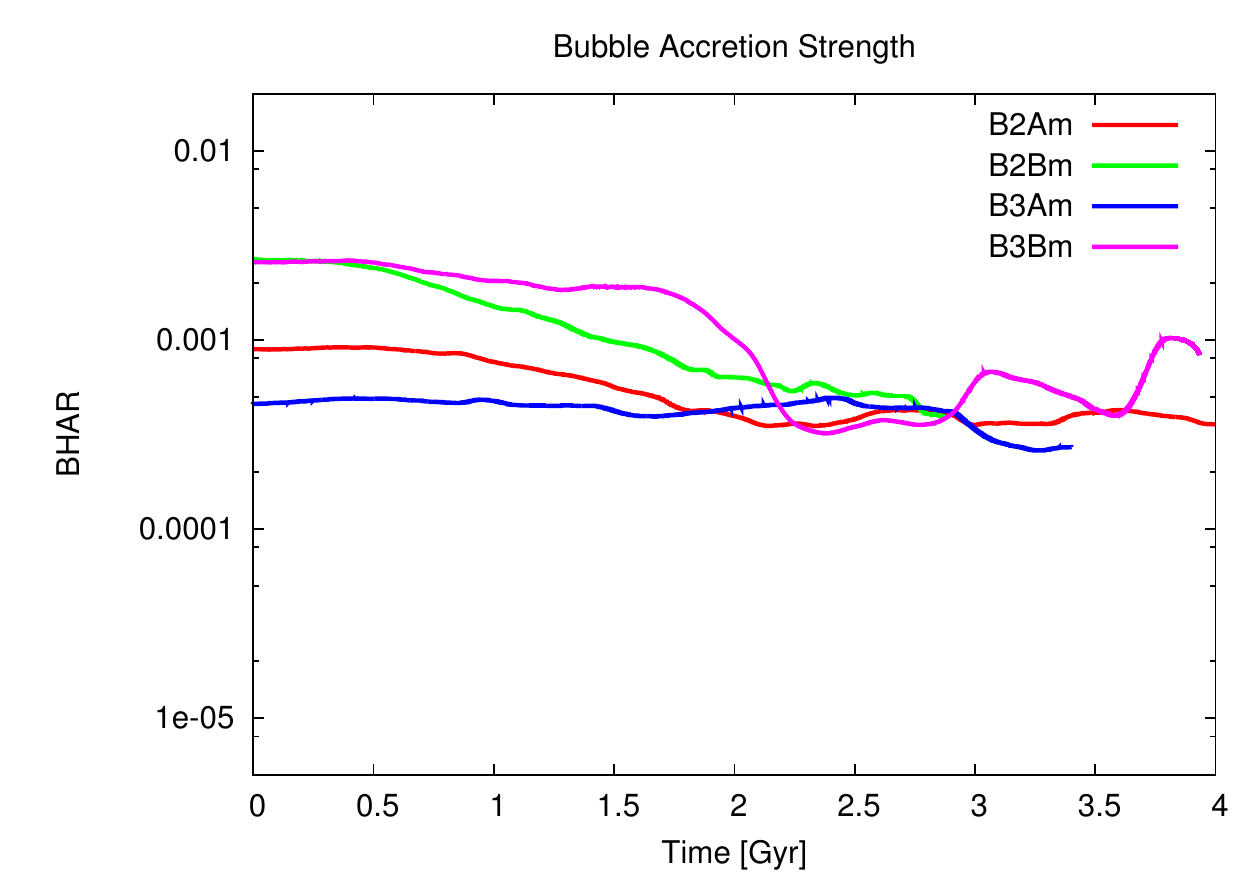}}
  \caption[Bubble model survey - black hole accretion rate.]
           {Affecting the black hole accretion rate by varying the 
           bubble model parameters. 
            Shown is BHAR ($\dot{M}/\dot{M}_{\rm Edd}$) versus time.}
\label{fig:agn_survey_bubble_accrate}
\end{figure*}

Our parameter changes significantly affect the accretion rate, as shown
 in~\figref{\ref{fig:agn_survey_bubble_accrate}}.  First, changing the
resolution has a stronger effect on the bubbles than the jets, mostly because
we can reduce the resolution much further than in the jet models and still
adequately resolve the bubbles. At the same resolution as the jets
($0.5$-$1.0$~kpc) we see very little difference. However, at lower resolutions
($4$-$8$~kpc), we begin to see much lower accretion rates as time progresses,
reaching an order of magnitude in difference at $5$~Gyr. At lower resolutions,
we also delay the onset of the formation of highly complex fields by over a
billion years, so the simulations can evolve further. 

When we vary the accretion strength parameter $\alpha$, we see much more
complicated behavior than in the jets. Whereas the jets maintained
self-regulated behavior, because the continuous jet-based feedback was coupled
more tightly to the immediate surroundings of the black hole, the bubbles are
much more sensitive to changes in $\alpha$. As we increase $\alpha$, as in
models B2A and B2B, we see immediate increases in the BHAR. The bubbles, which
take time to form and are not as efficient as the jets at driving gas away from
the core, only slightly reduce the accretion rate below the amplified value.
For example, setting $\alpha=100$ produces a BHAR only $\sim 80$~times stronger
rather than $100$. As we vary the resolution along with $\alpha$, as in models
B3A and B3B, we see that the changes to the accretion strength are dominant
over the changes to resolution.

\begin{figure*}
  \centering 
  {\includegraphics[width=\columnwidth]{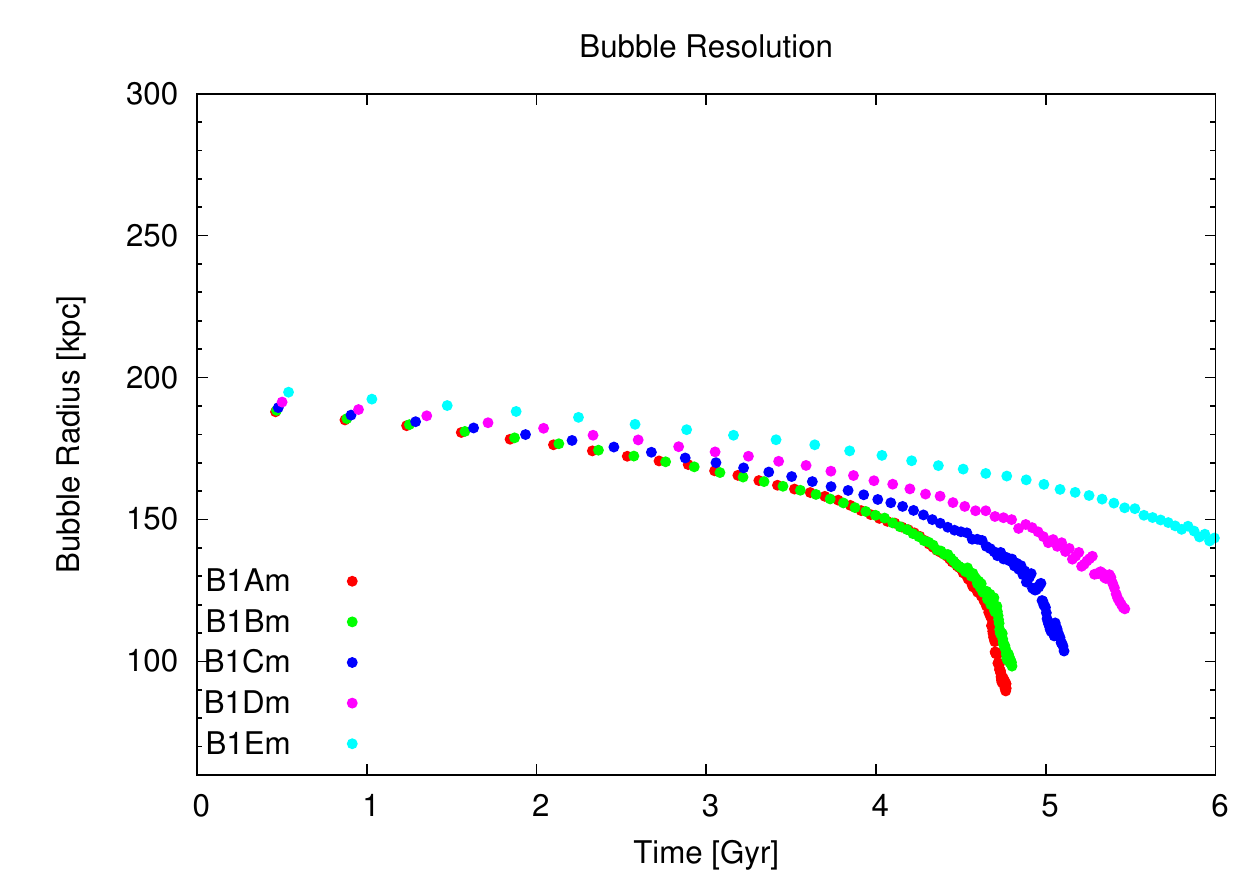}} 
  {\includegraphics[width=\columnwidth]{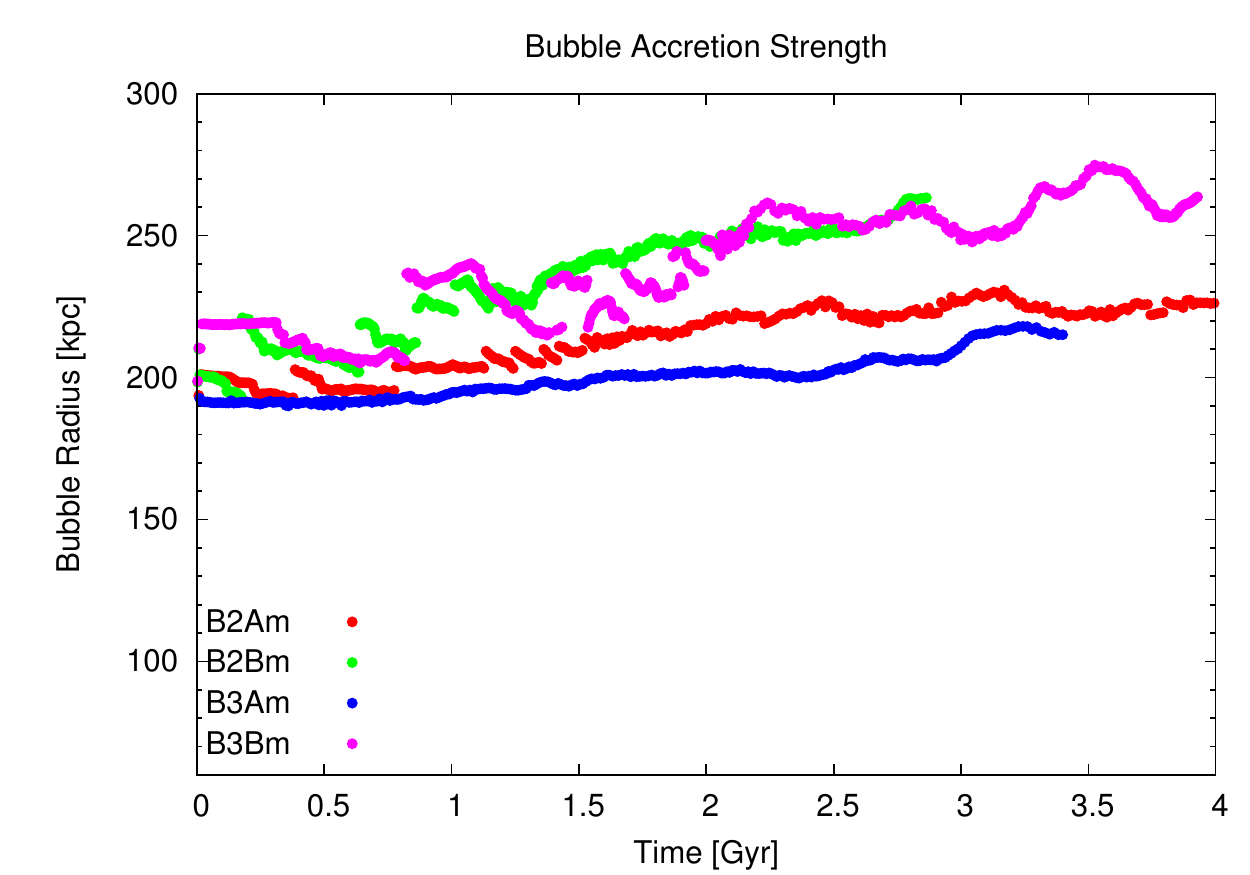}}
  \caption[Bubble model survey - effects of magnetization on bubble size.]
           {Affecting the bubble size by varying the bubble 
           model parameters. Shown is
            bubble radius at each injection event as a function of time.}
  \label{fig:agn_survey_bubble_bubsize}
\end{figure*}

The bubble sizes vary dramatically, as shown
 in~\figref{\ref{fig:agn_survey_bubble_bubsize}}. Note that as we lower the
resolution (and have correspondingly lower accretion rates) our bubbles become
slightly \emph{larger}. Since the bubble energy is fixed as a fractional
increase in the SMBH mass, at lower resolutions we begin to underestimate the
gas density, which both lowers the accretion rate and produces larger bubbles,
as~\eqref{\ref{eq:agn_rbub}} suggests.  Notice that is also takes longer to
produce bubbles at lower resolutions, taking almost twice as long in the
$8$~kpc case than in the $4$~kpc run.

Authors usually invoke an enhanced accretion strength parameter to account for
underestimating the gas density at lower resolutions.  However, for this
cluster our changes to $\alpha$ dominate over changes to resolution.  Note
especially models B3B and B1B. Model B3B has a resolution of $4$~kpc, and an
$\alpha$ of $300$ should account for the resulting under-resolution of the gas
density. Instead, model B3B produces many more bubbles with larger radii and
begins a periodic cycle within $1$~Gyr, which are features absent from model
B1B. We also notice some discontinuities, such as in model B3B, where strong
bubble events temporarily halt the cooling of gas.  All of the runs with
increased $\alpha$ produce unphysically-large bubbles ($> 200$~kpc) with
extremely high formation rates (one bubble per $10$-$20$~Myr), suggesting that
increasing $\alpha$ may not always be appropriate.

\begin{figure*}
  \centering 
  {\includegraphics[width=\columnwidth]{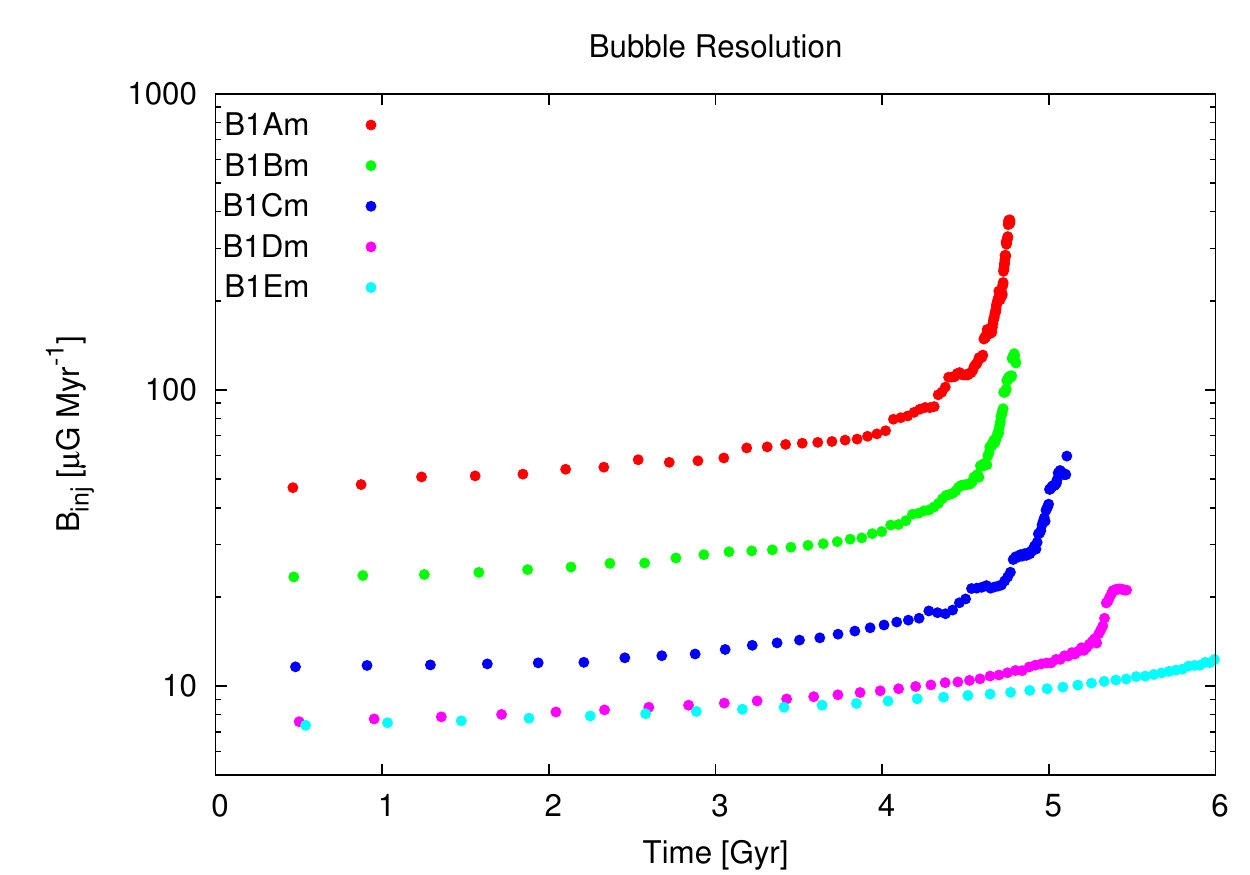}} 
  {\includegraphics[width=\columnwidth]{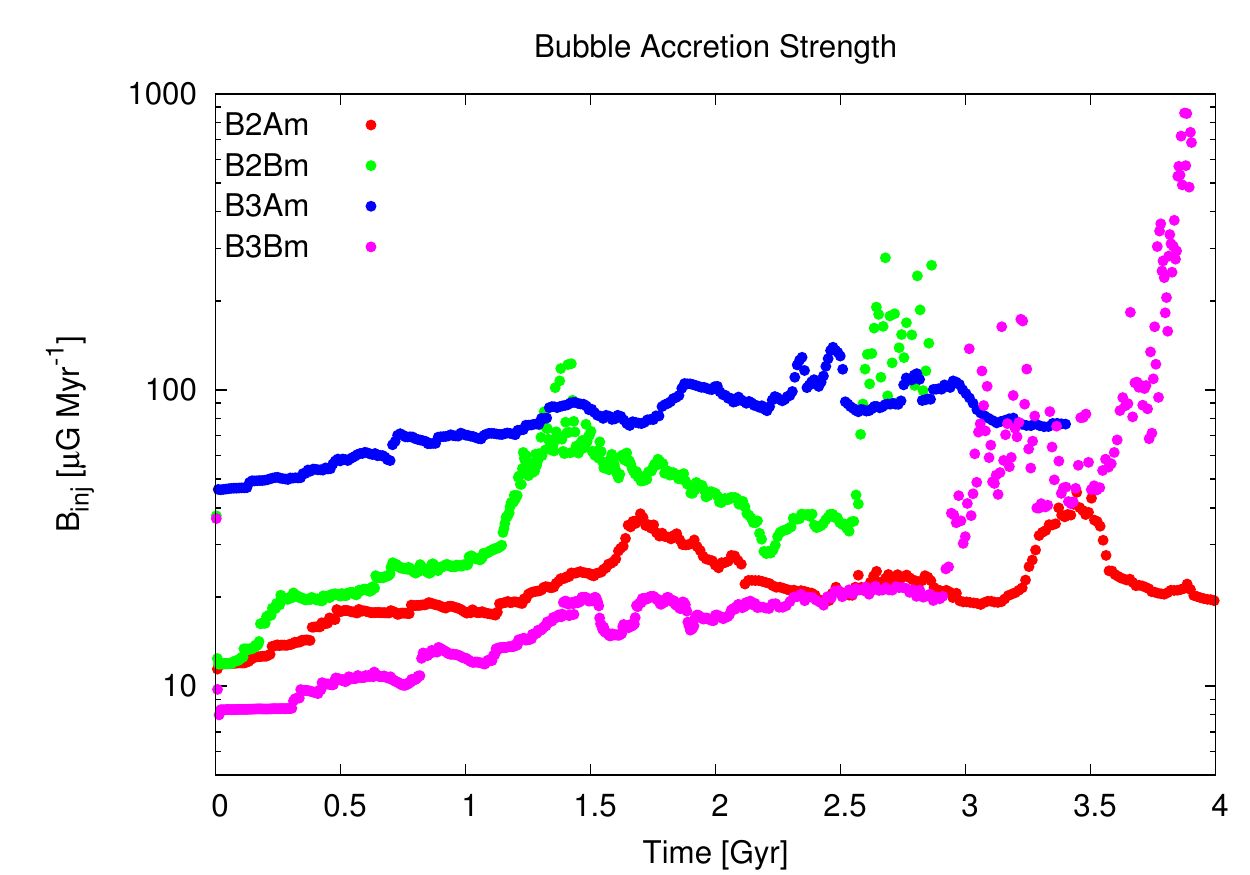}}
  \caption[Bubble model survey - rate of injected magnetic field strength.]
           {Affecting the rate of injected magnetic field strength 
           by varying the bubble model parameters.}
\label{fig:agn_survey_bubble_maginj}
\end{figure*}

Despite the small changes to the accretion rate and bubble size when changing
the resolution, we see dramatic differences in the magnetic injection rate, as
shown in~\figref{\ref{fig:agn_survey_bubble_maginj}}. Even though the injected
energy is the same for each bubble event (\eqref{\ref{eq:agn_feedbackBub}}),
this energy is distributed over different bubble volumes. Hence, small changes
in the bubble radius lead to large differences in the injected magnetic field
strength. The smallest bubbles, which occur with $\Delta x=0.5$~kpc in model
B1A, have the largest injected field strength, with a $B_{\rm inj}$ for the
first bubble over a factor of two larger than that of model B2A, which has a
resolution of $1.0$~kpc. The trends continue down to $4$~kpc. However, at this
resolution the bubbles are already so large that at lower resolutions there is
little difference in the injected field strength.  For all these models the
injection rate steadily increases as gas cools and accretes onto the core,
eventually producing magnetic fields which are too strong and tangled (which
drives the timestep lower) to simulate further.

With the boosted-$\alpha$ models we again see large differences in $B_{\rm
inj}$. However, instead of the steadily increasing magnetic field strength of
the $\alpha=1$ models, small cyclic variations in the accretion rate and bubble
size lead to similar, but amplified, patterns in the rate of injected field
strength.  This is most evident in model B2B, where the injection rate changes
by a factor of $\sim 5$ with a period of roughly $1$~Gyr. 

\begin{figure*}
  \centering   
  {\includegraphics[width=\columnwidth]{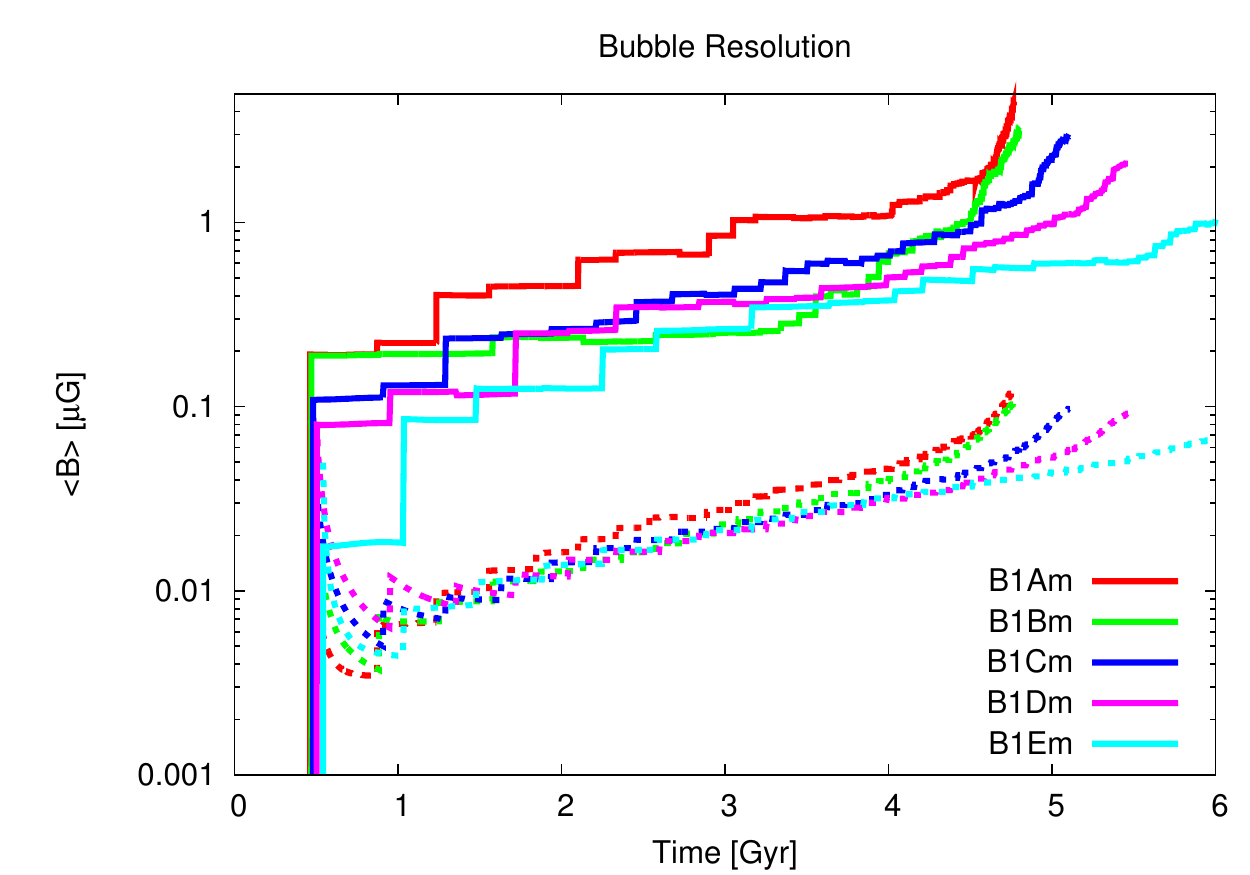}} 
  {\includegraphics[width=\columnwidth]{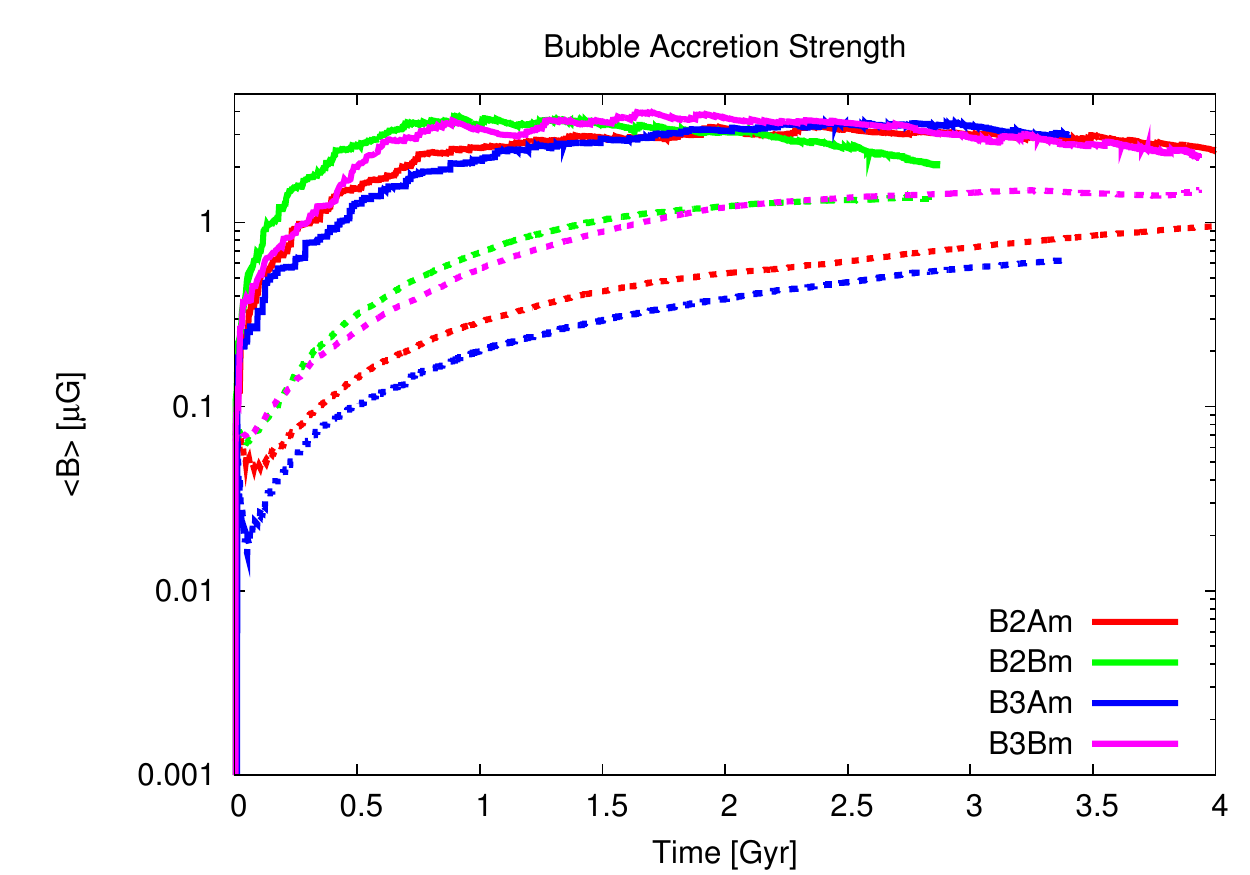}}
  \caption[Bubble model survey - average magnetic field strength.]
           {Affecting the average magnetic field strength 
           by varying the bubble model parameters. 
           Solid lines are density-weighted 
           average fields within $R_{\rm core}$ and dotted lines are 
           volume-weighted fields within $R_{\rm 200}$.}
\label{fig:agn_survey_bubble_aveb}
\end{figure*}

The smaller bubbles in the high-resolution runs are more effective at
magnetizing the cluster, as we show
in~\figref{\ref{fig:agn_survey_bubble_aveb}}.  Reduced numerical dissipation as
we increase resolution also plays a role; however, the changes in the accretion
rate due to resolution, and hence the resulting feedback energy, dominate over
any effects of numerical resistivity.  This figure displays the average
magnetic field strength within the core and within the entire cluster.  Model
B1A is able to generate $1 \mg$ fields within the core in $3$~Gyr, whereas
model B1D, with $16$~times lower resolution, takes twice as long. All of these
models produce the same general trends: steadily-increasing magnetic fields
both in the core and entire cluster from $1$ to $4$~Gyr followed by a sharp
rise as the accretion intensifies. 

We again see the strong dependence on accretion strength parameter: the
high-$\alpha$ runs generate $> 1 \mg$ fields within the core and $> 0.1 \mg$
fields averaged within the entire cluster in less than $1$~Gyr.  Indeed, models
B2B and B3B, with $\alpha=300$, are able to produce $\sim 1 \mg$ fields within
$R_{\rm 200}$.  The higher the value of $\alpha$, independent of resolution,
the greater the magnetization.  The time-dependent behavior of the average
field within the cores for these models looks more like the jets than the
fiducial bubble case.

\begin{figure*}
  \centering   
  {\includegraphics[width=\columnwidth]{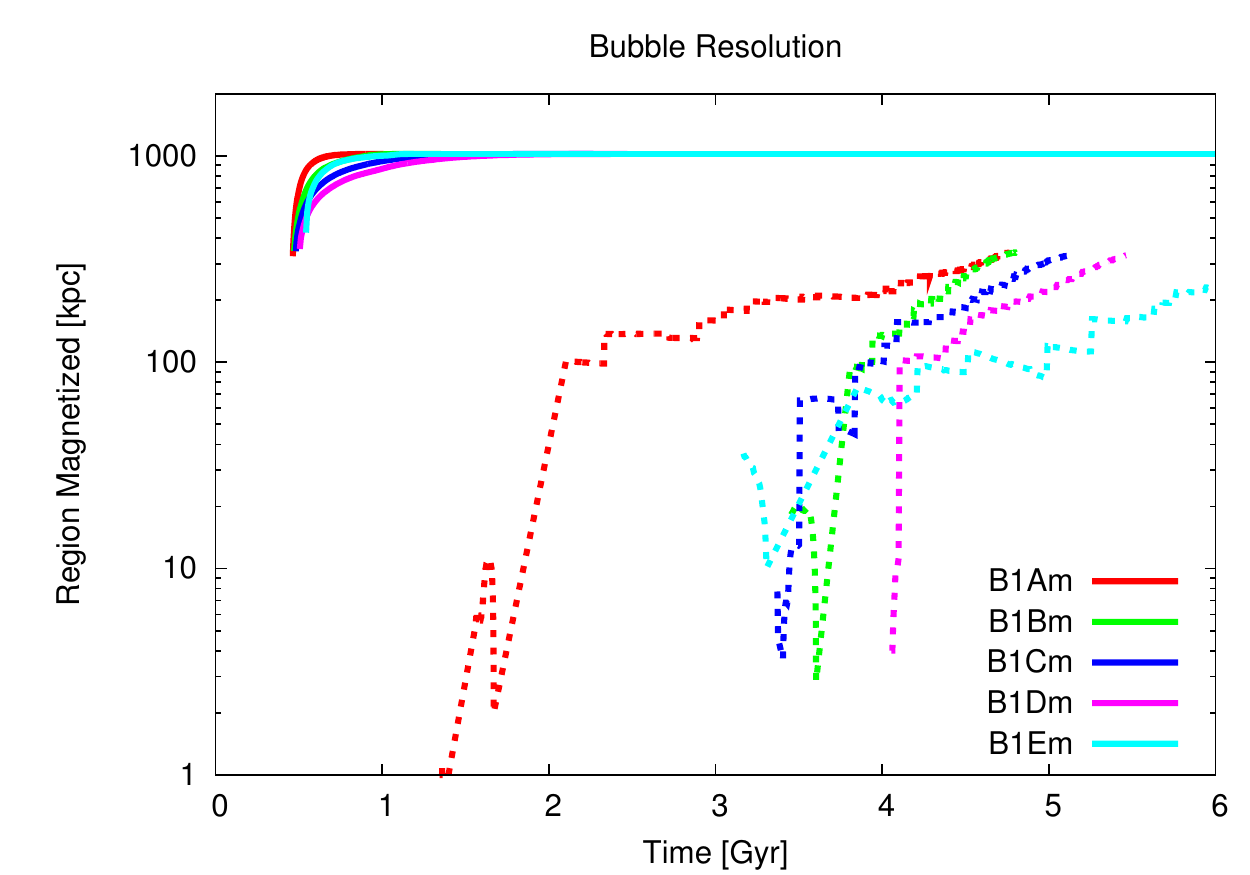}} 
  {\includegraphics[width=\columnwidth]{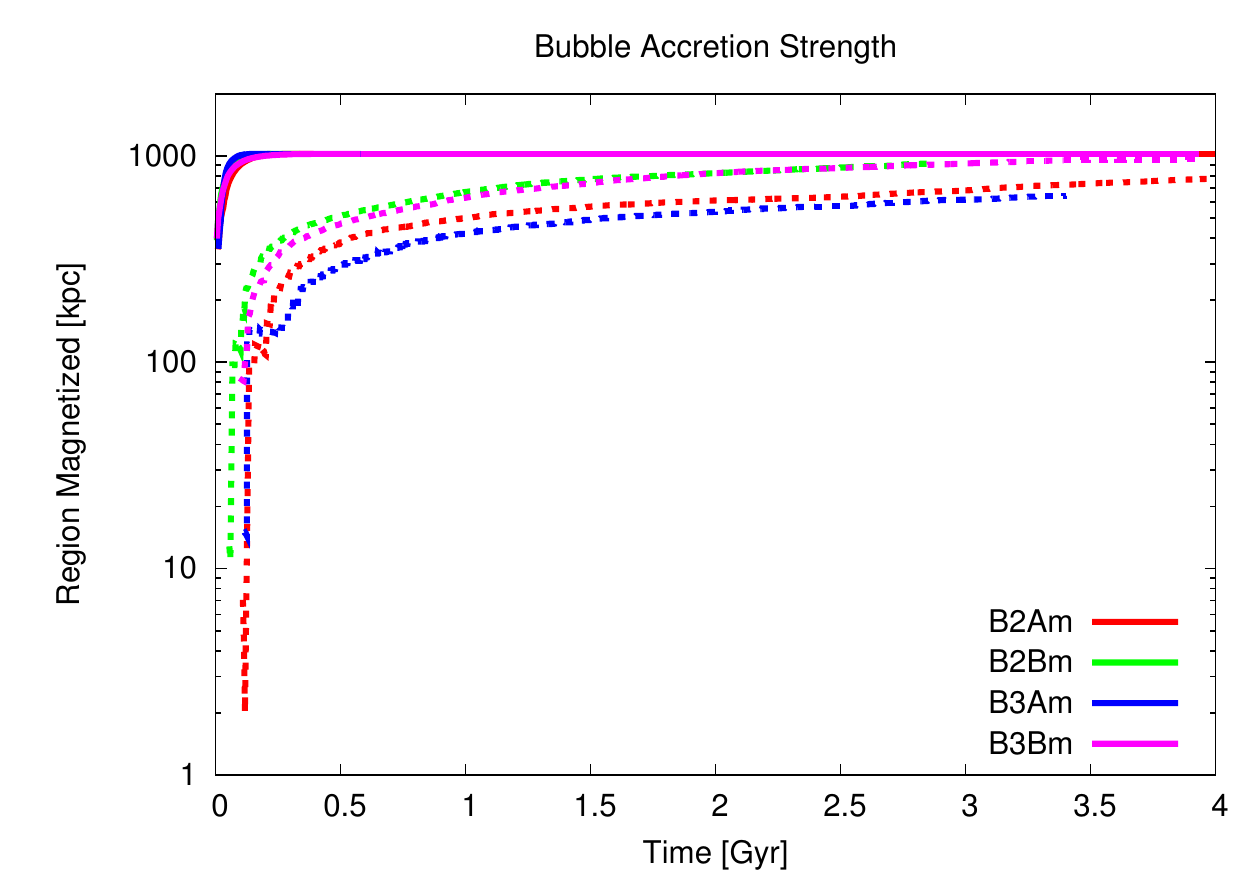}}
	\caption[Bubble model survey - magnetized volume.]
           {Affecting the volume magnetized by varying the bubble model 
           parameters. Shown is the cube root 
            of the total volume containing fields of strength at least 
            $10^{-12}$~G (solid lines) and $10^{-6}$~G (dotted lines).}
\label{fig:agn_survey_bubble_magvol}
\end{figure*}

In all bubble cases, regardless of changes to the resolution or accretion
strength, the cluster quickly magnetizes, as we see
in~\figref{\ref{fig:agn_survey_bubble_magvol}}.  However, the high-$\alpha$
models not only magnetize the cluster with weak fields, but also push strong
($> 1 \mg$) fields almost to the cluster edge. The $\alpha=1$ cases, however,
are much less effective at generating these strong fields, with most models
only creating these fields after $3$~Gyr.  These models are also only able to
push these fields to a meager $\sim 200$~kpc radius.

\begin{figure*}
  \centering  
  {\includegraphics[width=\columnwidth]{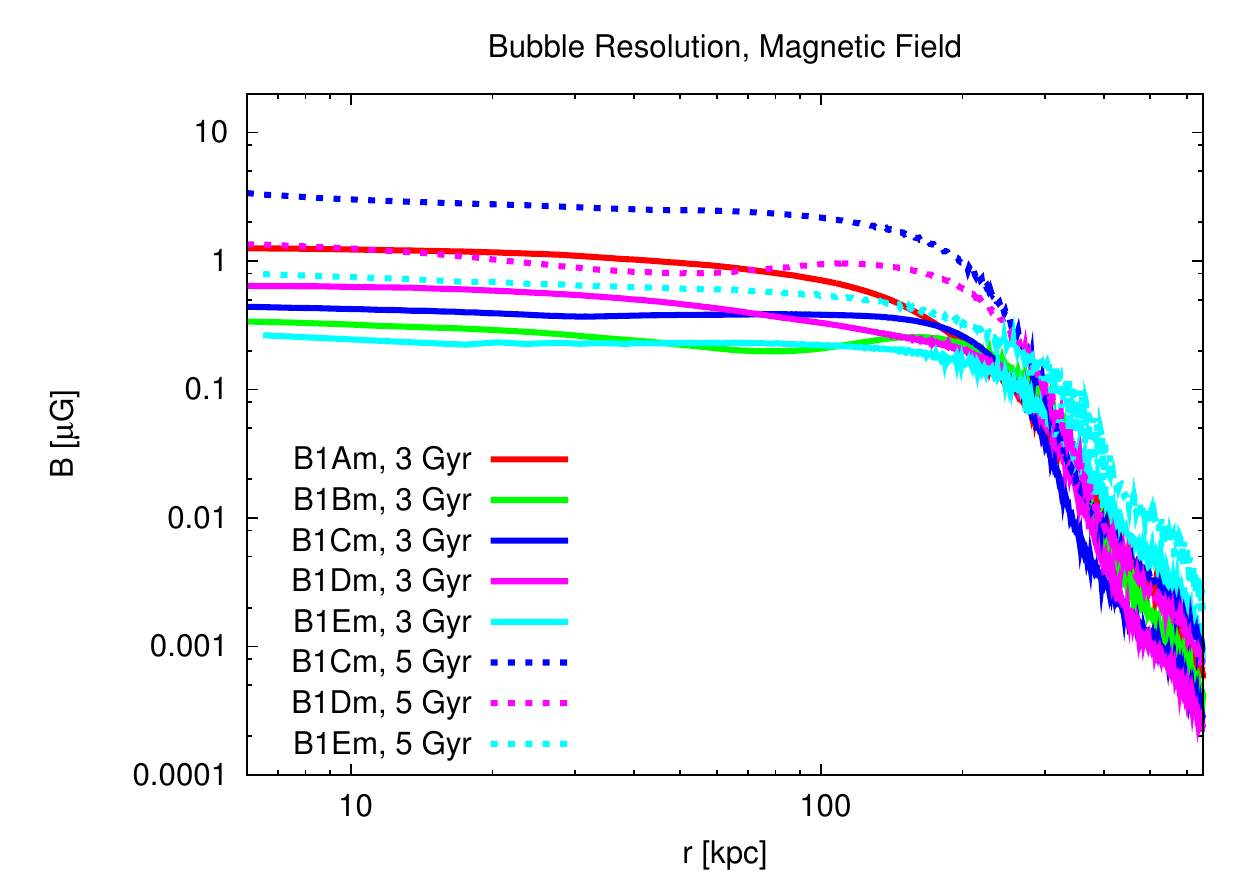}} 
  {\includegraphics[width=\columnwidth]{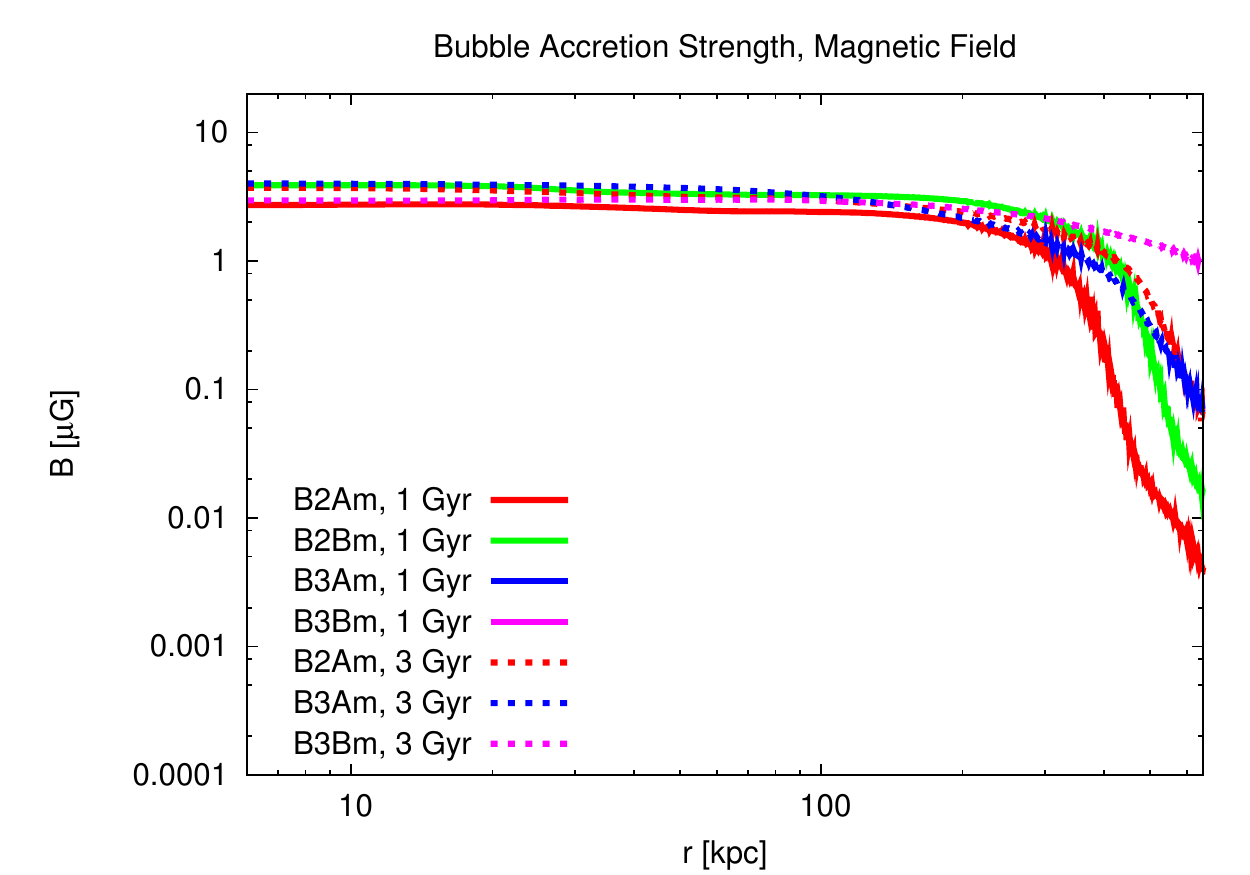}}
  \caption[Bubble model survey - magnetic field radial profiles.]
           {Affecting the magnetic profiles 
            by varying the bubble model parameters. 
            Shown are volume-weighted magnetic fields as a function 
           of radius.}
\label{fig:agn_survey_bubble_profile_mag}
\end{figure*}

This is also reflected in the radial profiles of the magnetic fields, which we
show in ~\figref{\ref{fig:agn_survey_bubble_profile_mag}}. All the
high-$\alpha$ models create a region of strong fields out to $\sim 400$~kpc
within only $1$~Gyr.  This is due to the strength and frequency of these
bubbles. After another $2$~Gyr of evolution the fields in this same region have
doubled in strength.  On the other hand, varying the resolution does not
produce significant differences until $5$~Gyr of evolution.  In all these
cases, the fields remain relatively constant out to a radius of $200$~kpc
before steeply falling, and changes to the resolution to not affect this
behavior. While higher resolution does result in stronger fields, a factor of
$16$ difference in resolution only produces a factor of $2$ difference in the
resulting magnetic fields, even after billions of years of evolution.

\section{Discussion \& Conclusion}
\label{sec:agn_conclusion}

We have conducted a systematic study of the growth and evolution of magnetic
fields due to the self-regulated feedback of a central active galactic nuclei
within an isolated mock cluster. This study is the first to link the energy
available for injected magnetic fields to an accretion rate measured on the
simulation mesh. We have implemented magnetic field injection using
representative models of continuous small-scale jets and sporadic large-scale
bubbles.  We have examined the effects of magnetizing the jet and bubble
outflows on the accretion properties of the black hole and the thermodynamic
quantities of the surrounding cluster gas.  We have compared the resulting
magnetization of the cluster using a fiducial jet and bubble model in terms of
the growth, evolution, and topological structure of the magnetic fields.
Finally, we have performed a parameter survey of the jet and bubble models to
determine the relationship between subgrid model parameter choices and the
resulting magnetic outflow properties.

The relative effects of adding magnetic fields into the injection depend on the
choice of subgrid model. Continuous jets, where the feedback is applied to the
central few zones, suffer a large decrease in the accretion rate due to the
tension forces inherent in the injected magnetic fields. Fixed-position
bubbles, with large radii, take roughly $3$~Gyr for noticeable differences to
emerge, while randomly-placed bubbles are almost entirely unaffected. While it
is thought that the evolution of bubbles is dominated by the presence of
magnetic fields, we do not have enough spatial or temporal resolution to fully
study the evolution of a single bubble. In general, strong magnetic fields
placed near the central black hole are very effective at pushing gas away from
the core and preventing cooled gas from accreting.

All models quickly permeate the entire cluster with fields of strength greater
than $10^{-12}$~G and produce strong fields ($> 10^{-6}$~G) in the central
core. While jets produce slightly stronger fields in the innermost regions of
the cluster, the bubbles more quickly disperse relatively strong fields out to
larger radii. Additionally, the bubble models produce flat radial profiles out
to $\sim 300$~kpc, beyond which the field strength drops rapidly. Although the
jets produce steeper radial profiles, they can take up to $4$~Gyr to
sufficiently magnetize the inner regions of the cluster.

The jets and fixed bubbles maintain regular torus-like shapes in the magnetic
field that persist over several billion years with strong fields concentrated
in the core, although eventually the fixed bubbles begin to disperse.  While we
did not explicitly calculate correlation lengths, all of our models produce
fields lines which are not sufficiently tangled.  Thus, if we are to consider
the jet or fixed bubble subgrid models as realistic approaches, we must require
some external source of turbulence to tangle the generated magnetic fields.
For example, galaxy wakes, which can amplify seed fields in less than
$5$~Gyr~\citep{Subramanian2006}, can easily tangle the magnetic fields produced
even at early times.  Also, observations suggest that jets can
precess~\citep[e.g,][]{Storchi1997,Caproni2004}, which would naturally lead to
more tangled fields.  The randomly-placed bubbles immediately produce
large-scale highly tangled magnetic fields out to relatively large radii.
However, the bubbles have difficulties generating magnetic fields with strength
greater than a microGauss, something that is relatively easily to accomplish
with the fixed bubbles and especially the jets.

We found the jet models to be relatively insensitive to many changes in the
underlying model. Since the jets require high resolution in the first place
(i.e., $< 1.0$~kpc) further improvements to the resolution leave the jet
outflow properties unchanged. We found that it is necessary to keep the
accretion radius smaller than the injection region size, but changes to the
depletion radius are less important since the accretion rate is so low. We see
the biggest differences in the jet outflows when changing jet size.  On the
other hand, changing the accretion strength parameter has little effect, since
stronger jet outflows simply push more gas away from the black hole, reducing
the measured accretion rate to nearly its value in the fiducial case.

The bubble models are much more sensitive to changes in the accretion strength
parameter, since the injection region is much larger. Even though increasing
$\alpha$ is generally used to compensate for lower resolution, we found that
the combination of lower resolution with enhanced $\alpha$ do \emph{not} mimic
the behavior of a higher resolution simulation with lower accretion strength
parameter. Higher values of $\alpha$ lead to very large, rapidly-forming
bubbles. Thus, we advise caution when choosing a value for $\alpha$, especially
when considering magnetic feedback subgrid models. A value of $\alpha$ chosen
to match, for example, the observed $\msig$ relation may produce unrealistic
bubbles.  Furthermore, for this particular cluster, which has a relatively flat
inner density profile, we only began to see significant differences when
lowering the resolution past $\sim 4$~kpc.  Mature cool-core clusters, more
typically studied in isolated simulations, have steeper inner radial density
profiles, and thus we expect to see a stronger dependence on $\alpha$. However,
this cluster represents a younger cluster, before it has developed a cool core,
and better represents a typical cluster when AGN activity first begins seeding
magnetic fields in its atmosphere.

While the jets appear to be more robust, they have significant difficulties in
quickly and efficiently magnetizing the cluster.  They must require some
additional mechanisms operating in the intracluster medium to spread and tangle
the injected fields.  On the other hand, the randomly-placed bubbles, which
automatically generate rich, tangled magnetic field morphologies, are more
sensitive to subgrid model choices. Additionally, our fiducial case - with
parameters taken directly from~\citet{Sijacki2007} - produces very large
bubbles very rapidly, which suggests that this model is not appropriate for
general cluster atmospheres.  Since the fixed bubbles have a predetermined
radius they do not face this problem. While they also produce somewhat ordered
fields, they spread these fields out to larger volumes, which makes these
fields more susceptible to tangling from cluster mergers and galaxy motions.
They can also strongly magnetize the cluster very quickly - in less than
$1$~Gyr.  

While we have performed a relatively exhaustive analysis, many questions
remain.  The diversity of the model behavior exhibited under our analysis
suggests significant theoretical uncertainty in the effects of AGN-based
magnetic fields - we must carefully select subgrid models such that they
reproduce cluster observables as well as faithfully represent the unresolved
scales and physics.  To achieve this, these magnetic field models must be
tested in more realistic clusters to determine their ability to sufficiently
magnetize the clusters absent any assumptions about the strength of the
injected magnetic fields.  Secondly, we must use these resimulated clusters to
more fully evaluate these magnetized outflows in terms of observables, such as
AGN duty cycle and cluster X-ray surface brightness profiles.  We must study
more topologies of magnetic injection, such as random fields, to determine if
the resulting magnetic outflows are robust against changes to the injected
field structure.  Finally, we must perform fully cosmological simulations to
ensure that the given AGN subgrid models, when injected magnetic fields are
included, reproduce key observables, such as the relationship between
supermassive black hole mass and host galaxy bulge velocity dispersion.  Only
then can we be confident that our magnetic subgrid models accurately represent
realistic aspects of AGN evolution, and use them to evaluate the ability of
active galactic nuclei to generate cluster-wide magnetic fields.

\section*{Acknowledgments} The authors acknowledge support under a DOE
Computational Science Graduate Fellowship (DE-FG02-97ER25308) and a
Presidential Early Career Award from the U.S. Department of Energy, Lawrence
Livermore National Laboratory (contract B532720).  Computing resources were
supplied by an allocation provided by the National Science Foundation
(TG-AST040034N) on Kraken at the National Institute for Computational Sciences,
as well as the Oak Ridge Leadership Computing Facility at the Oak Ridge
National Laboratory, which is supported by the Office of Science of the U.S.
Department of Energy under Contract No.  DE-AC05-00OR22725.  The software used
in this work was in part developed by the DOE-supported Flash Center for
Computational Science at the University of Chicago.  Visualizations were
created using the DOE-supported VisIt program.

\bibliography{ms}		

\begin{thebibliography}{69}
\expandafter\ifx\csname natexlab\endcsname\relax\def\natexlab#1{#1}\fi

\bibitem[{Allen {et~al.}(2006)Allen, Dunn, Fabian, Taylor, \&
  Reynolds}]{Allen2006}
Allen, S.~W., Dunn, R. J.~H., Fabian, A.~C., Taylor, G.~B., \& Reynolds, C.~S.
  2006, MNRAS, 372, 21

\bibitem[{{Balbus}(2000)}]{Balbus2000}
{Balbus}, S.~A. 2000, ApJ, 534, 420

\bibitem[{{Balbus}(2001)}]{Balbus2001}
---. 2001, ApJ, 562, 909

\bibitem[{Bamba {et~al.}(2008)Bamba, Ohta, \& Tsujikawa}]{Bamba2008}
Bamba, K., Ohta, N., \& Tsujikawa, S. 2008, Phys. Rev. D, 78

\bibitem[{Bandara {et~al.}(2009)Bandara, Crampton, \&
  Simard}]{bandara_relationship_2009}
Bandara, K., Crampton, D., \& Simard, L. 2009, ApJ, 704, 1135

\bibitem[{Battaglia {et~al.}(2010)Battaglia, Bond, Pfrommer, Sievers, \&
  Sijacki}]{Battaglia2010}
Battaglia, N., Bond, J.~R., Pfrommer, C., Sievers, J.~L., \& Sijacki, D. 2010,
  ApJ, 725, 91

\bibitem[{{Battefeld} {et~al.}(2008){Battefeld}, {Battefeld}, {Wesley}, \&
  {Wyman}}]{Battefeld2008}
{Battefeld}, D., {Battefeld}, T., {Wesley}, D.~H., \& {Wyman}, M. 2008, JCAP,
  2, 1

\bibitem[{Baym {et~al.}(1996)Baym, B\"{o}deker, \& McLerran}]{Baym1996}
Baym, G., B\"{o}deker, D., \& McLerran, L. 1996, Phys. Rev. D, 53, 662

\bibitem[{{Biermann}(1950)}]{BiermannL.1950}
{Biermann}, L. 1950, Zeitschrift Naturforschung Teil A, 5, 65

\bibitem[{{Bondi}(1952)}]{BondiH.1952}
{Bondi}, H. 1952, MNRAS, 112, 195

\bibitem[{Booth \& Schaye(2009)}]{booth_cosmological_2009}
Booth, C.~M. \& Schaye, J. 2009, MNRAS, 398, 53

\bibitem[{Brunetti {et~al.}(2007)Brunetti, Venturi, Dallacasa, Cassano, Dolag,
  Giacintucci, \& Setti}]{Brunetti2007}
Brunetti, G., Venturi, T., Dallacasa, D., Cassano, R., Dolag, K., Giacintucci,
  S., \& Setti, G. 2007, ApJ, 670, L5

\bibitem[{{Caproni} \& {Abraham}(2004)}]{Caproni2004}
{Caproni}, A. \& {Abraham}, Z. 2004, ApJ, 602, 625

\bibitem[{Carilli \& Taylor(2002)}]{Carilli2002a}
Carilli, C.~L. \& Taylor, G.~B. 2002, ARA\&A, 40, 319

\bibitem[{Cattaneo \& Teyssier(2007)}]{Cattaneo2007a}
Cattaneo, A. \& Teyssier, R. 2007, MNRAS, 376, 1547

\bibitem[{Chandran \& Maron(2004)}]{Chandran2004a}
Chandran, B. D.~G. \& Maron, J.~L. 2004, ApJ, 602, 170

\bibitem[{Colbert {et~al.}(1996)Colbert, Baum, Gallimore, O'Dea, \&
  Christensen}]{Colbert1996}
Colbert, E. J.~M., Baum, S.~A., Gallimore, J.~F., O'Dea, C.~P., \& Christensen,
  J.~A. 1996, ApJ, 467, 551

\bibitem[{{Colgate} \& {Li}(2000)}]{Colgate2000}
{Colgate}, S.~A. \& {Li}, H. 2000, 195, 255

\bibitem[{Contopoulos {et~al.}(2009)Contopoulos, Christodoulou, Kazanas, \&
  Gabuzda}]{Contopoulos2009a}
Contopoulos, I., Christodoulou, D.~M., Kazanas, D., \& Gabuzda, D.~C. 2009,
  ApJ, 702, L148

\bibitem[{Daly \& Loeb(1990)}]{Daly1990}
Daly, R.~A. \& Loeb, A. 1990, ApJ, 364, 451

\bibitem[{{Debuhr} {et~al.}(2010){Debuhr}, {Quataert}, {Ma}, \&
  {Hopkins}}]{DeBuhr2010}
{Debuhr}, J., {Quataert}, E., {Ma}, C.-P., \& {Hopkins}, P. 2010, MNRAS, 406,
  L55

\bibitem[{{Di Matteo} {et~al.}(2008){Di Matteo}, Colberg, Springel, Hernquist,
  \& Sijacki}]{DiMatteo2008}
{Di Matteo}, T., Colberg, J., Springel, V., Hernquist, L., \& Sijacki, D. 2008,
  ApJ, 676, 33

\bibitem[{{Dolag} \& {Schindler}(2000)}]{Dolag2000}
{Dolag}, K. \& {Schindler}, S. 2000, A\& AP, 364, 491

\bibitem[{Dubey {et~al.}(2008)Dubey, Reid, \& Fisher}]{Dubey2008}
Dubey, A., Reid, L.~B., \& Fisher, R. 2008, Physica Scripta, T132, 014046

\bibitem[{{Dubois} {et~al.}(2009){Dubois}, {Devriendt}, {Slyz}, \&
  {Silk}}]{Dubois2008a}
{Dubois}, Y., {Devriendt}, J., {Slyz}, A., \& {Silk}, J. 2009, MNRAS, 399, L49

\bibitem[{Dubois {et~al.}(2010)Dubois, Devriendt, Slyz, \&
  Teyssier}]{Dubois2010}
Dubois, Y., Devriendt, J., Slyz, A., \& Teyssier, R. 2010, MNRAS, 409, 985

\bibitem[{{Dubois} {et~al.}(2011){Dubois}, {Devriendt}, {Slyz}, \&
  {Teyssier}}]{Dubois2011}
{Dubois}, Y., {Devriendt}, J., {Slyz}, A., \& {Teyssier}, R. 2011, ArXiv
  e-prints

\bibitem[{{Dursi} \& {Pfrommer}(2008)}]{Dursi2008}
{Dursi}, L.~J. \& {Pfrommer}, C. 2008, ApJ, 677, 993

\bibitem[{Falceta-Gon\c{c}alves
  {et~al.}(2010{\natexlab{a}})Falceta-Gon\c{c}alves, Caproni, Abraham,
  Teixeira, \& {de Gouveia Dal Pino}}]{Falceta-Goncalves2010b}
Falceta-Gon\c{c}alves, D., Caproni, A., Abraham, Z., Teixeira, D.~M., \& {de
  Gouveia Dal Pino}, E.~M. 2010{\natexlab{a}}, ApJ, 713, L74

\bibitem[{Falceta-Gon\c{c}alves
  {et~al.}(2010{\natexlab{b}})Falceta-Gon\c{c}alves, {de Gouveia Dal Pino},
  Gallagher, \& Lazarian}]{Falceta-Goncalves2010}
Falceta-Gon\c{c}alves, D., {de Gouveia Dal Pino}, E.~M., Gallagher, J.~S., \&
  Lazarian, A. 2010{\natexlab{b}}, ApJ, 708, L57

\bibitem[{Fan(2006)}]{Fan2006}
Fan, X. 2006, New Astronomy Reviews, 50, 665

\bibitem[{Fender {et~al.}(1999)}]{Fender1999}
Fender, R. {et~al.} 1999, ApJ, 519, L165

\bibitem[{Fryxell {et~al.}(2000)}]{Fryxell2000b}
Fryxell, B. {et~al.} 2000, ApJs, 131, 273

\bibitem[{Gallo {et~al.}(2003)Gallo, Fender, \& Pooley}]{Gallo2003}
Gallo, E., Fender, R.~P., \& Pooley, G.~G. 2003, MNRAS, 344, 60

\bibitem[{Gardini(2007)}]{Gardini2007}
Gardini, A. 2007, A\&A, 464, 143

\bibitem[{Gaspari {et~al.}(2011)Gaspari, Melioli, Brighenti, \&
  D'Ercole}]{Gaspari2011}
Gaspari, M., Melioli, C., Brighenti, F., \& D'Ercole, A. 2011, MNRAS, 411, 349

\bibitem[{Gourgouliatos {et~al.}(2010)Gourgouliatos, Braithwaite, \&
  Lyutikov}]{Gourgouliatos2010b}
Gourgouliatos, K.~N., Braithwaite, J., \& Lyutikov, M. 2010, MNRAS, 409, 1660

\bibitem[{Hopkins {et~al.}(2006)Hopkins, Narayan, \& Hernquist}]{Hopkins2006}
Hopkins, P.~F., Narayan, R., \& Hernquist, L. 2006, ApJ, 643, 641

\bibitem[{Jones \& {De Young}(2005)}]{Jones2005}
Jones, T.~W. \& {De Young}, D.~S. 2005, ApJ, 624, 586

\bibitem[{{Kirkpatrick} {et~al.}(2011){Kirkpatrick}, {McNamara}, \&
  {Cavagnolo}}]{Kirkpatrick2011}
{Kirkpatrick}, C.~C., {McNamara}, B.~R., \& {Cavagnolo}, K.~W. 2011, ApJ, 731,
  L23+

\bibitem[{Koide {et~al.}(1999)Koide, Shibata, \& Kudoh}]{Koide1999}
Koide, S., Shibata, K., \& Kudoh, T. 1999, ApJ, 522, 727

\bibitem[{Lesch \& Birk(1998)}]{Lesch1998}
Lesch, H. \& Birk, G.~T. 1998, Physics of Plasmas, 5, 2773

\bibitem[{Li {et~al.}(2006)Li, Lapenta, Finn, Li, \& Colgate}]{Li2006a}
Li, H., Lapenta, G., Finn, J.~M., Li, S., \& Colgate, S.~A. 2006, ApJ, 643, 92

\bibitem[{McNamara \& Nulsen(2007)}]{McNamara2007}
McNamara, B. \& Nulsen, P. 2007, ARA\&A, 45, 117

\bibitem[{Miniati {et~al.}(2001)Miniati, Jones, Kang, \& Ryu}]{Miniati2001}
Miniati, F., Jones, T.~W., Kang, H., \& Ryu, D. 2001, ApJ, 562, 233

\bibitem[{Morsony {et~al.}(2010)Morsony, Heinz, Br\"{u}ggen, \&
  Ruszkowski}]{Morsony2010}
Morsony, B.~J., Heinz, S., Br\"{u}ggen, M., \& Ruszkowski, M. 2010, MNRAS, 407,
  1277

\bibitem[{Narayan \& Medvedev(2001)}]{Narayan2001a}
Narayan, R. \& Medvedev, M.~V. 2001, ApJ, 562, L129

\bibitem[{Navarro {et~al.}(1996)Navarro, Frenk, \& White}]{Navarro1996}
Navarro, J.~F., Frenk, C.~S., \& White, S. D.~M. 1996, ApJ, 462, 563

\bibitem[{{O'Neill} \& {Jones}(2010)}]{ONeill2010}
{O'Neill}, S.~M. \& {Jones}, T.~W. 2010, ApJ, 710, 180

\bibitem[{Parrish {et~al.}(2009)Parrish, Quataert, \& Sharma}]{Parrish2009a}
Parrish, I.~J., Quataert, E., \& Sharma, P. 2009, ApJ, 703, 96

\bibitem[{Peterson \& Fabian(2006)}]{Peterson2006}
Peterson, J. \& Fabian, A. 2006, Physics Reports, 427, 1

\bibitem[{Pfrommer {et~al.}(2007)Pfrommer, En{\ss}lin, Springel, Jubelgas, \&
  Dolag}]{Pfrommer2007}
Pfrommer, C., En{\ss}lin, T.~A., Springel, V., Jubelgas, M., \& Dolag, K. 2007,
  MNRAS, 378, 385

\bibitem[{Pope(2007)}]{Pope2007}
Pope, E. C.~D. 2007, MNRAS, 381, 741

\bibitem[{Robinson {et~al.}(2004)}]{Robinson2004a}
Robinson, K. {et~al.} 2004, ApJ, 601, 621

\bibitem[{Ruszkowski {et~al.}(2007)Ruszkowski, En{\ss}lin, Br\"{u}ggen, Heinz,
  \& Pfrommer}]{Ruszkowski2007}
Ruszkowski, M., En{\ss}lin, T.~A., Br\"{u}ggen, M., Heinz, S., \& Pfrommer, C.
  2007, MNRAS, 378, 662

\bibitem[{Shukurov {et~al.}(2006)Shukurov, Subramanian, \&
  Haugen}]{Shukurov2006}
Shukurov, A., Subramanian, K., \& Haugen, N. E.~L. 2006, Astronomische
  Nachrichten, 327, 583

\bibitem[{Sijacki {et~al.}(2007)Sijacki, Springel, {Di Matteo}, \&
  Hernquist}]{Sijacki2007}
Sijacki, D., Springel, V., {Di Matteo}, T., \& Hernquist, L. 2007, MNRAS, 380,
  877

\bibitem[{Skillman {et~al.}(2008)Skillman, O'Shea, Hallman, Burns, \&
  Norman}]{Skillman2008}
Skillman, S.~W., O'Shea, B.~W., Hallman, E.~J., Burns, J.~O., \& Norman, M.~L.
  2008, ApJ, 689, 1063

\bibitem[{Sternberg {et~al.}(2007)Sternberg, Pizzolato, \&
  Soker}]{Sternberg2007}
Sternberg, A., Pizzolato, F., \& Soker, N. 2007, ApJ, 656, L5

\bibitem[{{Sternberg} \& {Soker}(2008)}]{Sternberg2008}
{Sternberg}, A. \& {Soker}, N. 2008, MNRAS, 389, L13

\bibitem[{{Storchi-Bergmann} {et~al.}(1997){Storchi-Bergmann}, {Eracleous},
  {Ruiz}, {Livio}, {Wilson}, \& {Filippenko}}]{Storchi1997}
{Storchi-Bergmann}, T., {Eracleous}, M., {Ruiz}, M.~T., {Livio}, M., {Wilson},
  A.~S., \& {Filippenko}, A.~V. 1997, ApJ, 489, 87

\bibitem[{Subramanian {et~al.}(2006)Subramanian, Shukurov, \&
  Haugen}]{Subramanian2006}
Subramanian, K., Shukurov, A., \& Haugen, N. E.~L. 2006, MNRAS, 366, 1437

\bibitem[{Sutherland \& Dopita(1993)}]{Sutherland1993}
Sutherland, R.~S. \& Dopita, M.~A. 1993, ApJS, 88, 253

\bibitem[{Sutter \& Ricker(2010)}]{Sutter2010}
Sutter, P.~M. \& Ricker, P.~M. 2010, ApJ, 723, 1308

\bibitem[{Voit \& Donahue(2005)}]{Voit2005a}
Voit, G.~M. \& Donahue, M. 2005, ApJ, 634, 955

\bibitem[{Widrow(2002)}]{Widrow2002a}
Widrow, L. 2002, Reviews of Modern Physics, 74, 775

\bibitem[{Xu {et~al.}(2008)Xu, Li, Collins, Li, \& Norman}]{Xu2008b}
Xu, H., Li, H., Collins, D., Li, S., \& Norman, M.~L. 2008, ApJ, 681, L61

\bibitem[{Xu {et~al.}(2010)Xu, Li, Collins, Li, \& Norman}]{Xu2010a}
Xu, H., Li, H., Collins, D.~C., Li, S., \& Norman, M.~L. 2010, ApJ, 725, 2152

\bibitem[{Yang {et~al.}(2011)Yang, Sutter, \& Ricker}]{Yang2011}
Yang, H.-Y., Sutter, P., \& Ricker, P. 2011, ApJ (submitted)

\end{thebibliography}
\bibliographystyle{apj}	
\nocite{*}

\end{document}